\documentclass[twocolumn,traditabstract]{aa}
\usepackage{natbib}
\bibpunct{(}{)}{;}{a}{}{,}
\usepackage[english]{babel}
\usepackage{algorithm}
\usepackage{algorithmic}
\usepackage{theorem}
\usepackage{amssymb}
\usepackage{amsmath}
\usepackage{graphics}
\usepackage{subfigure}
\usepackage{url}
\usepackage{hyperref}
\usepackage{epsfig}
\usepackage{fullpage}
\graphicspath{./figures/} 

\DeclareMathOperator*{\argmin}{argmin}

\begin{document}

\title{Astronomical Image Denoising Using Dictionary Learning}

\author{S. Beckouche 
\and J.L. Starck
\and J. Fadili
}
\maketitle

\abstract{
Astronomical images suffer a constant presence of multiple defects that are consequences of the intrinsic properties of the acquisition equipments, and atmospheric conditions. One of the most frequent defects in astronomical imaging is the presence of  additive noise which makes a denoising step mandatory before processing data. During the last decade, a particular modeling scheme, based on sparse representations, has drawn the attention of an ever growing community of researchers. Sparse representations offer a promising framework to many image and signal processing tasks, especially denoising and restoration applications. At first, the harmonics, wavelets, and similar bases and overcomplete representations have been considered as candidate domains to seek the sparsest representation. A new generation of algorithms, based on data-driven dictionaries, evolved rapidly and compete now with the off-the-shelf fixed dictionaries. While designing a dictionary beforehand leans on a guess of the most appropriate representative elementary forms and functions, the dictionary learning framework offers to construct the dictionary upon the data themselves, which provides us with a more flexible setup to sparse modeling and allows to build more sophisticated dictionaries.  In this paper, we introduce the Centered Dictionary Learning (CDL) method and we study its performances for astronomical image denoising. We show how CDL outperforms wavelet or classic dictionary learning denoising techniques on astronomical images, and we give a comparison of the effect of these different algorithms on the photometry of the denoised images. }




\section{Introduction}

\subsection{Overview of sparsity in astronomy}
The wavelet transform (WT) has been extensively used in astronomical data analysis during the last ten years, and this holds for
all astrophysical domains,  from study of the sun through Cosmic Microwave Background (CMB) analysis \citep{starck:book06}. 
X-ray and Gamma-ray sources  catalogs are generally based on wavelets  \citep{xmmlss,fermi2012_catalog}. 
Using multiscale approaches such as the wavelet transform, an image can be decomposed into components at different scales, and the wavelet transform is therefore well-adapted to the study of astronomical data \citep{starck:book06}. Furthermore, since noise in physical sciences is not always Gaussian, modeling in wavelet space of many kinds of noise such as Poisson noise has been a key motivation for the use of wavelets in astrophysics \citep{starck:schmitt10}.
If wavelets represent well isotropic features, they are far from optimal for analyzing anisotropic objects such as filaments, jets, etc.. This has motivated the construction of a collection of basis functions generating possibly overcomplete dictionaries, eg. cosine, wavelets, curvelets  \citep{starck:sta02_3}. 
More generally, we assume that the data $X$ is a superposition of atoms from a  {\bf dictionary} $D$ such that $X=D\alpha$, where $\alpha$ are the synthesis coefficients of $X$ from $D$. 
The best data decomposition is the one which leads to the sparsest representation, i.e. few coefficients have a
large magnitude, while most of them are close to zero \citep{starck:book10}. Hence,
for some astronomical data sets containing edges (planetary images, cosmic strings, etc.), curvelets should be preferred to wavelets. 
But for a signal composed of a  sine, the Fourier dictionary is optimal from a sparsity standpoint since all information is contained in a single coefficient. 
Hence, the representation space that we use in our analysis can be seen as a prior we have on our observations. The larger the dictionary is, the better data analysis will be, but also the larger the computation time to derive the coefficients $\alpha$ in the dictionary will be. For some specific dictionaries limited to a given set of functions (Fourier, wavelet, etc.) we however
have very fast implicit operators allowing us to compute the coefficients with a complexity of ${\cal O}(N\log N)$, which makes these dictionaries very attractive.
But what can we do if our data are not well represented by these fixed existing dictionaries ? Or if we do not know the morphology of features contained in our data ?
Is there a way to optimize our data analysis by constructing a dedicated dictionary ?
To answer these questions, a new field has recently emerged, called {\bf Dictionary Learning} (DL). Dictionary learning techniques offer to learn an adaptive dictionary  $D$ directly from the data (or from a set of exemplars that we 
believe to well represent the data). DL is at the interface of machine learning, optimization and harmonic analysis. 
\subsection{Contributions}
In this paper, we show how classic dictionary learning for denoising behaves with astronomical images. We introduce a new variant, the Centered Dictionary Learning (CDL), developed to process more efficiently point-like features that are extremely common in astronomical images. Finally, we perform a study comparing how wavelet and dictionary learning denoising methods behave regarding sources photometry, showing that dictionary learning is better at preserving the sources flux. 
\subsection{Paper organization}
This paper is organized as follows. Section 2 presents the sparsity regularization problem where we introduce notations and the paradigm of dictionary for sparse coding. We introduce in Section 3 the methods for denoising by dictionary learning and we introduce the CDL technique. We give in Section 4 our results on astronomical images and we conclude with some perspectives in Section 5.

\section{Sparsity regularization}
\subsection{Notations}
We use the following notations. Uppercase letters are used for matrices notation and lowercase for vectors. Matrices are written column-wise $ D=[d_1,\ldots ,d_m] \in \mathbb{R}^{n \times m}$. If $D$ is a dictionary matrix, the columns $d_i \in \mathbb{R}^n$ represent the atoms of the dictionary. We define the  $\ell_p$ pseudo-norm ($p > 0$) of a vector $x \in \mathbb{R}^n$ as  $\left\| x \right\|_p=  \left (\sum_{i=1}^{n} | x_i | ^p\right )^{1/p}$. As an extension, the $\ell_{\infty}$ norm is defined as $ \left\| x \right\|_{\infty} = \max_{1 \leq i \leq n} \left\lbrace x_i \right\rbrace$, and the pseudo-norm  $\ell_0 $ stands for the number of non-zero entries of a vector: $ \left\| x \right\|_0 = \# \left\lbrace i, x_i \neq 0 \right\rbrace$. Given an image $Y$ of $Q\times Q$ pixels, a patch size $n = \tau \times \tau$ and an overlapping factor $\Delta \in [1,\dots,n]$, we denote by $R_{(i_1,i_2)}(Y)$ the patch extracted from $Y$ at the central position $i = (i_1,i_2)\in [0,\dots,Q/\Delta]^2$ and converts it into a vector of $\mathbb{R}^n$, such that $\forall j_1,j_2 \in [-\tau/2,\dots,\tau/2]$, 
\begin{equation}
Y(i_1\Delta+j_1,i_2\Delta+j_2). = R_i(Y)[\tau j_1+j2],
\end{equation}
which corresponds to stacking the extracted square patch into a column vector. 
Given a patch $R_{i,j}(Y)\in\mathbb{R}^n$, we define the centering operator $C_{i,j}$ as the translation operator 
\begin{equation}
C_{i,j}R_{i,j}[l] = \left\{ 
  \begin{array}{l l}
   R_{i,j}[l+\delta_{i,j}] &  \text{ if } 1\leq l \leq n-\delta_{i,j}\\
   R_{i,j}[l+\delta_{i,j}-n] &  \text{ if  } n-\delta_{i,j}< l \leq n\\
  \end{array} \right.
  \label{centering_operator}
\end{equation}
and $\delta_{i,j}$ is the smallest index verifying
\begin{equation}
\left\{ 
  \begin{array}{l l}
   C_{i,j}R_{i,j}[n/2] = \underset{l}{\max}\{R_{i,j}[l]\}\quad \text{if $n$ is even}\\
   C_{i,j}R_{i,j}[(n-1)/2] = \underset{l}{\max}\{R_{i,j}[l]\}\quad \text{if $n$ is odd}\\
  \end{array} \right.
\end{equation}
The centering operator translates the original vector values to place the maximum values in the central index position. In case where the original vector has more than one entry that reach its maximum values, the smallest index with this value is placed at the center in the translated vector. Finally, to compare two images $M_1,M_2$, we use the Peak Signal to Noise Ratio $\mathrm{PSNR} = 10\log_{10} \left( \frac{\max(M_1,M_2)^2}{\mathrm{MSE}(M_1,M_2)} \right)$, where $\mathrm{MSE}(M_1,M_2)$ is the mean square error of the two images and $\max(M_1,M_2)$ is the highest value contained in $M_1$ and $M_2$.
\subsection{Sparse recovery}
A signal $ \alpha  = \left[ \alpha_1, \ldots ,\alpha_n \right] $, is said to be sparse when most of its entries $\alpha_i $ are equal to zero. When the observations do not satisfy the sparsity prior in the direct domain, computing their representation coefficients in a given dictionary might yield a sparser representation of the data. Overcomplete dictionaries, which contain more atoms than their dimension and thus are redundant, when coupled with sparse coding framework, have shown in the last decade to lead to more significant and expressive representations, which help to better interpret and understand the observations \citep{Starck_Fadili_2009, Starck_Murtagh_Fadili_2010}. Sparse coding concentrates around two main axes: finding the appropriate dictionary, and computing the encodings given this dictionary. 

Sparse decomposition requires the summation of the relevant atoms with their appropriate weights. However, unlike a transform coder that comes with an inverse transform, finding such sparse codes within overcomplete dictionaries is non-trivial, in particular because the decomposition of a signal on an overcomplete dictionary is not unique. The combination of a dictionary representation with sparse modeling has first been introduced in the pioneering work of \citet{Mallat_Zhang_1993}, where the traditional wavelet transforms have been replaced by the more generic concept of dictionary for the first time. 

We use in this paper a sparse synthesis prior. Given an observation $x \in \mathbb{R}^n $, and a sparsifying dictionary $D \in \mathbb{R}^{n \times k}$, sparse decomposition refers to finding an encoding vector $\alpha \in \mathbb{R}^k$ that represents a given signal $x$ in the domain spanned by the dictionary $D$, while minimizing the number of elementary atoms involved in synthesizing it:
\begin{equation}
\hat{\alpha}\in \argmin_{\alpha} \left\| \alpha  \right\|_0 \text{  s.t.  } x=D\alpha ~.
\label{main_problem}
\end{equation}
When the original signal is to be reconstructed only approximately, the equality constrain is replaced by an $\ell_2$ norm inequality constrain: 
\begin{equation}
\hat{\alpha}\in \argmin_{\alpha} \left\| \alpha \right\|_0 \text{ s.t. } \left\| x-D\alpha \right\|_2 \leq \varepsilon ~,
\label{sparse_denoising}
\end{equation}
where $\varepsilon$ is a threshold controlling the misfitting between the observation $x$ and the recovered signal $\hat{x}=D\hat{\alpha}$. 

The sparse prior can also be used from an analysis point of view \citep{elad_analysis_synthesis}. In this case, the computation of the signal coefficient is simply obtained by the sparsifying dictionary and the problem becomes

\begin{equation}
\hat{y} \in \argmin_{y} \left\| D^*y  \right\|_0 \text{  s.t.  } y = x
\end{equation}
or 

\begin{equation}
\hat{y} \in \argmin_{y} \left\| D^*y  \right\|_0 \text{ s.t. } \left\|x-y \right\|_2 \leq \varepsilon
\end{equation}

whether the signal $x$ is contaminated by noise or not. This approach has been explored more recently than the synthesis model and has thus far yielded promising results \citep{rubinstein_ksvd_analysis}. We chose to use the synthesis model for our work because it offers more guarantees as it has been proved to be an efficient model in many different contexts.

Solving \eqref{sparse_denoising} proves to be conceptually Np-hard and numerically intractable. Nonetheless, heuristic methods called greedy algorithms were developed to approximate the sparse solution of the $\ell_0$ problem, while considerably reducing the resources requirements. The process of seeking a solution can be divided into two effective parts: finding the support of the solution and estimating the values of the entries over the selected support \citep{Mallat_Zhang_1993}.
Once the support of the solution is found, estimating the signal coefficients becomes a straightforward and easier problem since a simple least-squares application can often provide the optimal solution regarding the selected support. This class of algorithms includes matching pursuit (MP), orthogonal matching pursuit (OMP), gradient pursuit (GP) and their variants. 

A popular alternative to the problem \eqref{sparse_denoising} is to use the $\ell_1-$norm instead of the $\ell_0$ to promote a sparse solution. Using the $\ell_1$ norm as a sparsity prior results on a convex optimization problem (Basis Pursuit De-Noising or Lasso) that is that is computationally tractable, finding
\begin{equation}
\hat{\alpha}\in \argmin_{\alpha} \left\| \alpha \right\|_1 \text{ s.t. }  \left\| x-D\alpha \right\|_2 \leq \varepsilon.
\label{l1_constrained}
\end{equation}
The optimization problem \eqref{l1_constrained} can also be written in its unconstrained penalized form:
\begin{equation}
\hat{\alpha}\in \argmin_{\alpha} \left\| x-D\alpha \right\|_2^2 + \lambda \left\| \alpha \right\|_1 
\label{lasso}
\end{equation}

where $\lambda$ is a Lagrange multiplier, controlling the sparsity of the solution \citep{Chen_lasso}. The larger $\lambda$ is, the sparser the solution becomes. Many frameworks have been proposed in this perspective, leading to multiple basis pursuit schemes.  Readers interested in an in-depth study of sparse decomposition algorithms can be referred to \citet{Starck_Murtagh_Fadili_2010, Elad_2010}.   
\subsection{Fixed dictionaries}
A data set can be decomposed in many dictionaries, but the best dictionary to solve \eqref{sparse_denoising} is the one with the sparsest (most economical) representation of the signal. In practice, it is convenient to use dictionaries with fast implicit transform (such as Fourier transform, wavelet transform, etc.) which allow us to directly obtain the coefficients and reconstruct the signal from these coefficients using fast algorithms running in linear or almost linear time (unlike matrix-vector multiplications). The Fourier, wavelet and discrete cosine transforms provide certainly the most well known dictionaries. 
 
Most of these dictionaries are designed to handle specific contents, and are restricted to signals and images that are of a certain type. For instance, Fourier represents well stationary and periodic signals, wavelets are good to analyze isotropic objects of different scales, curvelets are designed for elongated features, etc.. They can not guarantee sparse representations of new classes of signals of interest, that present more complex patterns and features. Thus, finding new approaches to design these sparsifying dictionaries becomes of the utmost importance. Recent works have shown that designing adaptive dictionaries and learning them upon the data themselves instead of using predesigned selections of analytically-driven atoms leads to state-of-the-art performances in various tasks such as image denoising \citep{Elad_Aharon_denoising}, inpainting \citep{Mairal_Bach_Ponce_Sapiro_2010}, source separation \citep{bobin-gmca-cmb,Bobin_bss_CMB} and so forth. 
\subsection{Learned dictionaries}
The problem of dictionary learning, in its non-overcomplete form (that is when the number of atoms in the dictionary is smaller or equal to the dimension of the signal to decompose), has been studied in depth and can be approached using many viable techniques, such as principal component analysis (PCA) and its variants, which are based on algorithms minimizing the reconstruction errors upon a training set of samples, while representing them as a linear combination of the dictionary elements \citep{Bishop_2007}.
Inspired from an analogy to the learning mechanism in the simple cells in the visual cortex, \citet{Olshausen_Field_1996} proposed a minimization process based on a cost function that balances between a misfitting term and a sparsity inducing term. The optimization process is performed by alternating the optimization with respect to the sparse encodings, and to the dictionary functions. Most of the overcomplete dictionary learning methods are based on a similar alternating optimization scheme, while using specific techniques to induce the sparsity prior and update the dictionary elements. This problem shares many similarities with the Blind Sources Separation problem \citep{Zibulevsky_BSS}, although in BSS the sources are assumed to be sparse in a fixed dictionary and the learning is performed on the mixing matrix. 

A popular approach is to learn patch-sized atoms instead of a dictionary of image-sized atoms. This allows a faster processing and makes the learning possible even with a single image to train on as many patch exemplars can be extracted from a single training image. Section 3 gives more details about the variational problem of patch learning and denoising. This patch-based approach lead to different learning algorithms such as MOD \citep{Engan_Aase_Husoy_1999}, Pojected Gradient Descent methods \citep{Lin_2007}, or K-SVD \citep{Aharon_Elad_Bruckstein_2006} that have proven efficient for image processing \citep{Elad_Aharon_denoising,Mairal_Bach_Ponce_Sapiro_2010,peyre_fadili_starck}.

\section{Denoising by centered dictionary learning}
\subsection{General variational problem}
The goal of denoising with dictionary learning is to build a suitable $n \times k $ dictionary $D$, a collection of atoms ${[d_i]}_{i = 1,\dots,P} \in \mathbb{R}^{N\times P}$, that offers a sparse representation to the estimated denoised image. As is it not numerically tractable to process the whole image as a large vector, \citet{Elad_Aharon_denoising,Mairal_Bach_Ponce_Sapiro_2010,peyre_fadili_starck} propose to break down the image into smaller patches and learn a dictionary of patch-sized atoms. When simultaneously learning a dictionary and denoising an image $Y$, the problem amounts to solving 
\begin{equation}
\left(\hat{X},\hat{A},\hat{D}\right)\in \argmin_{X,A,D}\mathcal{E} (X,A,D)
\end{equation}
where
\begin{equation}
\begin{split}
&\mathcal{E} (X,A,D) = \frac{\lambda}{2}\left\|Y-X\right\|_2^2 +\\
& \sum_{i,j}\left ( \frac{\mu_{i,j}}{2}\left\|C_{i,j}R_{i,j}(X)-D\alpha_{i,j}\right\|_2^2 + \left\|\alpha_{i,j}\right\|_1\right)\\
\end{split}
\label{denoising_energy}
\end{equation}

such that the learned dictionary $D$ is in $\mathcal{D}$, the set of dictionaries whose atoms are scaled to the unit $\ell_2$-ball
\begin{equation}
\forall j\in[1,\dots,k], \quad \left\|d_j\right\|^2 = \sum_{i=1}^N \left|d_j[i]\right|^2\leq1.
\label{bounded_atoms}
\end{equation}
Here, $Y$ is the noisy image, $X$ the estimated denoised image, $A=(\alpha_{i,j})_{i,j}$ is the sparse encoding matrix such that $\alpha_{i,j}$ is a sparse encoding of $R_{i,j}(X)$ in $D$ and $C_{i,j}$ is a centering operator defined by \eqref{centering_operator}. The parameters $\lambda$ and $(\mu_{i,j})_{i,j}$ balance the energy between sparsity prior, data fidelity between the learned dictionary and the training set, and denoising. The dictionary is constrained to   obey \eqref{bounded_atoms} to avoid classical scale indeterminacy in the bilinear model (the so-called equivalence class corresponds to scaling, change of sign and permutation). Indeed, if $(A,D)$ is a pair of sparsifying dictionary and coefficients, then the pair $\left(\nu A,\frac{1}{\nu}D\right)$, for any non-zero real $\nu$, leads to the same data fidelity. Thus, discarding the normalization constraint in the minimization problem \eqref{denoising_energy} would favor arbitrary small coefficients and arbitrary large dictionaries. 
It is also worth mentioning that the energy \eqref{denoising_energy} is not minimized with respect to the translation operators $(C_{i,j})_{i,j}$. Rather, we chose to use fixed translation operators that translate the patch such that the pixel of its maximum value is at its center. The rationale behind this is to increase the sensitivity of the algorithm to isotropic structures such as stars, which are ubiquitous in astronomical imaging. This will be clearly shown in the numerical results described in Section 4.

It is possible to learn a dictionary without denoising the image simultaneously, thus minimizing
\begin{equation}
 \sum_{i,j}\left (  \frac{1}{2}\left\|R_i(X)-D\alpha_i\right\|^2 + \lambda \left\|\alpha_i\right\|_1\right)
\end{equation}
with respect to $D$ and $A$.
This allows to learn a dictionary from a noiseless training set, or learn from a small noisy training set extracted from a large noisy image when it is numerically not tractable to process the whole image directly. Once the dictionary learned, an image can be denoised solving \eqref{sparse_denoising} as we show in Section 4. The classical scheme of dictionary learning for denoising dos not include the centering operators and has proven to be an efficient approach \citep{Elad_Aharon_denoising,peyre_fadili_starck}.

An efficient way to find a minimizer of \eqref{denoising_energy} is to use a alternating minimization scheme. The dictionary $D$, the sparse coding coefficient matrix $A$ and the denoised image $X$ are alternatively updated one at a time, the other being fixed. We give more details about each step and how we tuned the parameters below.

\subsection{Alternating minimization}
\subsubsection{Sparse coding}
We consider here that the estimated image $X$ and the dictionary $D$ are determined to minimize $\mathcal{E}$ with respect to $A$. Estimating the sparse encoding matrix $A$ comes down to solve \eqref{lasso}, that can be solved using iterative soft thresholding \citep{Daubechies_2004} or interior point solver \citep{Chen_lasso}. We chose to use the Orthogonal Matching Pursuit algorithm \citep{OMP}, a greedy algorithm that find an approximate solution of \eqref{sparse_denoising}. OMP yields satisfying results while being very fast and parameters simple to tune. When learning on a noisy image, we let OMP find the sparsest representation of a vector up to an error threshold set depending on the noise level. In the case of learning an image on a noiseless image, we reconstruct an arbitrary number of component of OMP.
\subsubsection{Dictionary update}
We consider that the encoding matrix $A$ and the training image $Y$ are fixed here, and we explain how the dictionary $D$ can be updated. The dictionary update consists in finding
\begin{equation}
\hat{D} \in \argmin_{D\in\mathcal{D}}\sum_{i,j} \frac{\mu_{i,j}}{2}\left\|C_{i,j}R_{i,j}(X)-D\alpha_{i,j}\right\|_2^2, 
\end{equation}
which can be rewritten in a matrix form as
\begin{equation}
\hat{D} \in \argmin_{D\in\mathcal{D}}\left\|P-DA\right\|_F^2
\label{dictionary_update}
\end{equation}
where each columns of $P$ contain a the patch $C_{i,j}R_{i,j}(X)$. We chose to use the Method of Optimal Directions that minimizes the mean square error of the residuals, introduced in \citet{Engan_Aase_Husoy_1999}. The MOD algorithm uses a single conjugate gradient step and gives the following dictionary update
\begin{equation}
D = \mathrm{Proj}_{\mathcal{D}}\left(P A^T\left(AA^T\right)^{-1}\right)
\label{mod_update}
\end{equation}
where $\mathrm{Proj}_{\mathcal{D}}$ is the projection on $\mathcal{D}$ such that for $D_2 = \mathrm{Proj}_{\mathcal{D}}\left(D_1\right)$, $d_{2i} = d_{1i}/\left\|d_{1i}\right\|_2$ for each atom $d_{2j}$ of $D_2$.The MOD algorithm is simple to implement and fast. An exact minimization is possible with an iterative projected gradient descend \citep{peyre_fadili_starck} but the process is slower and require precise parameter tuning. Another successful approach, the K-SVD algorithm, updates the atoms of the dictionary one by one, using for the update of a given atom only the patches that use significantly this atom in their sparse decomposition \citep{Aharon_Elad_Bruckstein_2006}.  


\subsubsection{Image update}
When $D$ and $A$ are fixed, the energy \eqref{denoising_energy} is a quadratic function of $X$ minimized by the closed -form solution
\begin{equation}
\begin{split}
\widehat{X} = & \left(\sum_{i,j}  \mu_{i,j}R^*_{i,j}R_{i,j}+\lambda \mathrm{Id}\right)^{-1}  \\
&\left(\sum_{i,j} \mu_{i,j}R^*_{i,j}C^*_{i,j}D\alpha_{i,j} + \lambda Y\right).\\
\end{split}
\label{image_update}
\end{equation}
Updating $X$ with \eqref{image_update} simply consists in applying on each patch the "de-centering" operator $C^*_{i,j}$ and reconstruct the image by averaging overlapping patches. 
\subsubsection{Algorithm summary}
The centered dictionary learning for denoising algorithm is summarized in Algorithm \ref{DLC_algorithm}. It takes as input a noisy image to denoise and an initial dictionary, and iterates the three steps previously described to yield a noiseless image, a dictionary, and an encoding matrix. 

\begin{algorithm}[position=!h]                 
	\caption{Alternating scheme for centered dictionary learning and denoising}          
	\begin{algorithmic}                    
		\STATE \textbf{Input: }noisy image  $Y \in \mathbb{R}^{Q \times Q}$, number of iterations $K$, assumed noise level $\sigma$
		\STATE \textbf{Output: } sparse coding matrix $A$, sparsifying dictionary $D$, denoised image $X$
		\STATE \textbf{Initialize} $D\in\mathbb{R}^{n\times p}$ with patches randomly extracted from $Y$, set $\alpha_{i,j} = 0$ for all $i,j$, set $X=Y$, compute centering operators $(C_{i,j})_{i,j}$ by locating the maximum pixel of each patch $(R_{i,j}(X))_{i,j}$
		\FOR{ $k=1$ to $K$}
			\STATE \textbf{Step 1: Sparse coding}			
			\STATE Compute the sparse encoding matrix $A$ of $(R_{i,j}(X))_{i,j}$ in $D$ solving \eqref{sparse_denoising} or \eqref{l1_constrained}
			\STATE \textbf{Step 2: Dictionary update}			
			\STATE Update dictionary $D$ solving \eqref{dictionary_update}
			\STATE \textbf{Step 3: Image update}
			\STATE Update denoised image $X$ using \eqref{image_update}
		\ENDFOR
	\end{algorithmic}
		\label{DLC_algorithm}                          
\end{algorithm}

\subsection{Parameters}
Algorithm~\ref{DLC_algorithm} requires several parameters. All images are $512\times512$ in our experiments. 

\paragraph{Patch size and overlap}
We use $n=9\times9$ patches for our experiments and take an overlap of $8$ pixels between two consecutive patches. A odd number of pixels is more convenient for patch centering, and this patch size has proven to be a good trade off between speed and robustness. A high overlap parameter allows to reduce block artifacts. 

\paragraph{Dictionary size}
We learn a dictionary of $p = 2n = 162$ atoms. A ratio $2$ between the size of the dictionary and the dimension of its atoms. It makes the dictionary redundant and allows to capture different morphologies, without inducing an unreasonable computing complexity. 

\paragraph{Training set size}
We extract $80n$ training patches when learning patches of $n$ pixels. Extracting more training samples allows to better capture the image morphology, and while it leads to very similar dictionaries, it allows a slightly sparser representation and a slightly better denoising. Reducing the size of the training set might lead to miss some features from the image to learn from, depending on the diversity of the morphology it contains. 

\paragraph{Sparse coding stop criterion: }We stop OMP when the sparse coding $x_s$ of a vector $x$ verifies
\begin{equation}
\left\|x_s-x\right\|_2 \leq C  \sigma\sqrt{n}
\end{equation}
and we use $C = 1.15$ as gain parameter, similarly to \citet{Elad_Aharon_denoising}. When learning on noiseless images, we stop OMP computation when it finds the $3$ first components of $x_s$. 

\paragraph{Training set}
We do not use every patch available in $Y$ as it would be too computationally costly, so we select a random subset of patch positions that we extract from $Y$. We extract $80n$ training patches and after learning, we perform a single sparse coding step with the learned dictionary on every noisy patch from $Y$ that are then averaged using \eqref{image_update}. Extracting more training sample does not have a significant effect on the learned dictionary in our examples. Reducing the size of the training set might lead to miss some features from the image to learn from, depending on the diversity of the morphology it contains. 

\section{Application to astronomical imaging}
\begin{figure*}[htb]
\centerline{
\vbox{
\hbox{
\subfigure[][]{%
\label{pics:galaxies_clean}
\includegraphics[width=1.3\columnwidth]{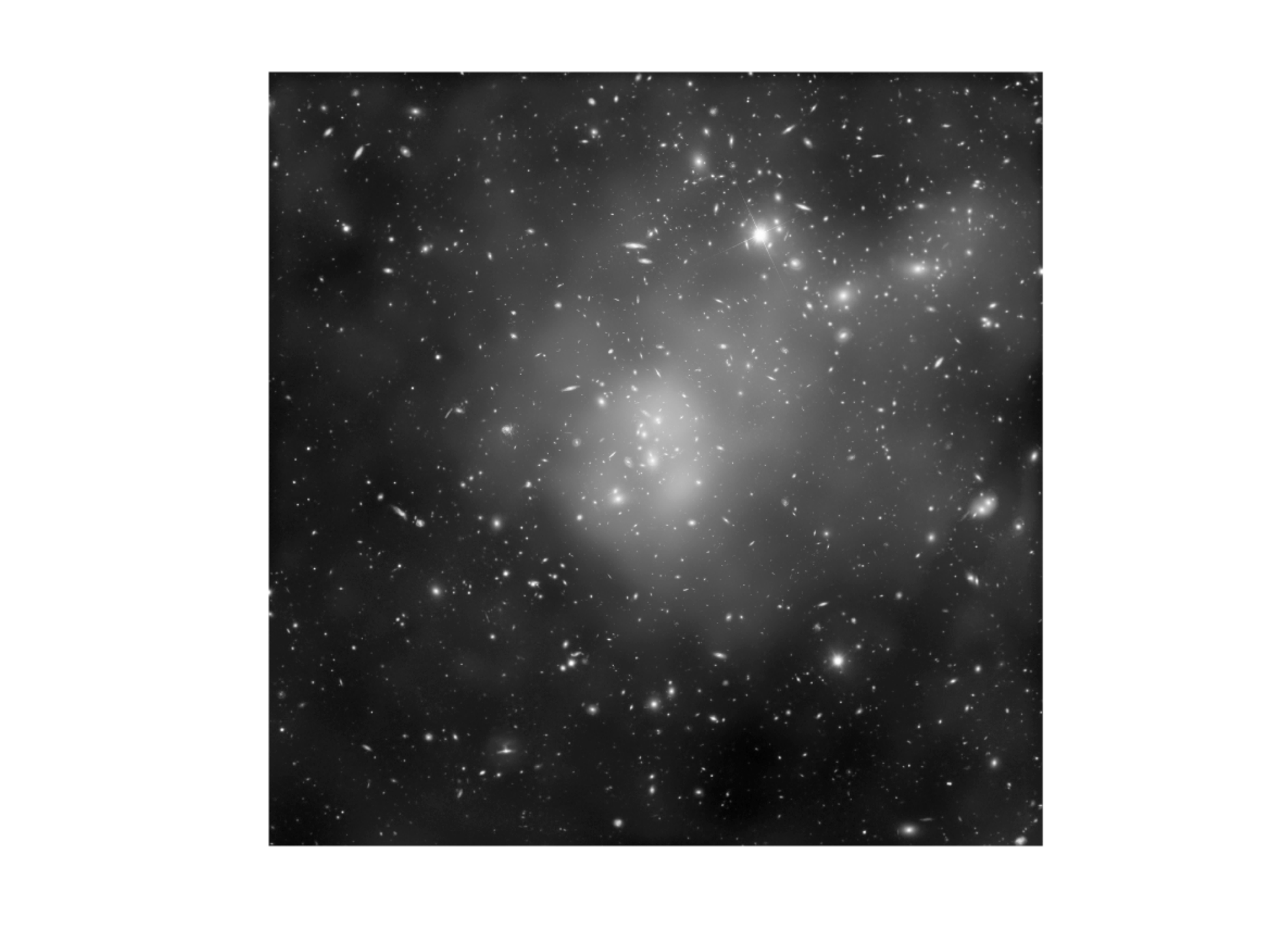}}
\subfigure[][]{%
\label{pics:stars_clean}
\includegraphics[width=1.3\columnwidth]{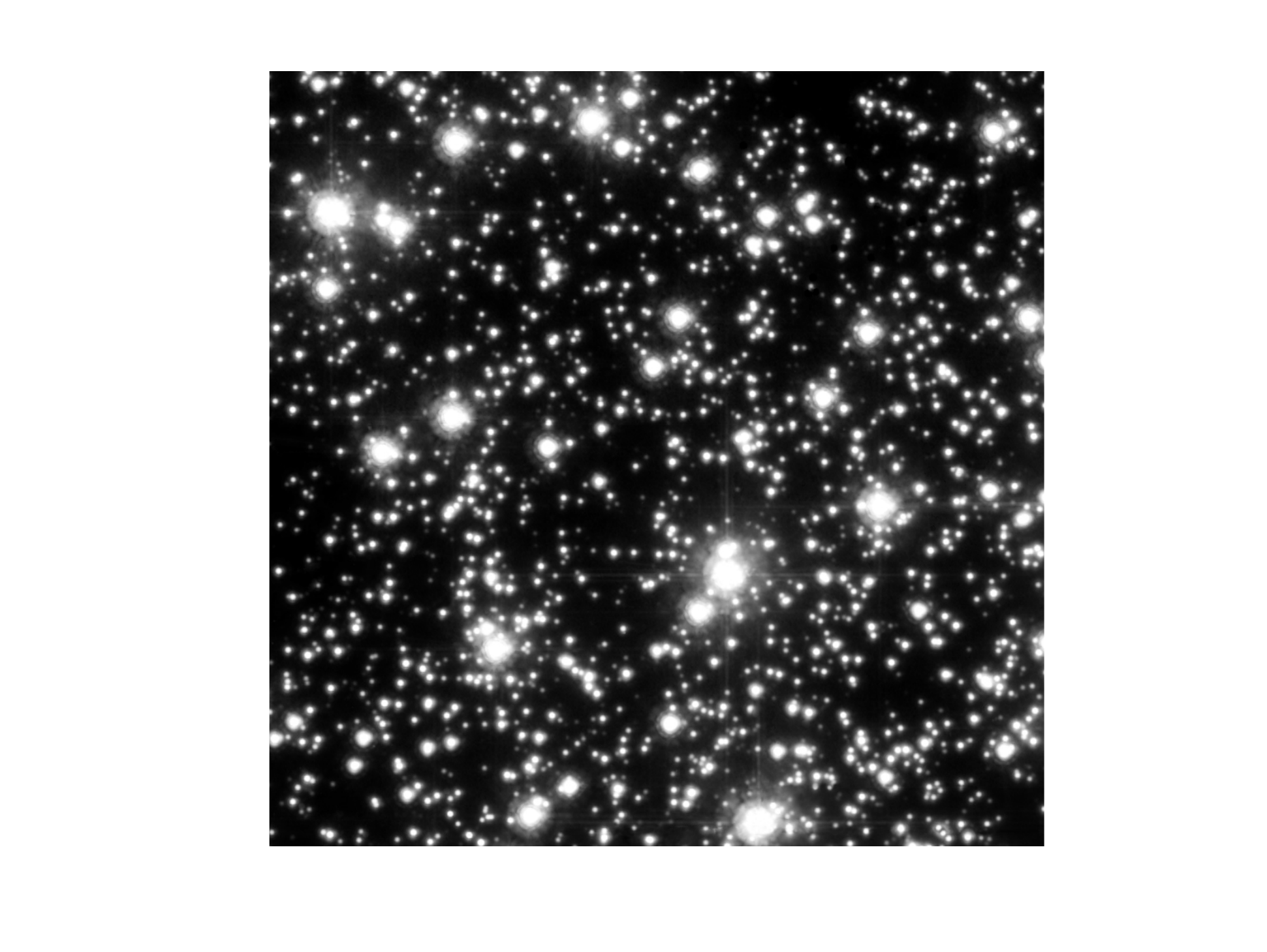}}
}
\hbox{
\subfigure[][]{%
\label{pics:nebula_clean}
\includegraphics[width=1.3\columnwidth]{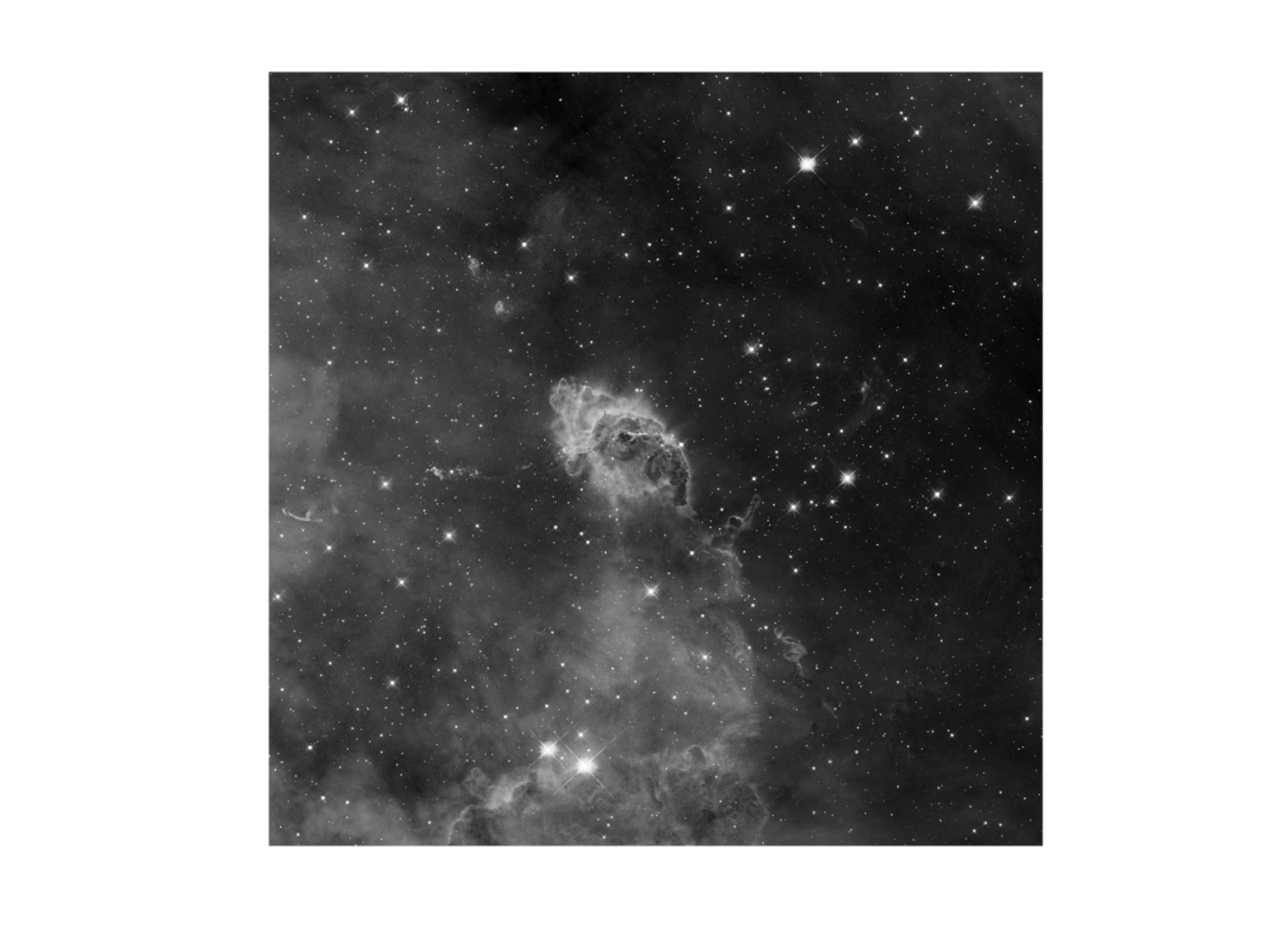}}
\subfigure[][]{%
\label{pics:cs_clean}
\includegraphics[width=1.3\columnwidth]{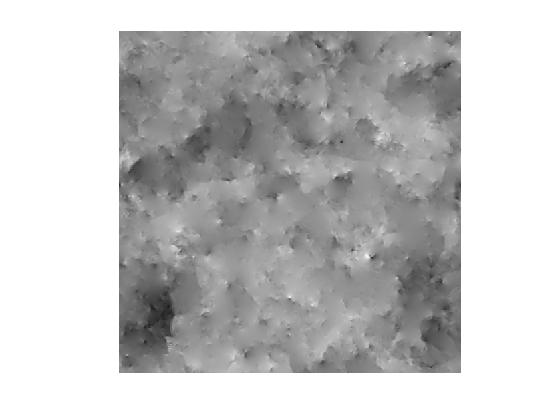}}
}}}
\caption{Hubble images used for numerical experiments. Figure \subref{pics:galaxies_clean} is the Pandora's Cluster Abell 2744, Figure \subref{pics:stars_clean} is an ACS image of 47 Tucanae, Figure \subref{pics:nebula_clean} is a image of WFC3 UVIS Full Field, and Figure \subref{pics:cs_clean} is a cosmic strings simulation.}
\label{fig:exp_images}
\end{figure*}

In this section, we report the main results of the experiments we conducted to study the performances of the presented strategy of centering dictionary learning and image denoising,  in the case of astronomical observations. We performed our tests on several Hubble images and cosmic string simulations (see Figure \ref{fig:exp_images}). Cosmic string maps are not standard astronomical images, but present the interest to have a complex texture and to be extremely difficult to detect. Wavelet filtering has been proposed to detect them \citep{wiaux:cordes} and it is interesting to investigate if DL could eventually be an alternative to wavelets for this purpose. It should however be clear that the level of noise that we are using here are not realistic, and this experiment has to be seen as a toy-model example rather than a cosmic string scientific study which would require to consider as well CMB and also more realistic cosmic string simulations. 
The three Hubble images are the Pandora's Galaxy Cluster Abell 2744, an ACS image of 47 Tucanae star field, and a WFC3 UVIS Full Field Nebula image. These images contain a mixture of isotropic and linear features, which make them difficult to process with the classical wavelets or curvelets-based algorithms.

We study two different case, where we perform dictionary learning and image denoising at the same time, and where the dictionary is learned on a noiseless image and used afterward to denoise a noisy image. 
We show for these two cases how DL is able to capture the natural features contained in the image, even in presence of noise, and how it outperforms wavelet-based denoising techniques. 

\subsection{Joint learning and denoising}
\begin{figure*}
\centerline{
\vbox{
\hbox{
\subfigure[][]{%
\label{pics:noisy1}
\centering
\includegraphics[width=1.3\columnwidth]{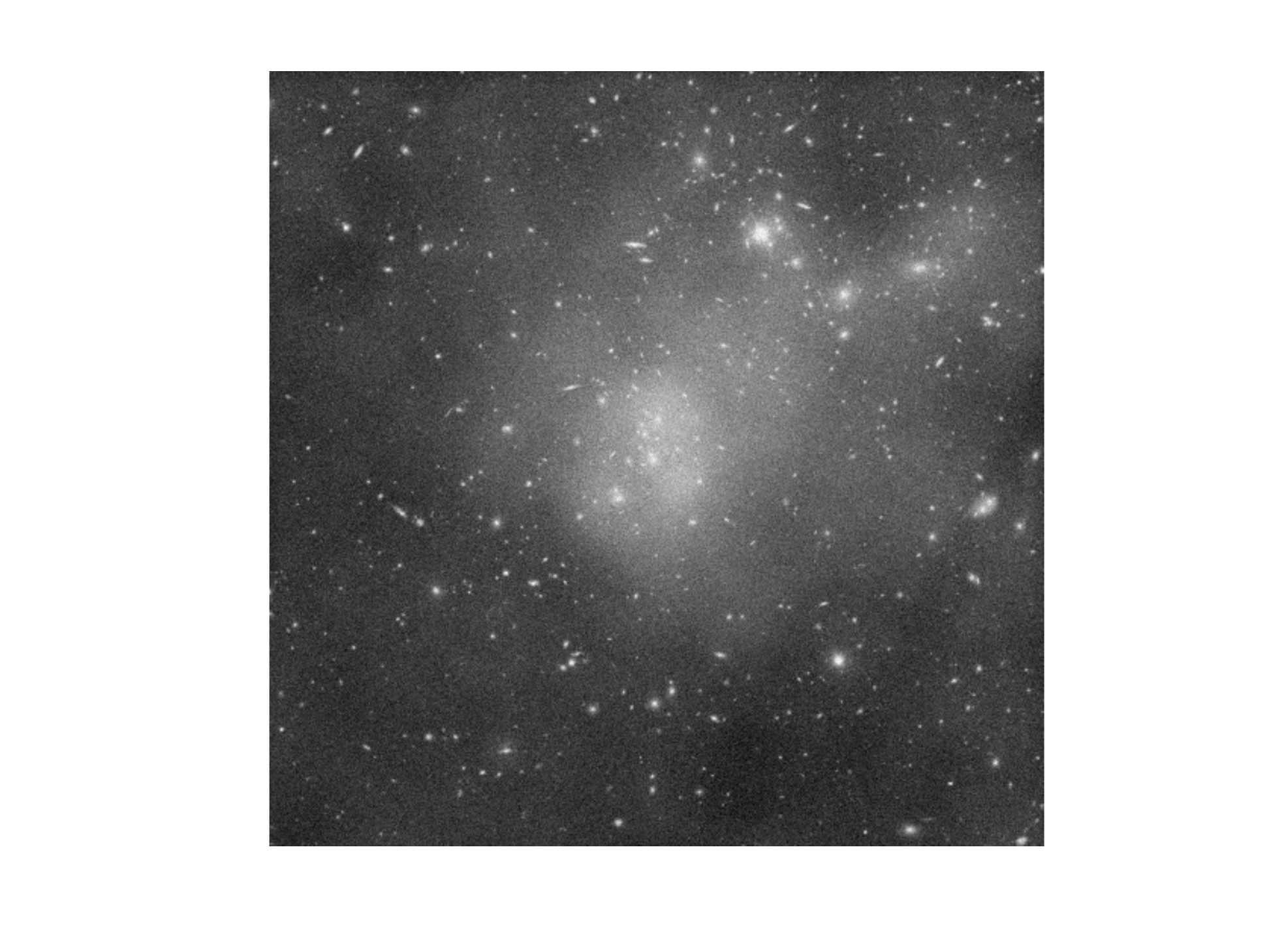}}
\subfigure[][]{%
\label{pics:dico1}
\centering
\includegraphics[width=1.3\columnwidth]{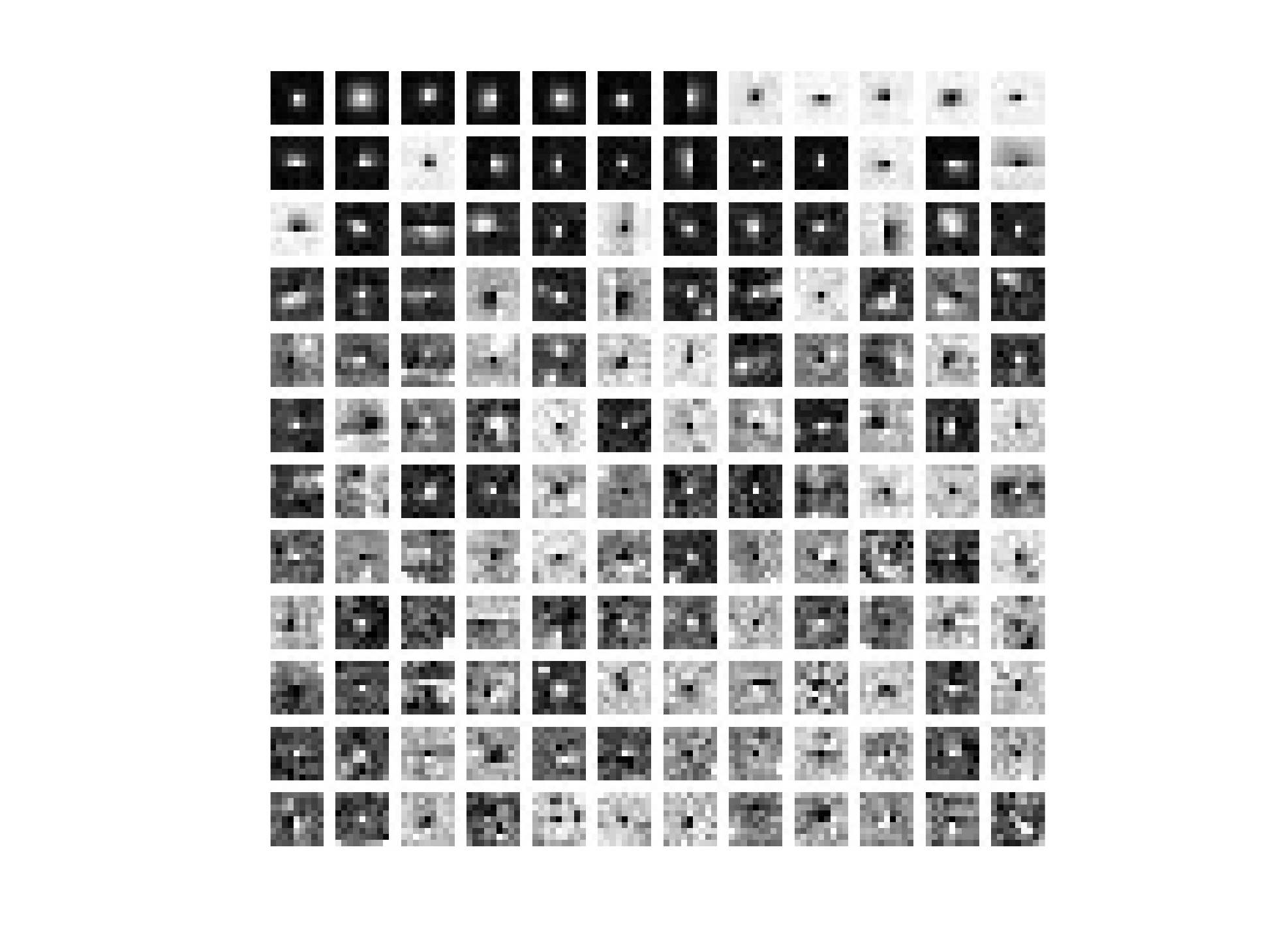}}
}
\hbox{
\subfigure[][]{%
\label{pics:mr1}
\centering
\includegraphics[width=1.3\columnwidth]{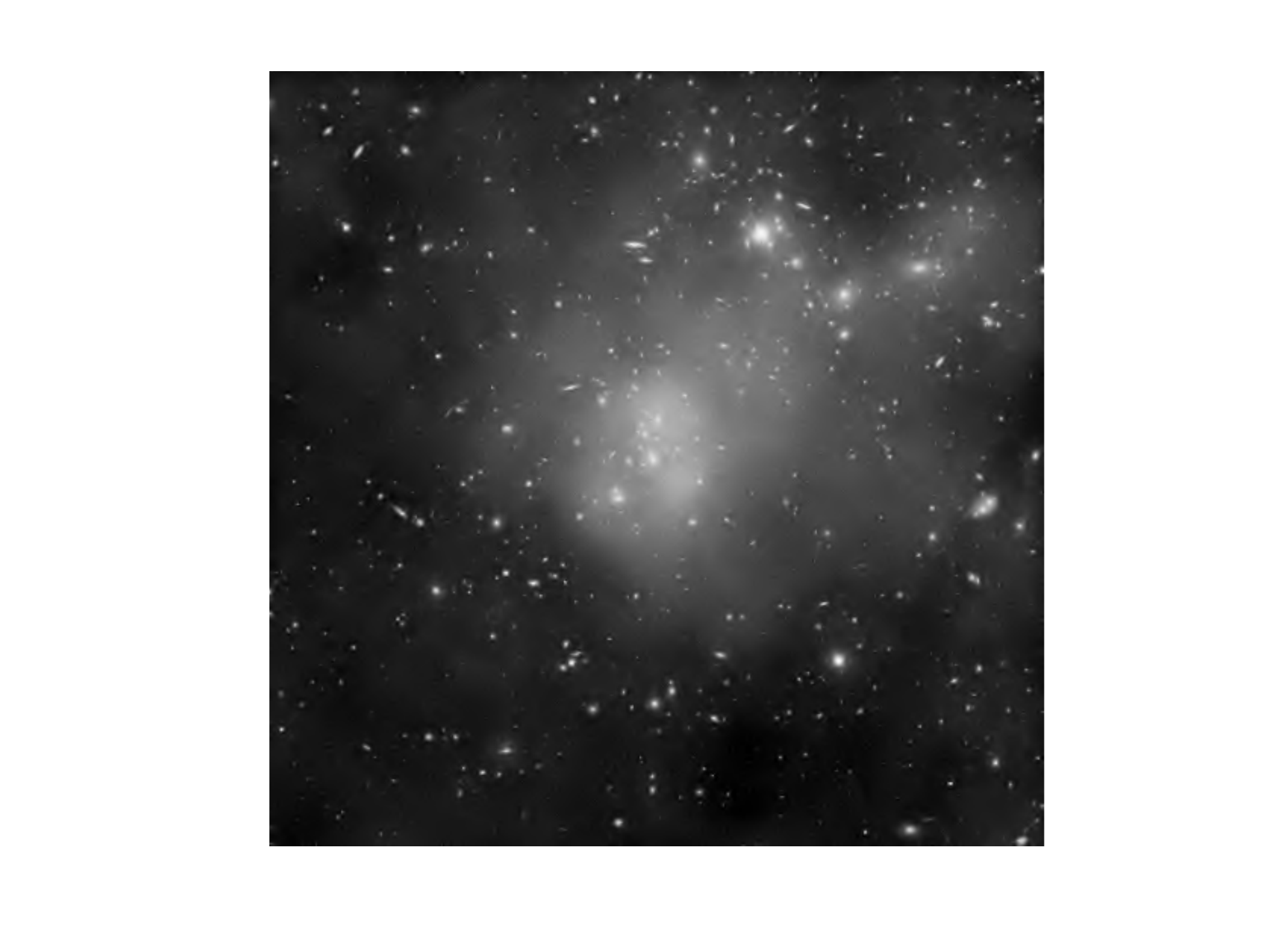}}
\subfigure[][]{%
\label{pics:dl1}
\centering
\includegraphics[width=1.3\columnwidth]{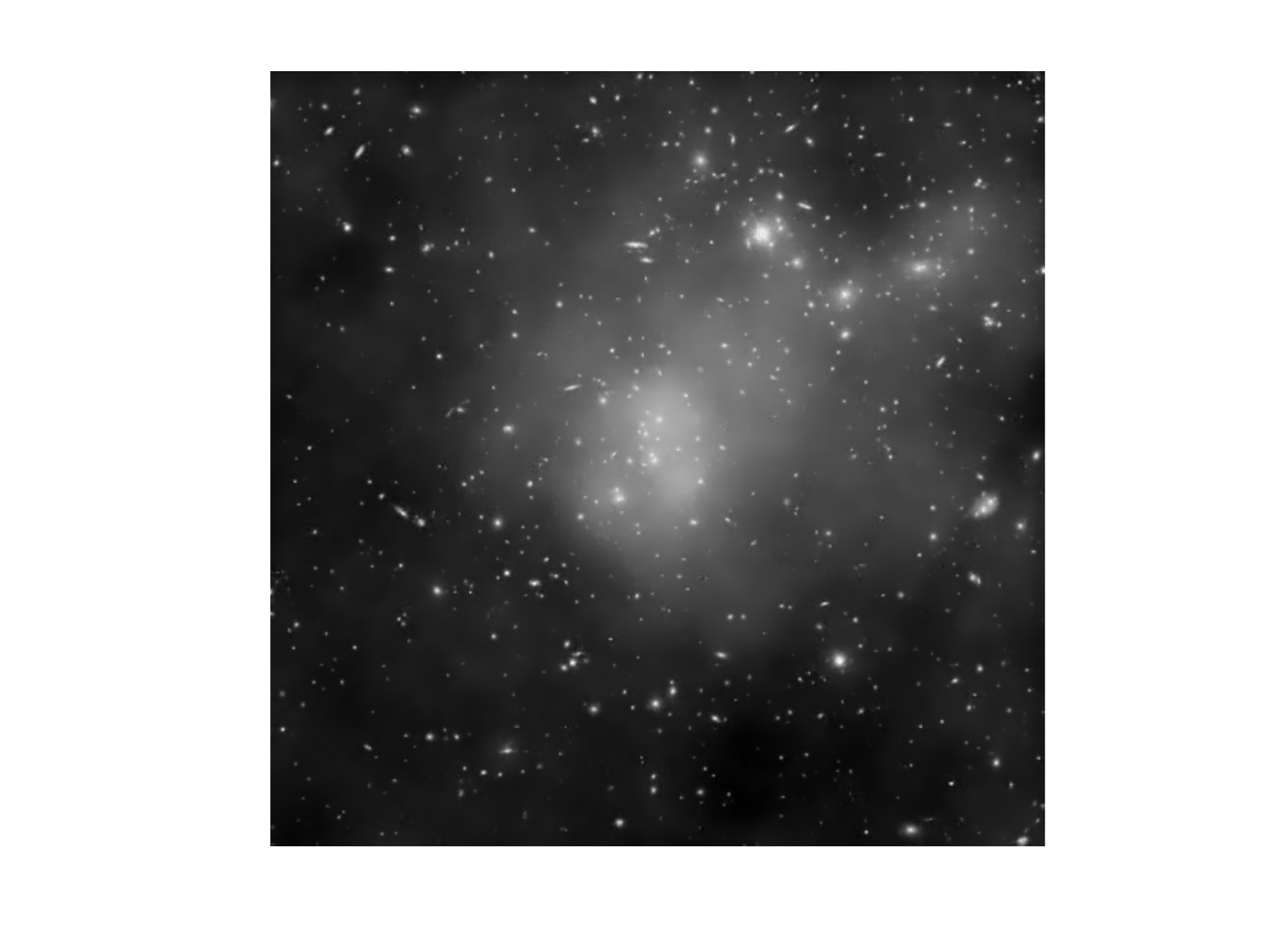}}
}}}
\caption{Results of denoising with galaxy cluster image. Figure \subref{pics:noisy1} shows the image noisy image, with a PSNR of 26.52 dB. The learned dictionary is shown Figure \subref{pics:dico1}. Figure \subref{pics:mr1} shows the result of the wavelet shrinkage algorithm that reaches a PSNR of 38.92 dB, Figure \subref{pics:dl1} shows the result of denoising using the dictionary learned on the noisy image, with a PSNR of 39.35 dB.}
\label{fig:galaxies_denoising}
\end{figure*}

\begin{figure*}
\centerline{
\vbox{
\hbox{
\subfigure[][]{%
\label{pics:noisy2}
\centering
\includegraphics[width=1.3\columnwidth]{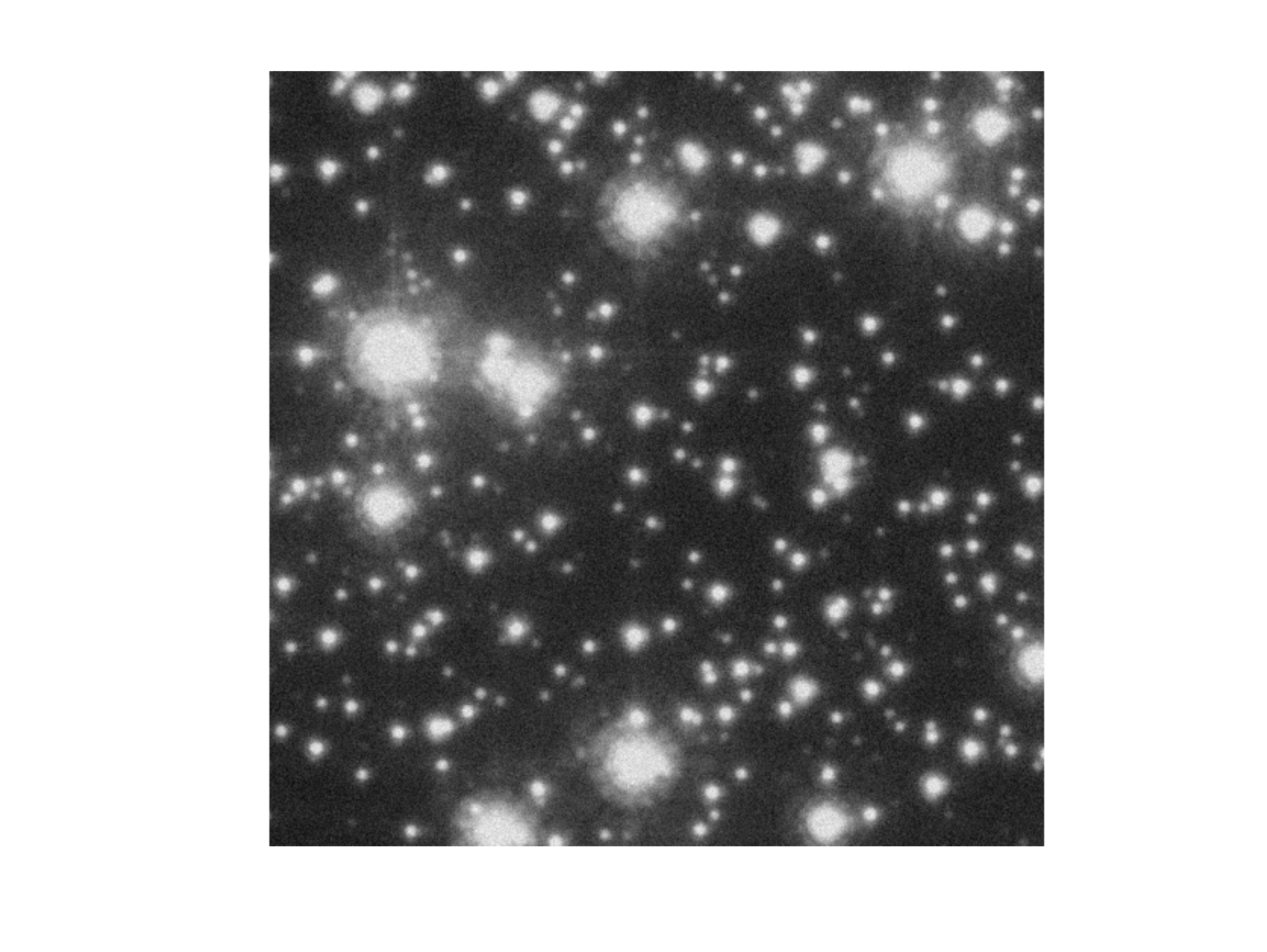}}
\subfigure[][]{%
\label{pics:dico2}
\centering
\includegraphics[width=1.3\columnwidth]{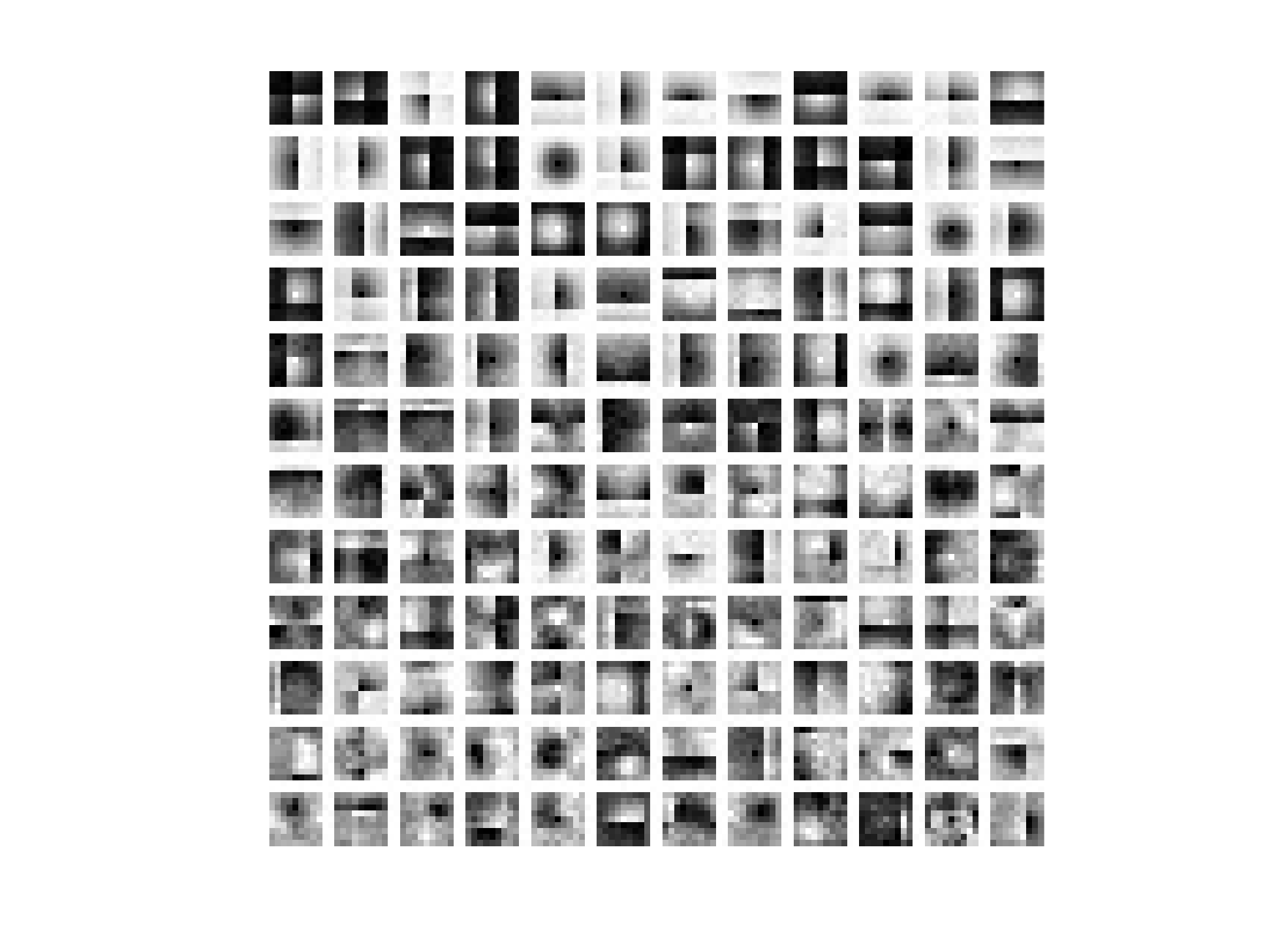}}
}
\hbox{
\subfigure[][]{%
\label{pics:mr2}
\centering
\includegraphics[width=1.3\columnwidth]{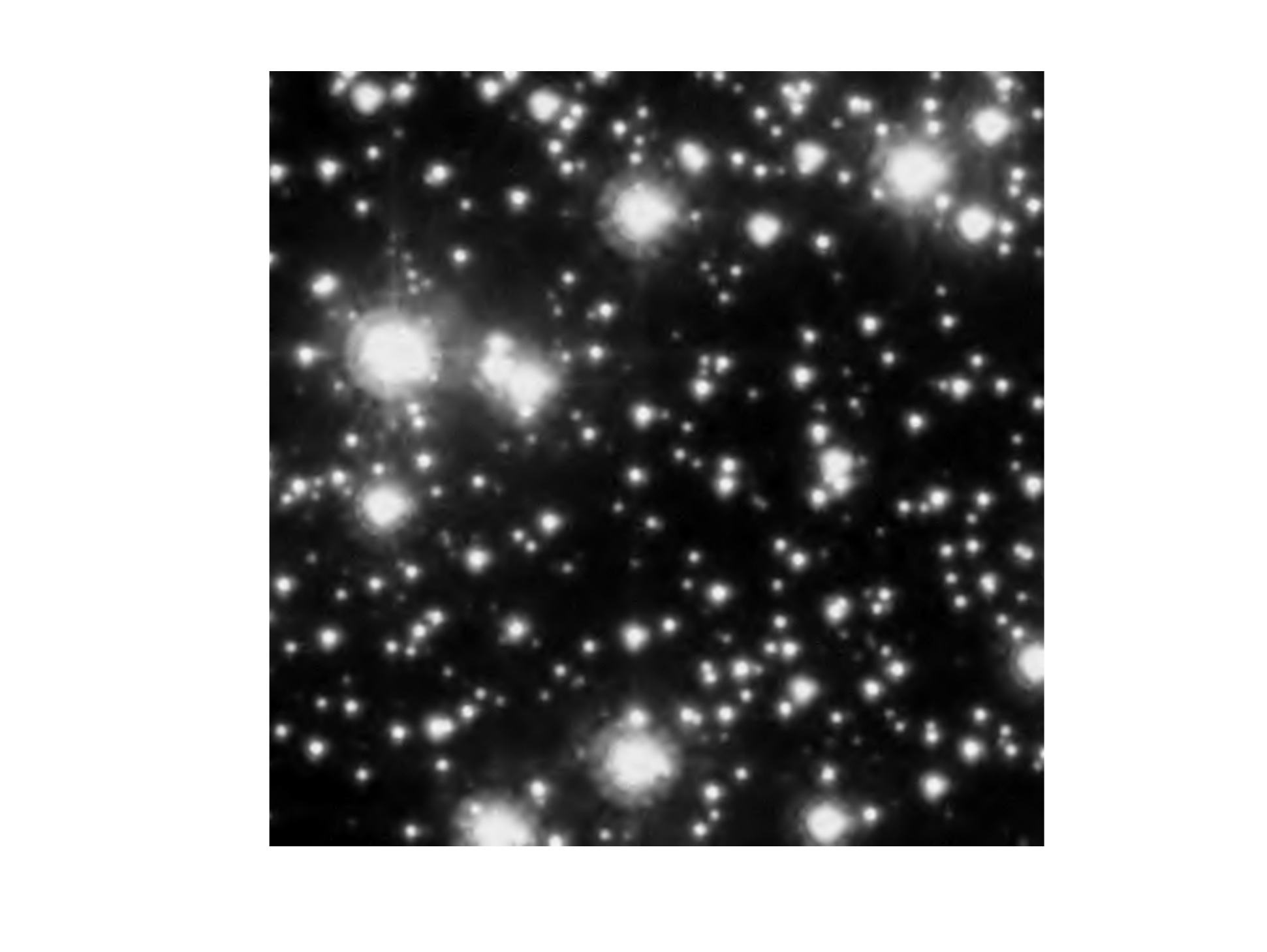}}
\subfigure[][]{%
\label{pics:dl2}
\centering
\includegraphics[width=1.3\columnwidth]{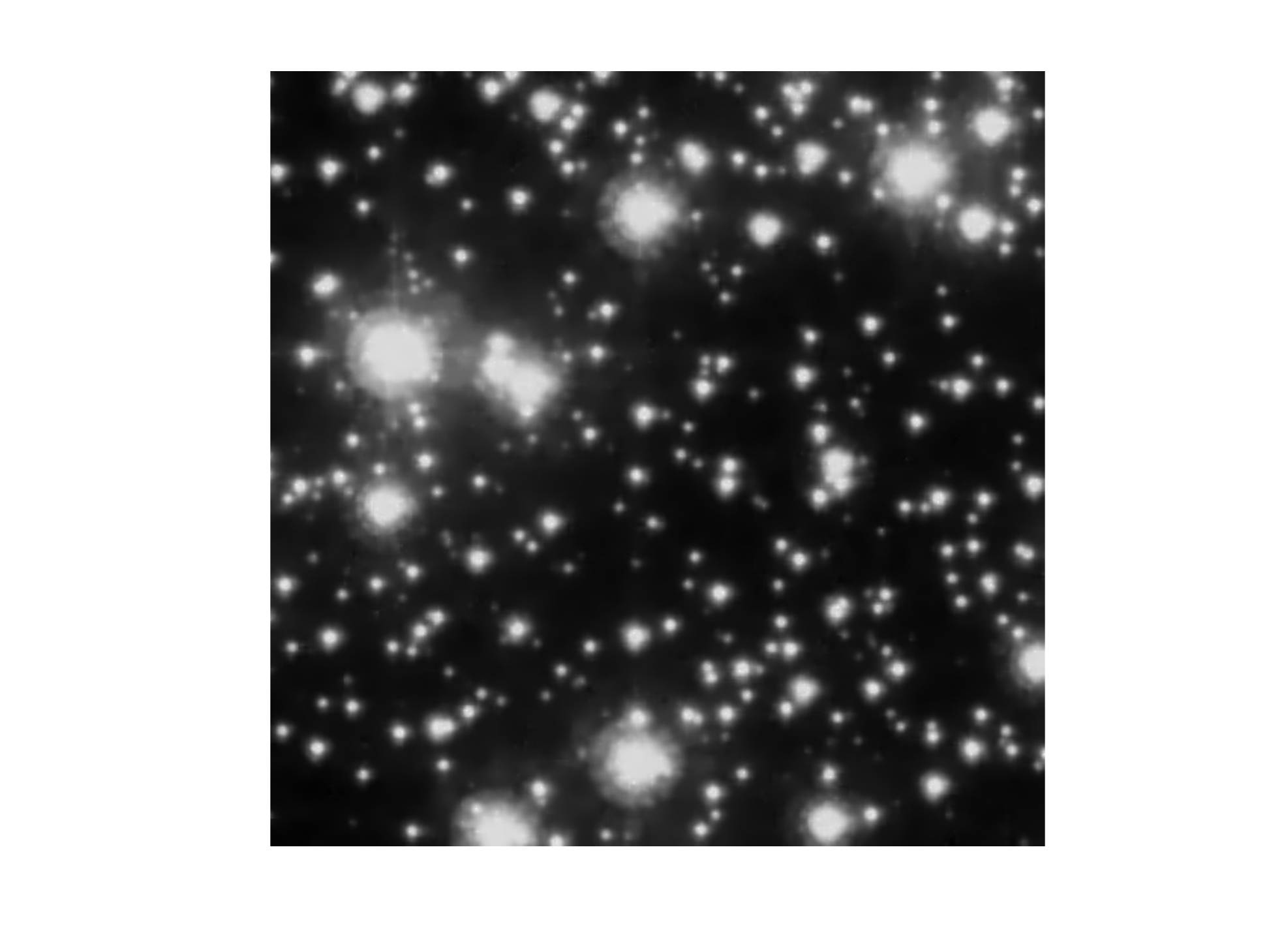}}
}}}
\caption{Results of denoising with star cluster image. Figure \subref{pics:noisy2} shows the image noisy image, with a PSNR of 27.42 dB. The learned dictionary is shown Figure \subref{pics:dico2}. Figure \subref{pics:mr2} shows the result of the wavelet shrinkage algorithm that reaches a PSNR of 37.28 dB, Figure \subref{pics:dl2} shows the result of denoising using the dictionary learned on the noisy image, with a PSNR of 37.87 dB.}
\label{fig:stars_denoising}
\end{figure*}

\begin{figure*}
\centerline{
\vbox{
\hbox{
\centering
\subfigure[][]{%
\label{pics:noisy3}
\centering
\includegraphics[width=1.3\columnwidth]{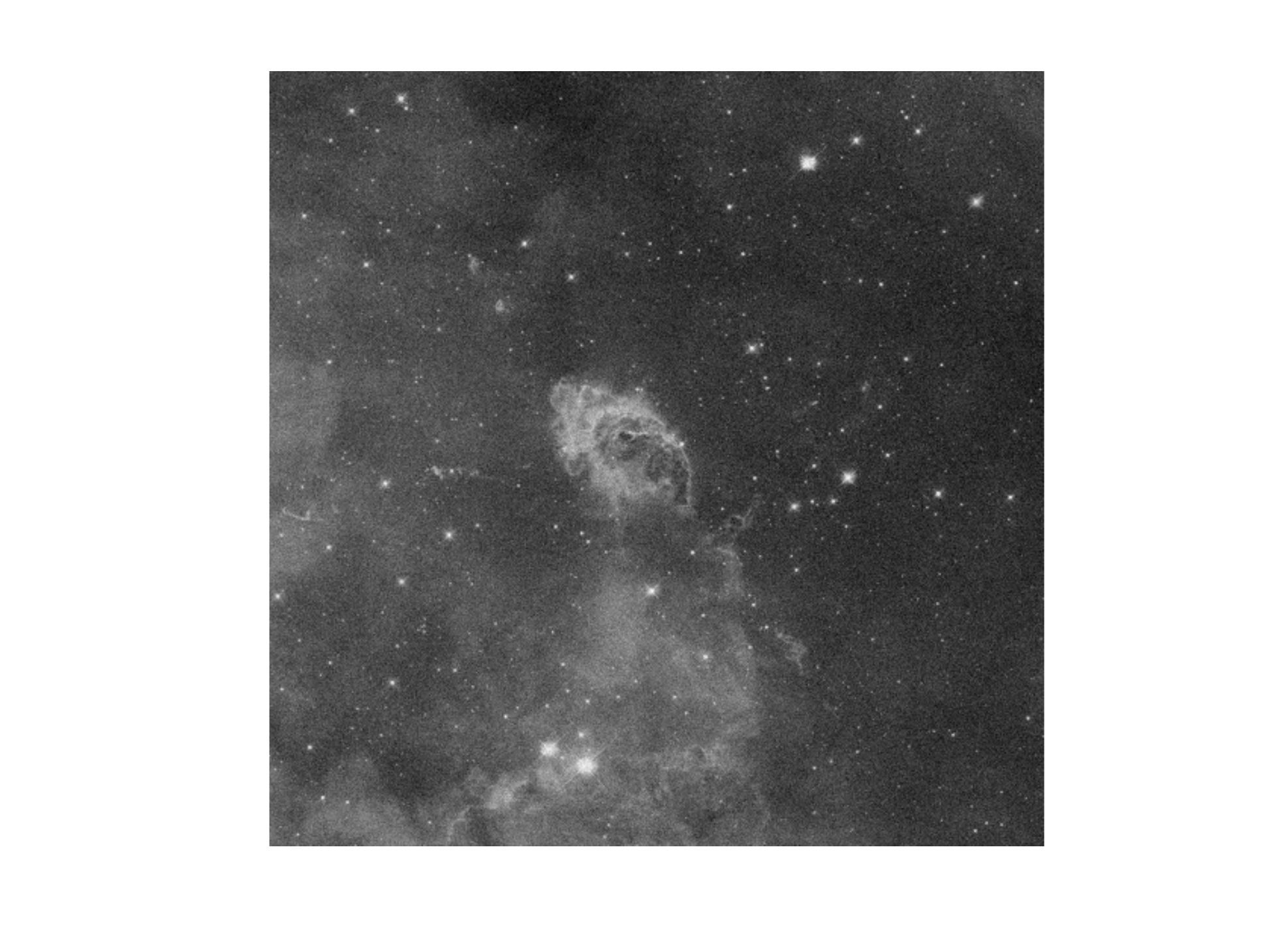}}
\subfigure[][]{%
\label{pics:dico3}
\centering
\includegraphics[width=1.3\columnwidth]{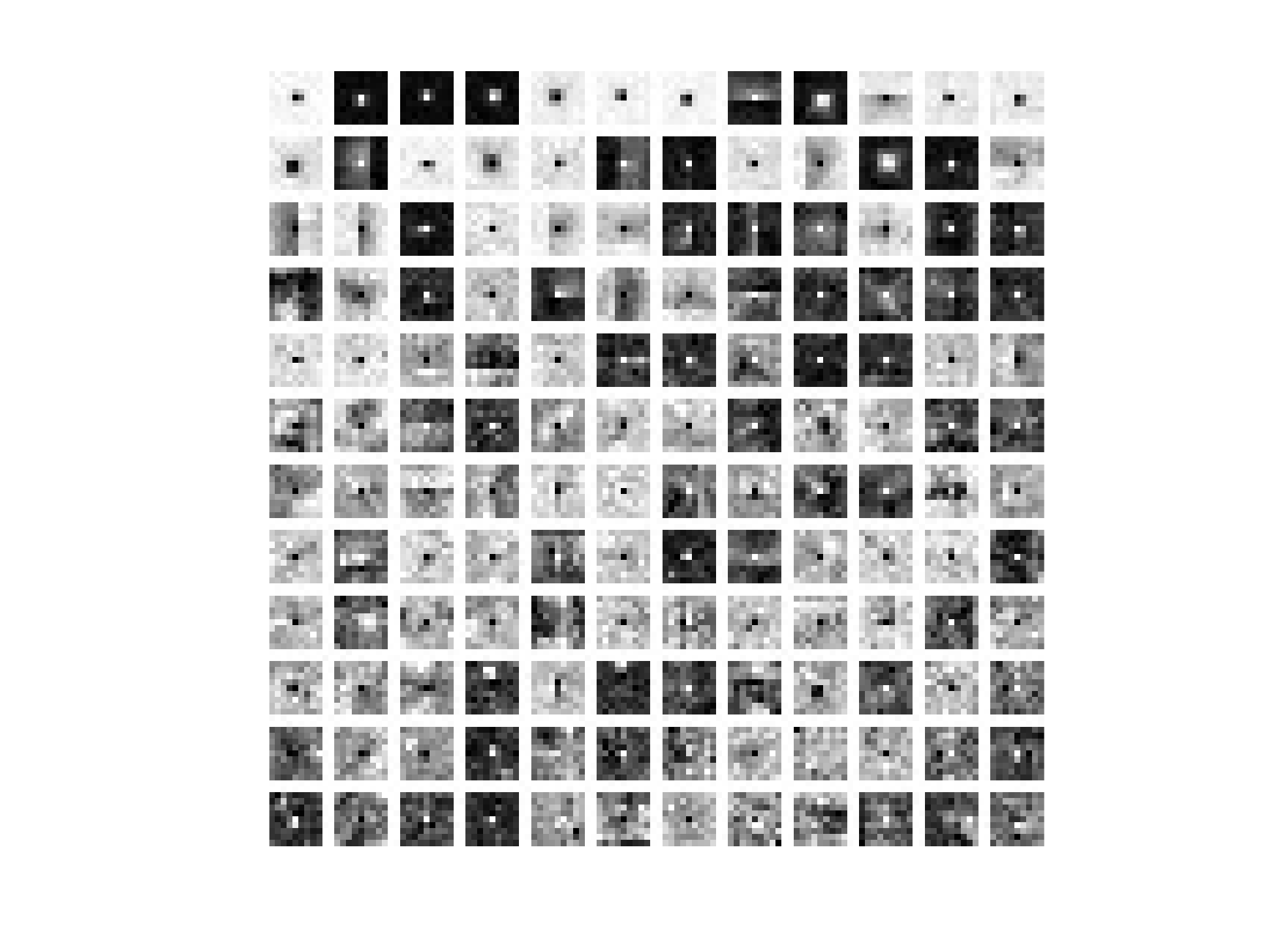}}
}
\hbox{
\subfigure[][]{%
\label{pics:mr3}
\centering
\includegraphics[width=1.3\columnwidth]{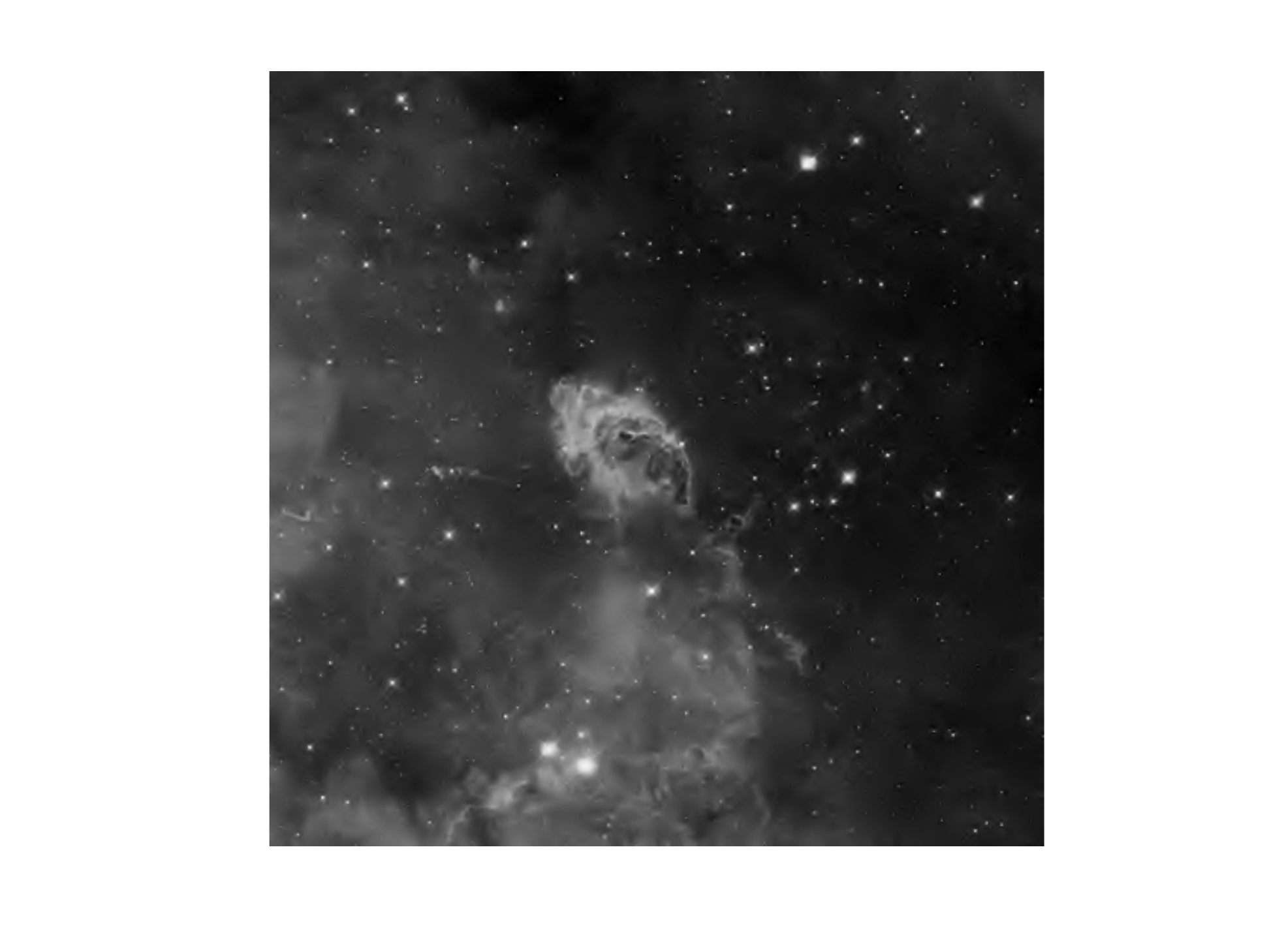}}
\subfigure[][]{%
\label{pics:dl3}
\centering
\includegraphics[width=1.3\columnwidth]{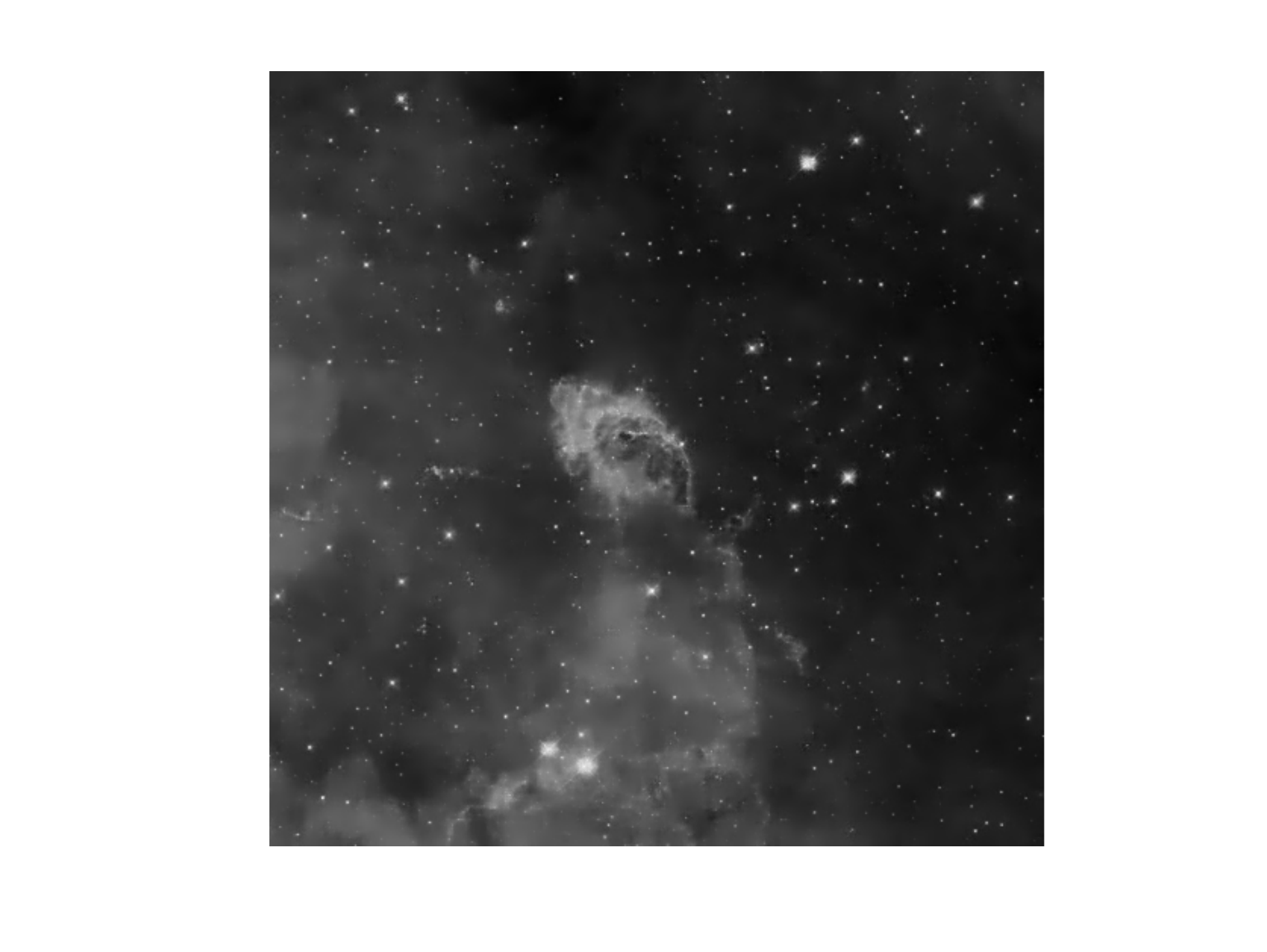}}
}}}
\caption{Results of denoising with nebula image. Figure \subref{pics:noisy3} shows the image used both for learning a noisy dictionary and denoising, with a PSNR of 26.67 dB. The learned dictionary is shown Figure \subref{pics:dico3}. Figure \subref{pics:mr3} shows the result of the wavelet shrinkage algorithm that reaches a PSNR of 33.61 dB, Figure \subref{pics:dl3} shows the result of denoising using the dictionary learned on the noisy image, with a PSNR of 35.24 dB.}
\label{fig:nebula_denoising}
\end{figure*}

We give several examples of astronomical images denoised with the method presented above. For all experiments, we show the noisy image, the learned dictionary and the denoised images, processed respectively with the wavelet shrinkage and the dictionary learning algorithms. We add a white Gaussian noise to a noiseless image. We then denoise them using Algorithm \ref{DLC_algorithm} and a wavelet shrinkage algorithm, and compare their performances in term of PSNR.
Figure \ref{fig:galaxies_denoising} shows the processing of a Hubble image of the Pandora galaxies cluster, Figure \ref{fig:stars_denoising} show our results on a star cluster image, and Figure \ref{fig:nebula_denoising} shows our results on a nebula image. The CDL proves to be superior to the wavelet-based denoising algorithm on each examples. The dictionary learning methods is able to capture the morphology of each kind of images and manages to give a good representation of point-like features.

\subsection{Separate learning and denoising}
\begin{figure*}
\centerline{
\hbox{
\subfigure{
\label{pics:source}
\includegraphics[width=1.3\columnwidth]{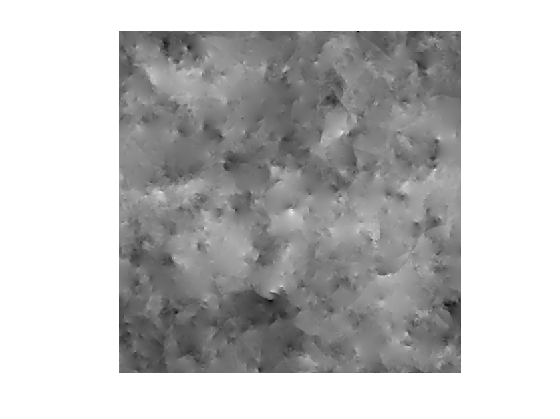}}
\subfigure{
\label{pics:training}
\includegraphics[width=1.3\columnwidth]{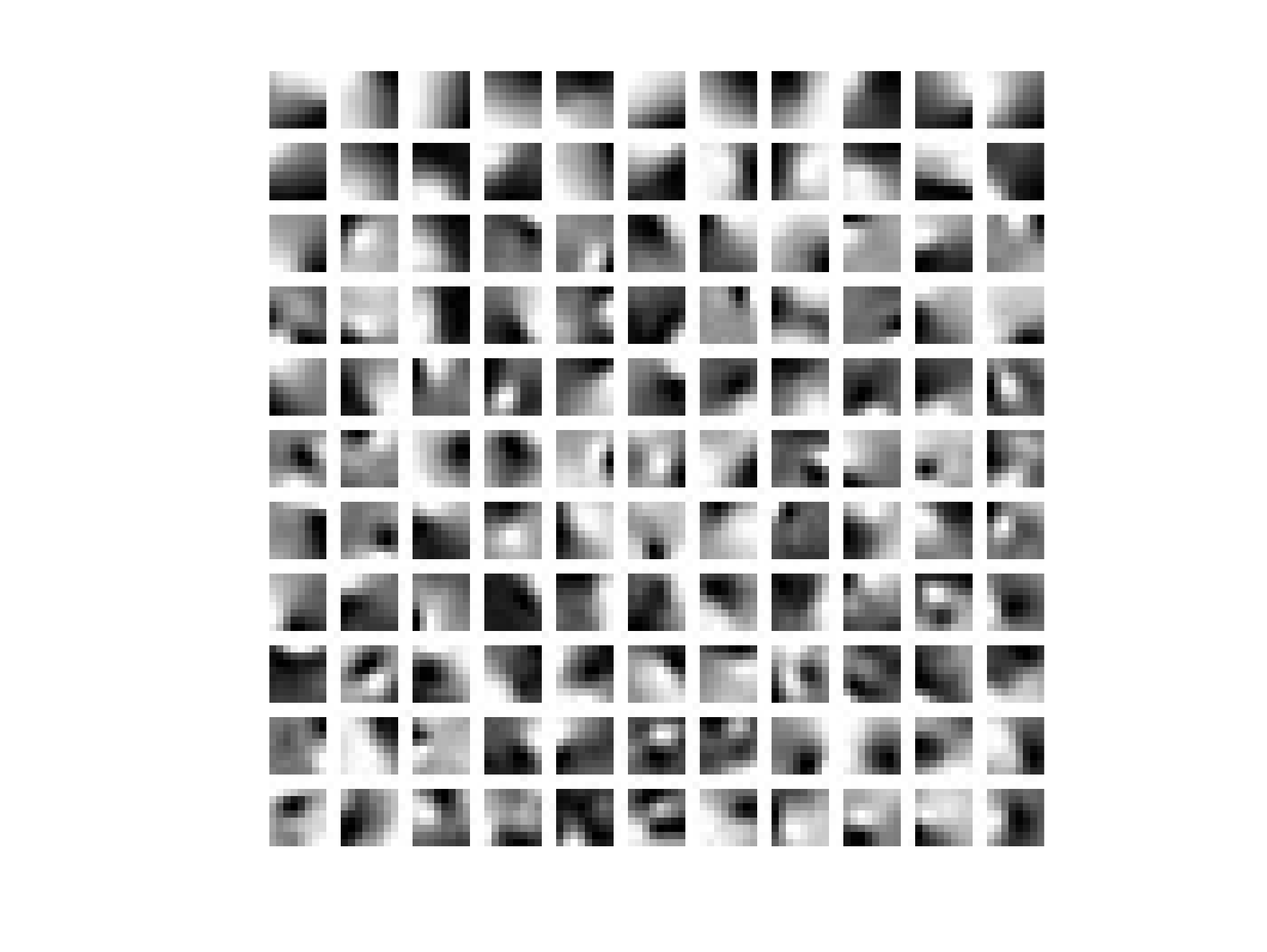}}
}}
\caption{Figure \subref{pics:source} shows a simulated cosmic string map (1"x1"), and Figure \subref{pics:training} shows the learned dictionary.}
\label{fig:cs_learning}
\end{figure*}

\begin{figure*}[htb]
\centerline{
\vbox{
\hbox{
\subfigure[][]{%
\label{pics:source0}
\centering
\includegraphics[width=1.3\columnwidth]{figures/ima_cs1}}
\subfigure[][]{%
\label{pics:noisy0}
\centering
\includegraphics[width=1.3\columnwidth]{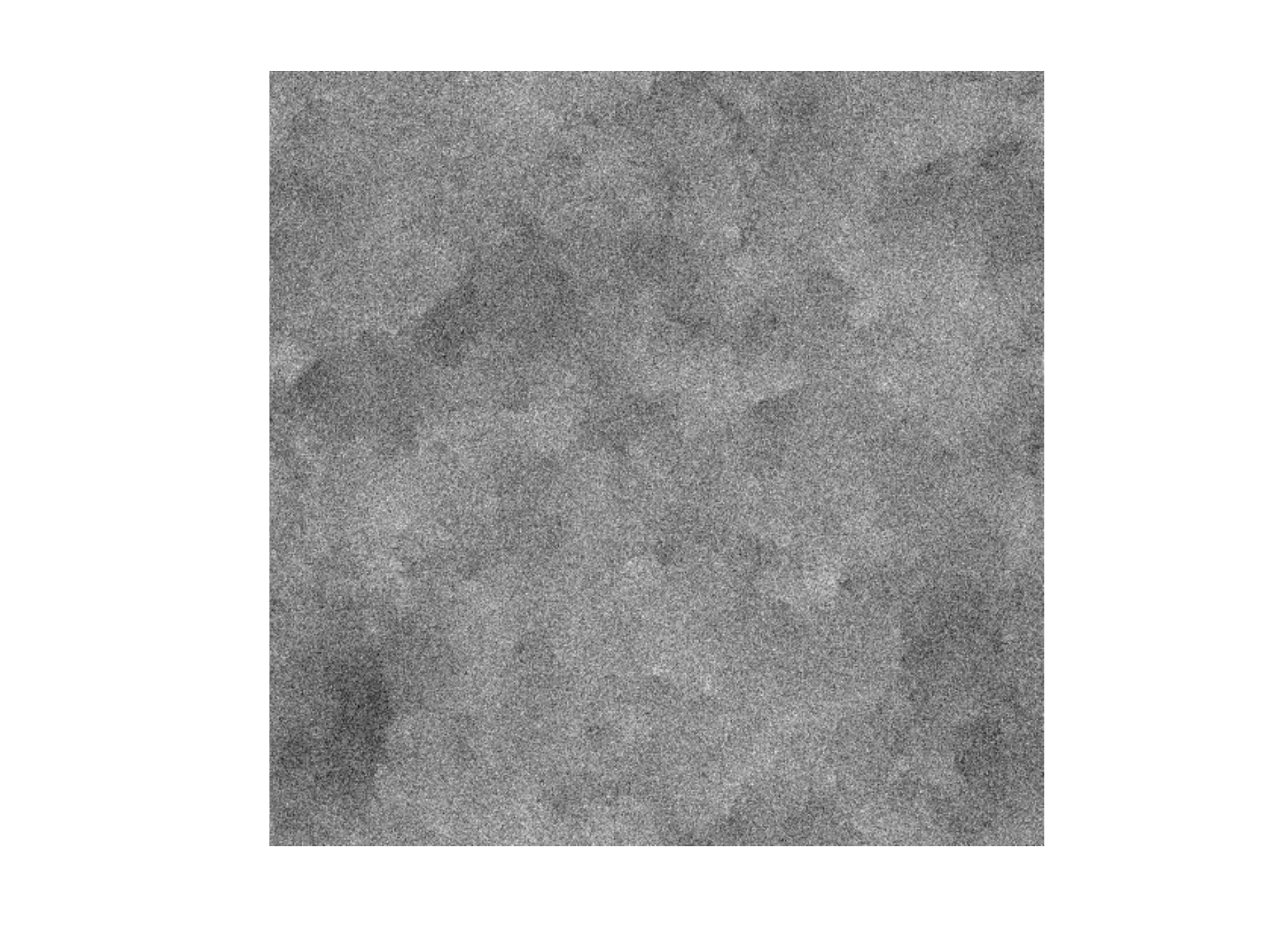}}
}
\hbox{
\subfigure[][]{%
\label{pics:mr0}
\centering
\includegraphics[width=1.3\columnwidth]{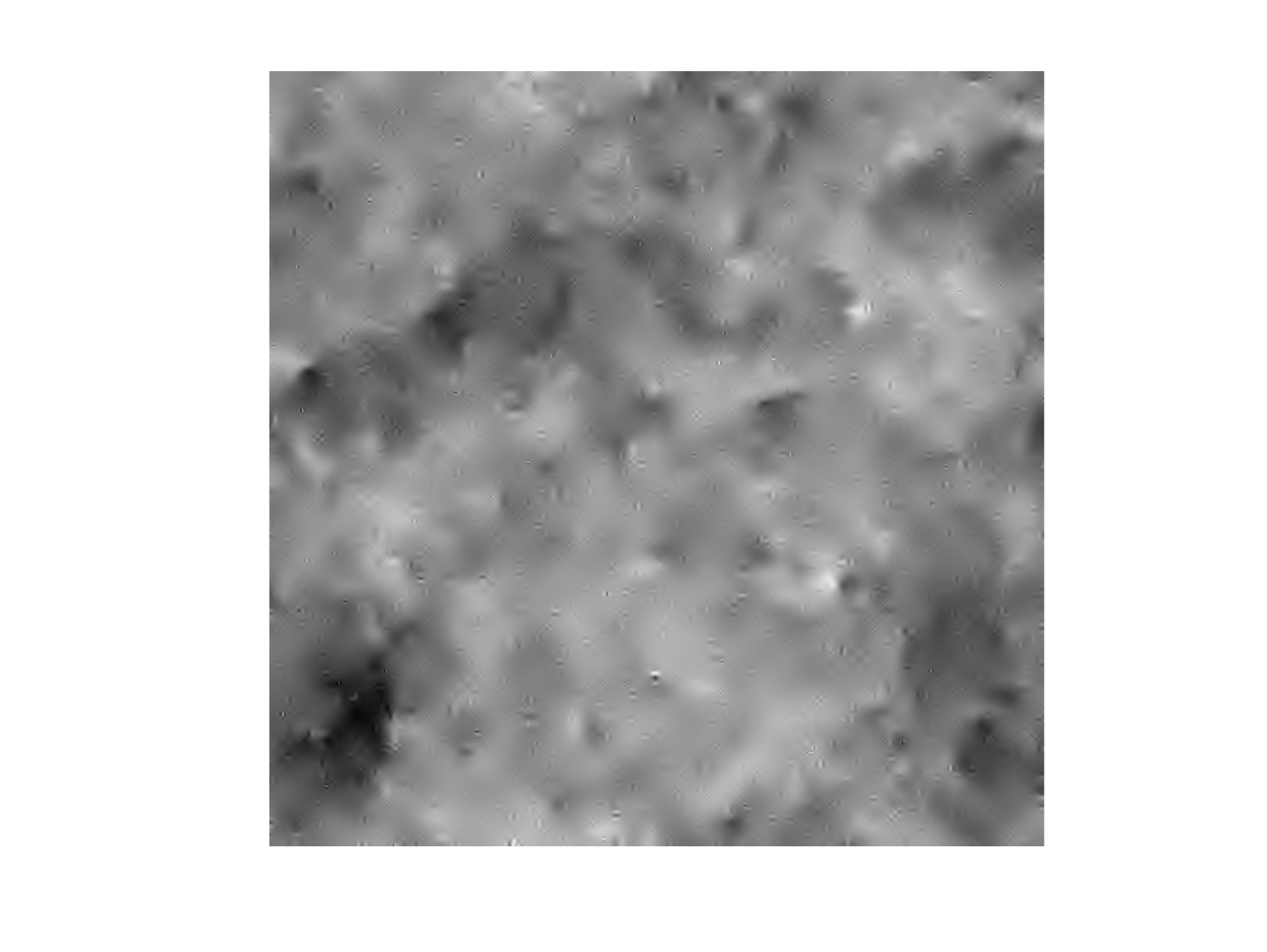}}
\subfigure[][]{%
\label{pics:dl0}
\centering
\includegraphics[width=1.3\columnwidth]{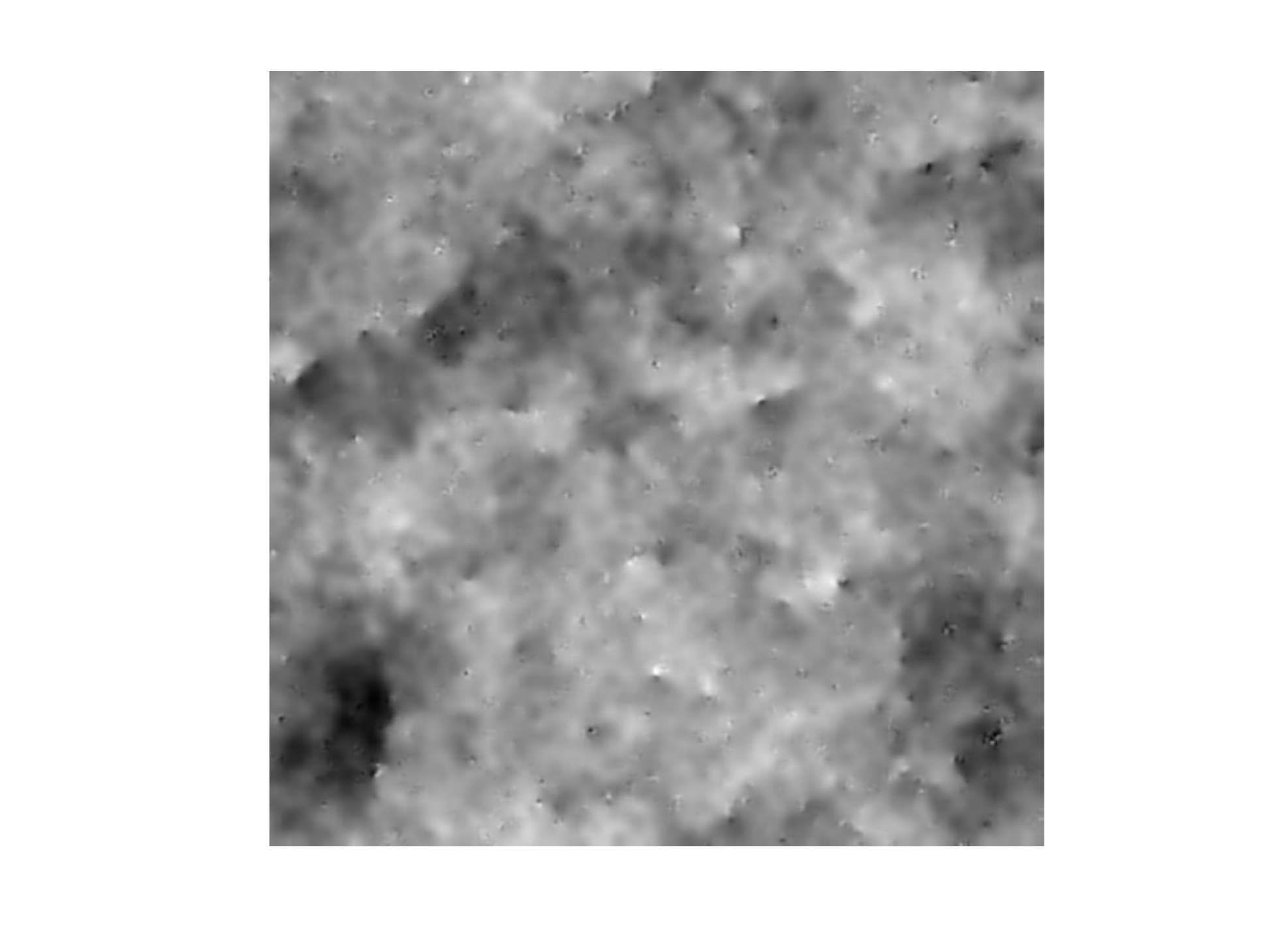}}
}}}
\caption{Example of cosmic string simulation denoising with a high noise level, using the learned dictionary from Figure~\ref{fig:cs_learning} and the wavelet algorithm. Figure \subref{pics:source0} is the source image, Figure \subref{pics:noisy0} shows the noisy image with a PSNR of 17.34 dB, Figure \subref{pics:mr0} shows the wavelet denoised version with a PSNR of 30.19 dB and Figure \subref{pics:dl0} shows the learned dictionary denoised version with a PSNR of 31.04 dB.}
\label{fig:cs_denoising}
\end{figure*}

\begin{figure*}
\centering
\includegraphics[width=0.99\columnwidth]{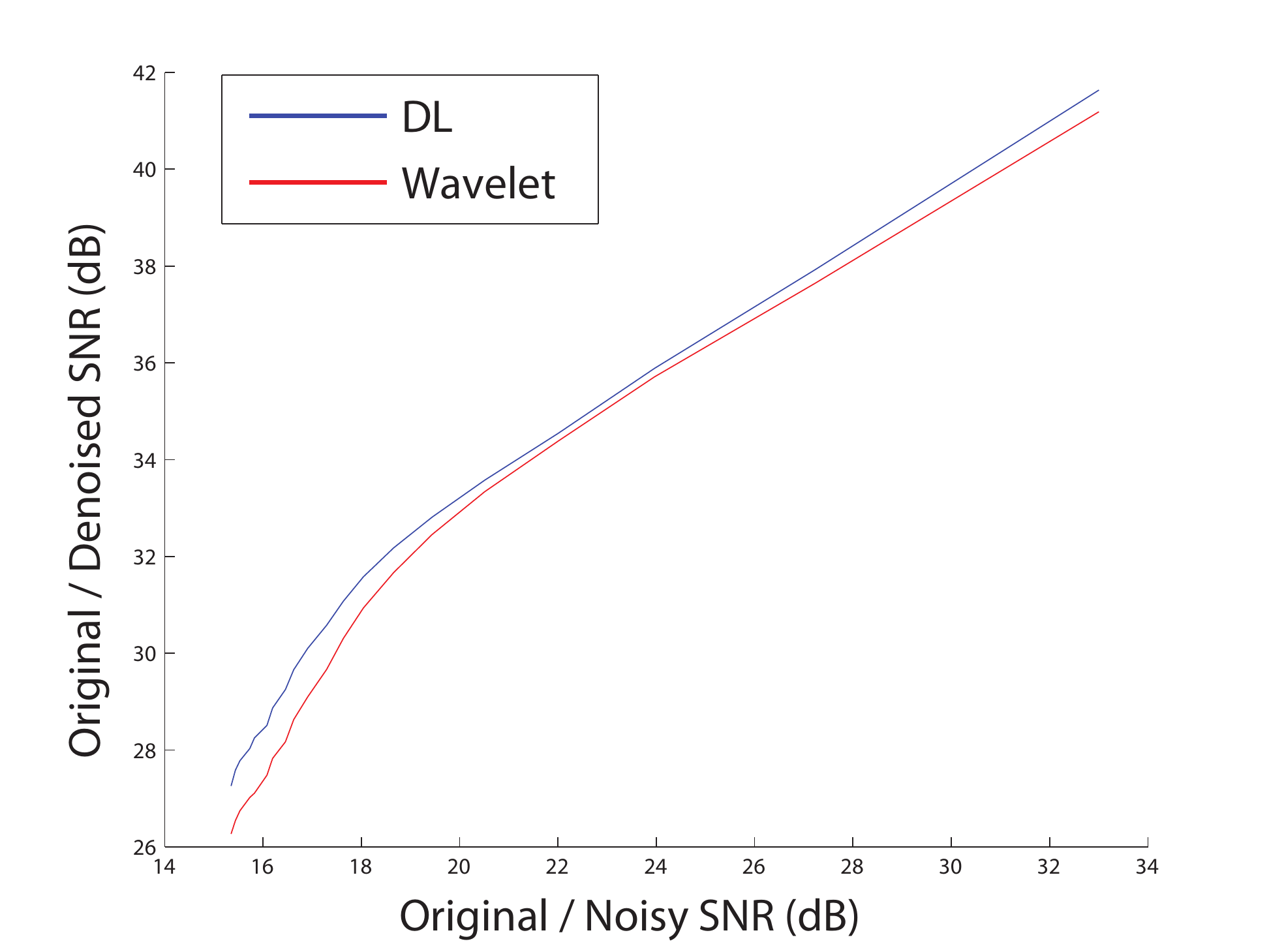}
\caption{Benchmark comparing the wavelet shrinkage algorithm to the dictionary learning denoising algorithm when dealing with various noise levels, using dictionary from Figure \ref{fig:cs_learning}. Each experiment is repeated 100 times and the results are averaged. We use the maximum value for the patch overlaping parameter. The sparse coding uses OMP and is set to reach an error margin $\left(C\sigma w\right)^2$ where $\sigma$ is the noise standard deviation and $C$ is a gain factor set to $1.15$. The wavelet algorithm uses $5$ scales of undecimated bi-orthogonal wavelets, with three bands per scale. The red and blue lines correspond respectively to wavelet and learned dictionary denoising. The horizontal axe is the PSNR between the noised and the source images, and the horizontal axe is the PSNR between the denoised and the source images.}
\label{fig:cs_denoising_benchmark}
\end{figure*}

We now apply the presented method to cosmic string simulations. We use a second image similar to the cosmic string simulation from Figure \ref{fig:exp_images} to learn a noiseless dictionary shown on Figure \ref{fig:cs_learning}. We add a high-level white Gaussian noise on the cosmic string simulation from Figure \ref{fig:exp_images} and we compare how classic DL and wavelet shrinkage denoising perform on Figure \ref{fig:cs_denoising}. We chose not to use CDL because the cosmic string images do not contain stars but more textured features. We give in Figure \ref{fig:cs_denoising_benchmark} a benchmark of the same process repeated for different noise levels. The PSNR between the denoised and source image is displayed as a function of the PSNR between the noisy and the original source image. The reconstruction score is higher for the dictionary learning denoising algorithm than for the wavelet shrinkage algorithm, for any noise level. This shows that the atoms computed during the learning are more sensitive to the features contained in the noisy image than wavelets. The dictionary learned was able to capture the morphology of the training image, which is similar to the morphology of the image to denoise. Hence, the coefficients of the noisy image's decomposition in the learned dictionary are more significant that its coefficient in the wavelet space, which leads to a better denoising.

\begin{figure*}
\centerline{
\hbox{
\subfigure{
\label{pics:star_gen}
\includegraphics[width=1\columnwidth]{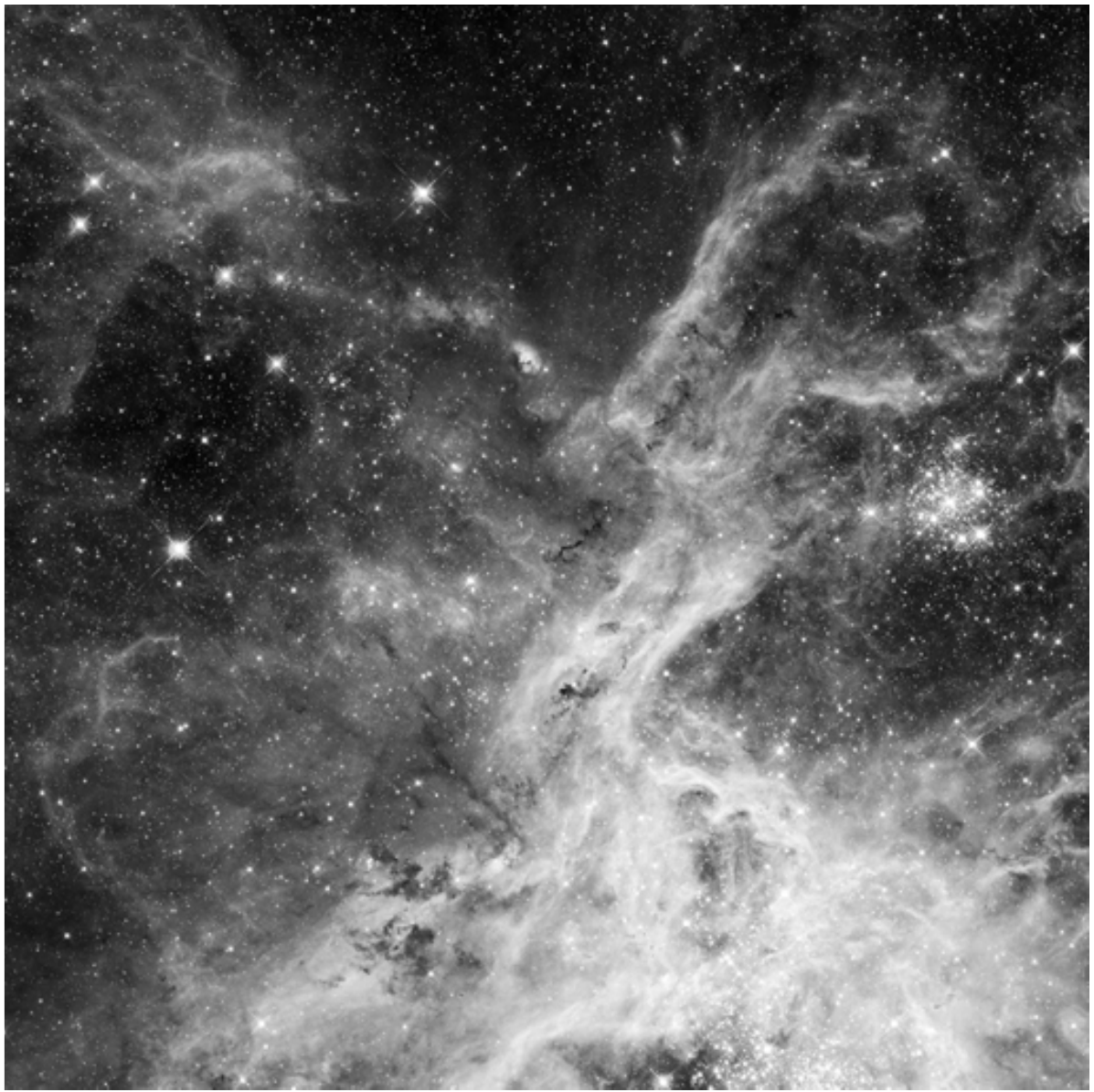}}
\subfigure{
\label{pics:lensing}
\includegraphics[width=0.965\columnwidth]{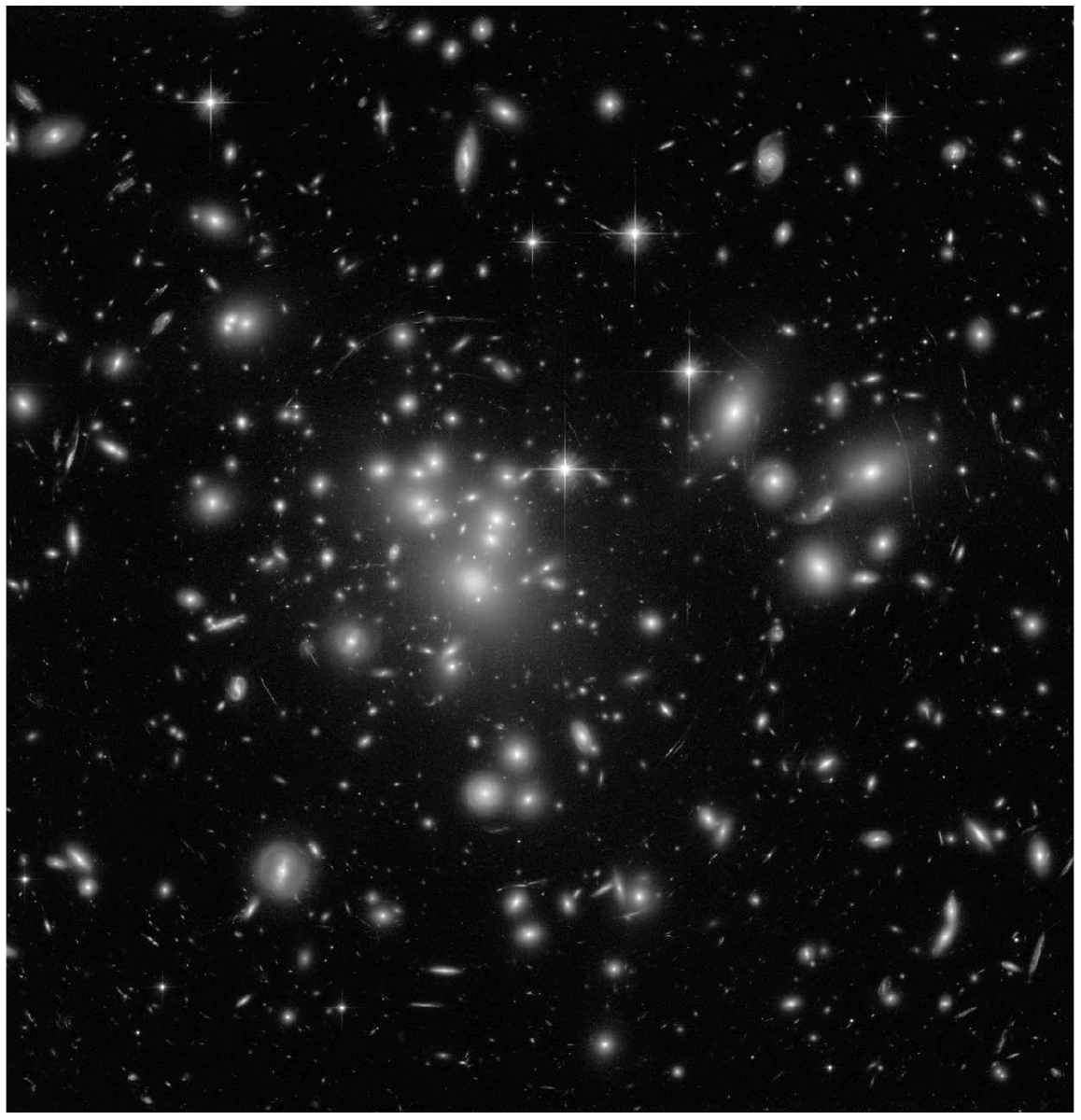}}
}}
\caption{Images used in CDL benchmarks. Figure \subref{pics:star_gen} is a Panoramic View of a Turbulent Star-making Region, and Figure \subref{pics:lensing} is an ACS/WFC image of Abell 1689.}
\label{fig:exp_images2}
\end{figure*}

\begin{figure*}[htb]
\centerline{
\vbox{
\hbox{
\subfigure[][]{%
\label{pics:second_nebula}
\includegraphics[width=1.3\columnwidth]{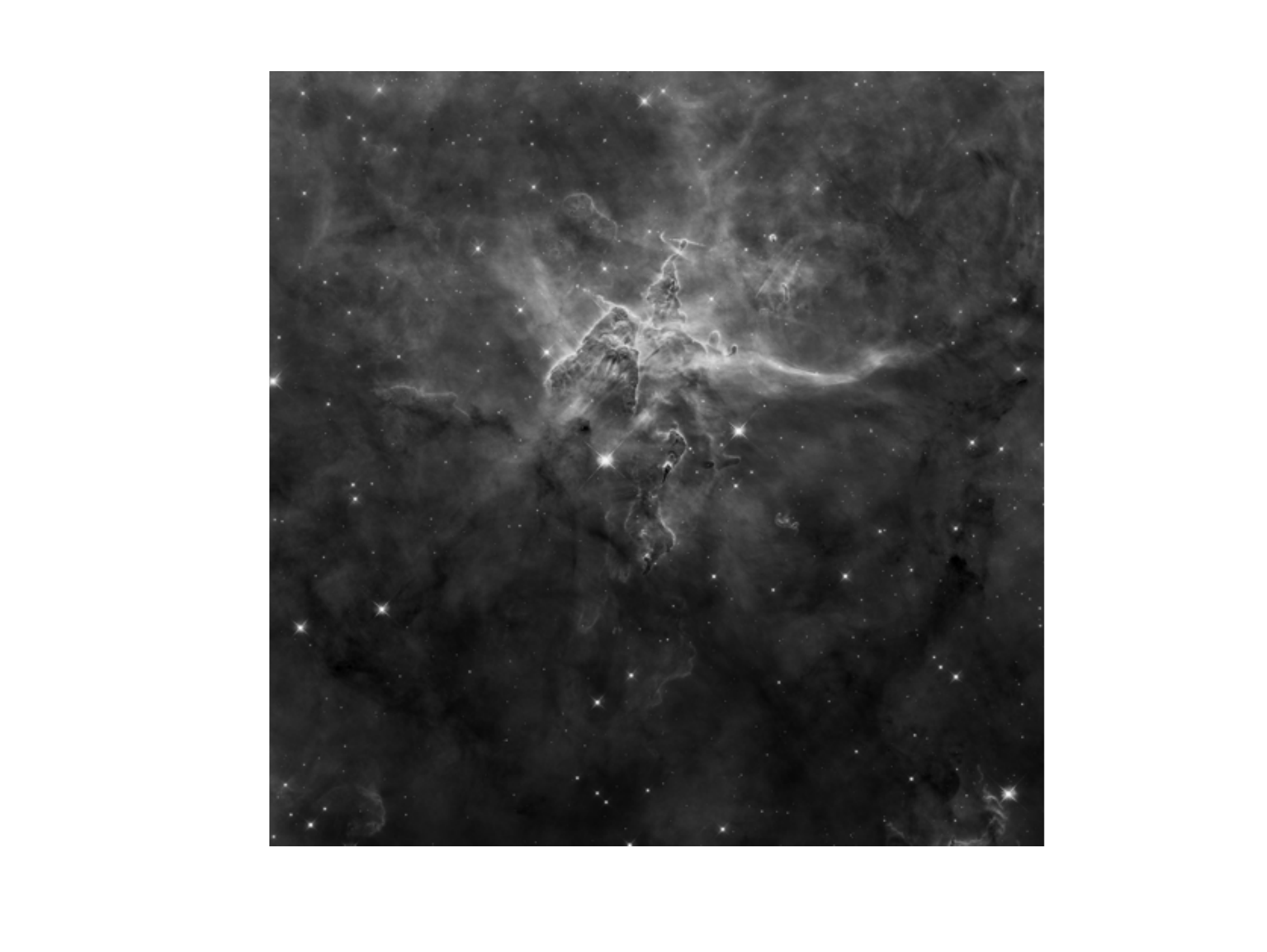}}
\subfigure[][]{%
\label{pics:second_galaxies}
\includegraphics[width=1.3\columnwidth]{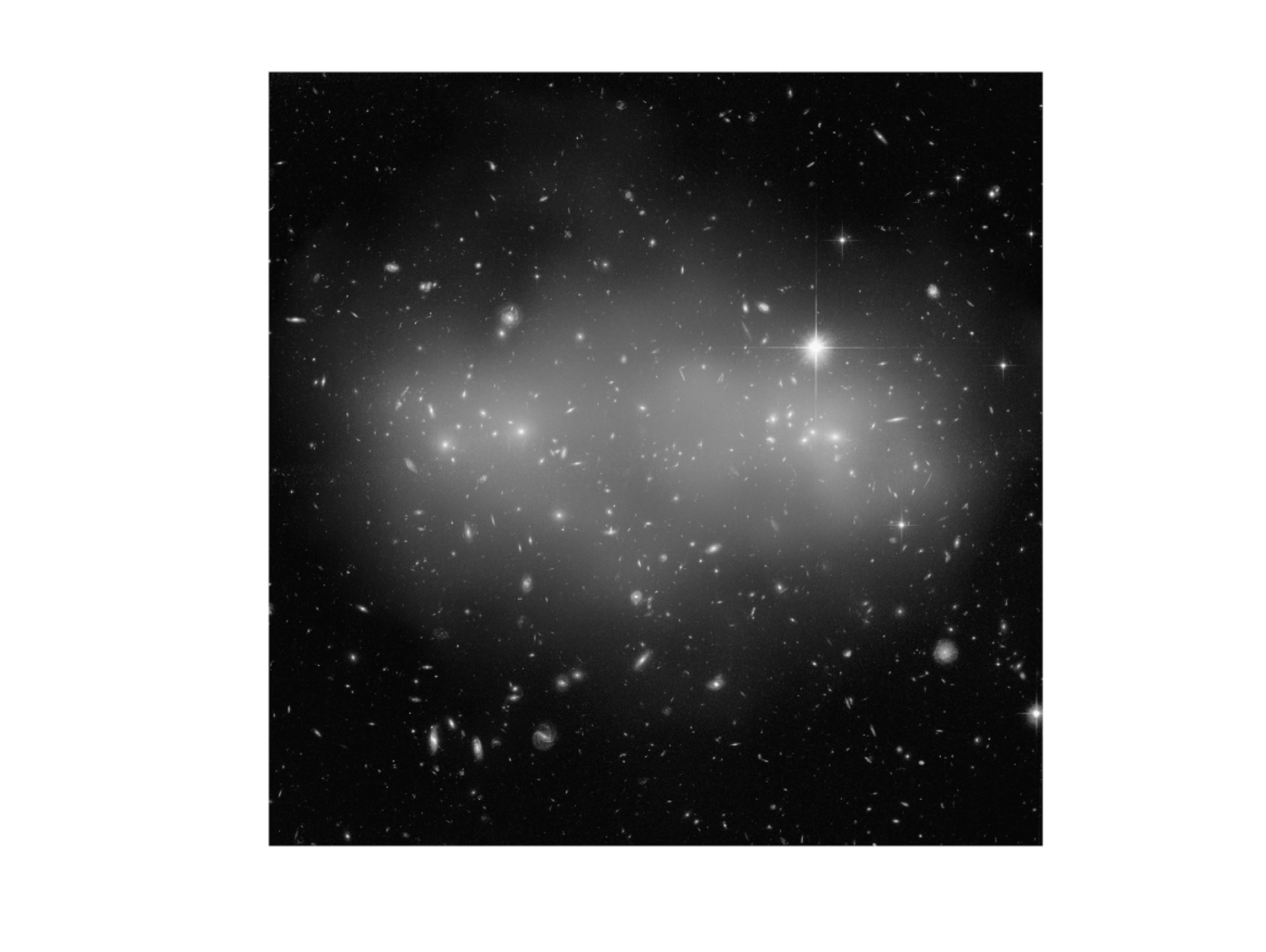}}
}
\hbox{
\subfigure[][]{%
\label{pics:second_star_gen}
\includegraphics[width=1.3\columnwidth]{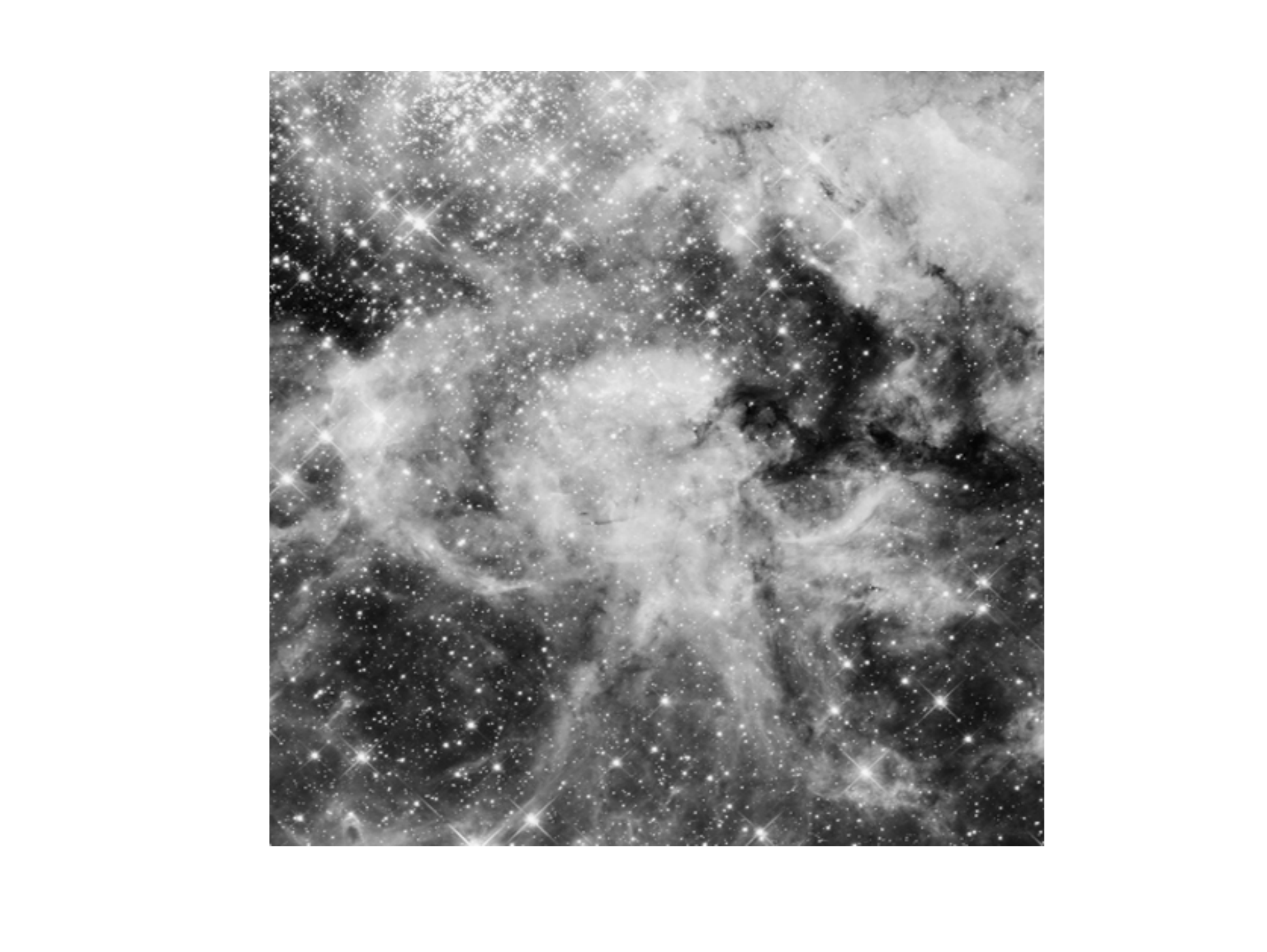}}
\subfigure[][]{%
\label{pics:second_lensing}
\includegraphics[width=1.3\columnwidth]{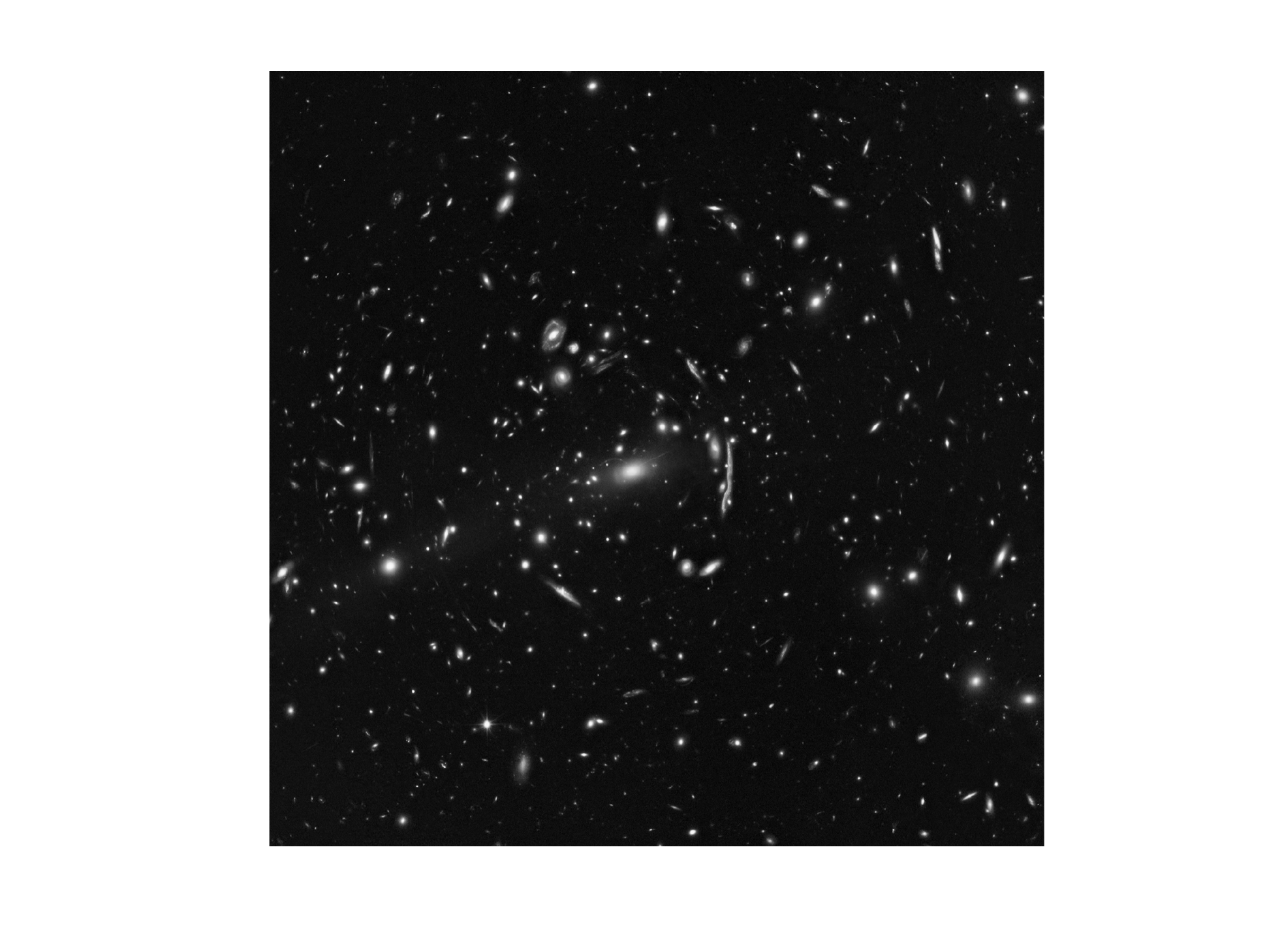}}
}}}
\caption{Hubble images used for noiseless dictionary learning. Figure \subref{pics:second_galaxies} is a Pandora's Cluster Abell, Figure \subref{pics:second_galaxies} is a galaxy cluster, Figure \subref{pics:second_star_gen} is a region in the Large Magellanic Cloud , and Figure \subref{pics:second_lensing} is a second Pandora's Cluster Abell.}
\label{fig:second_set}
\end{figure*}

\begin{figure*}
\centerline{
\hbox{
\subfigure{
\label{pics:dico}
\includegraphics[width=0.8\columnwidth]{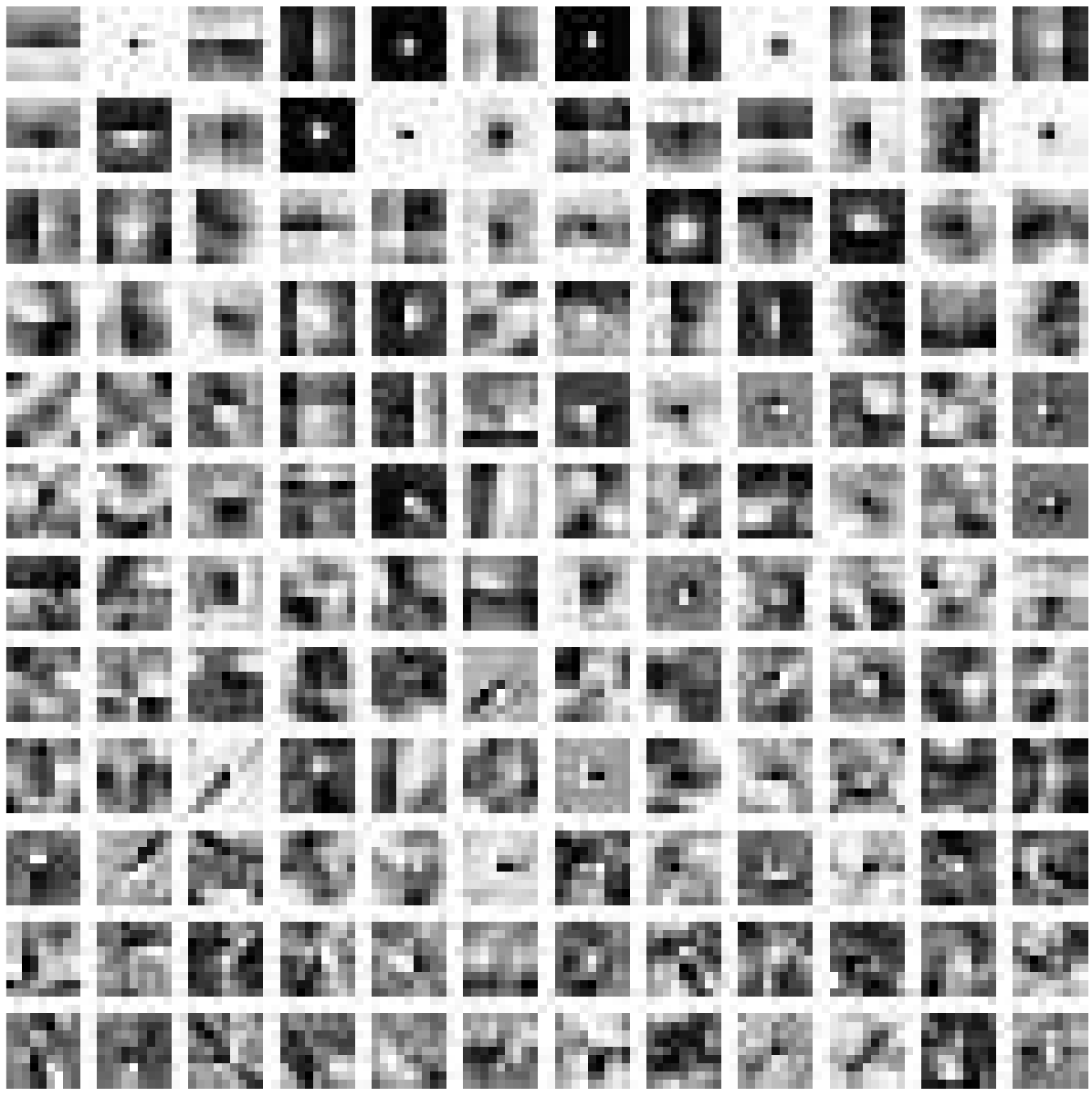}}
\subfigure{
\label{pics:curve}
\includegraphics[width=1\columnwidth]{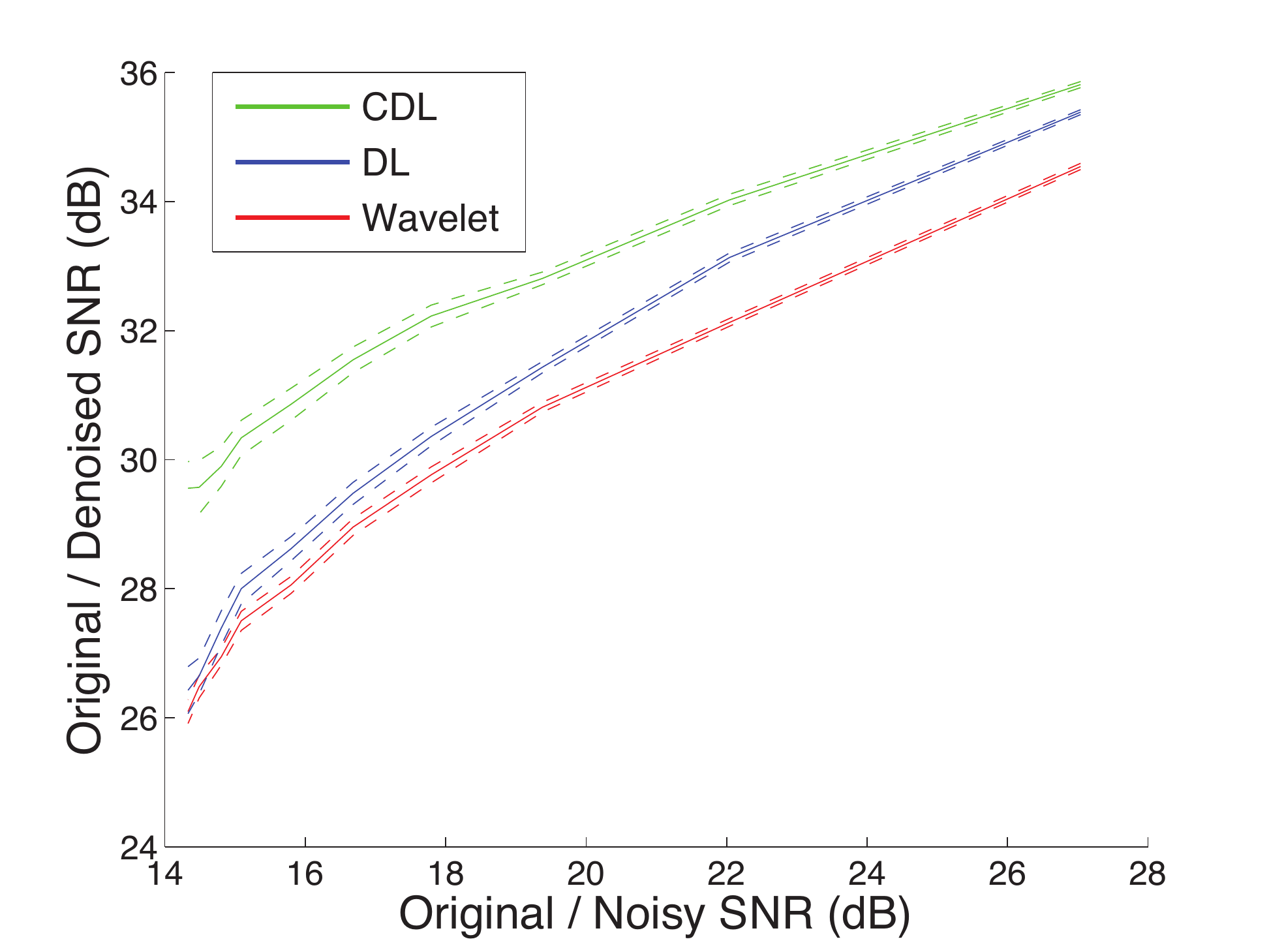}}
}}
\caption{Benchmark for nebula image from Figure \ref{fig:exp_images} comparing CDL to DL and wavelet denoising methods. \subref{pics:dico} shows a centered learned dictionary learned on a second, noiseless image and used for denoising. \subref{pics:curve} shows the PSNR curve for the three methods. Centered dictionary learning method is represented by the green curve, the classic dictionary learning in blue and the wavelet-based method in red. The horizontal axis represents the PSNR (dB) between the image before and after adding noise. For denoising, we use OMP with a stopping criterion fixed depending on the level of noise that was added. 100 experiments were repeated for each value of noise. }
\label{fig:centered_nebula}
\end{figure*}

\begin{figure*}
\centerline{
\hbox{
\subfigure{
\label{pics:dico}
\includegraphics[width=0.8\columnwidth]{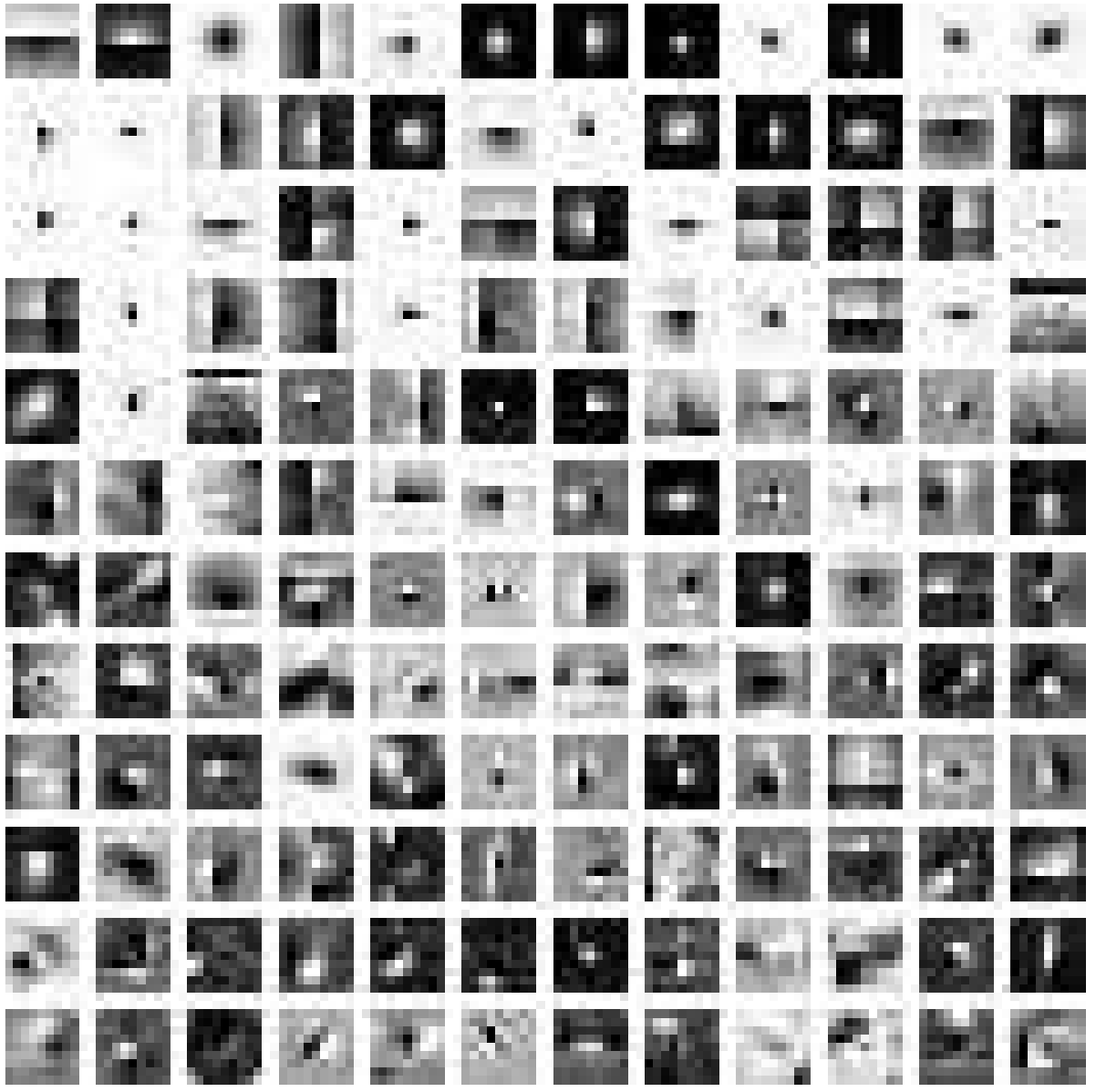}}
\subfigure{
\label{pics:curve}
\includegraphics[width=1\columnwidth]{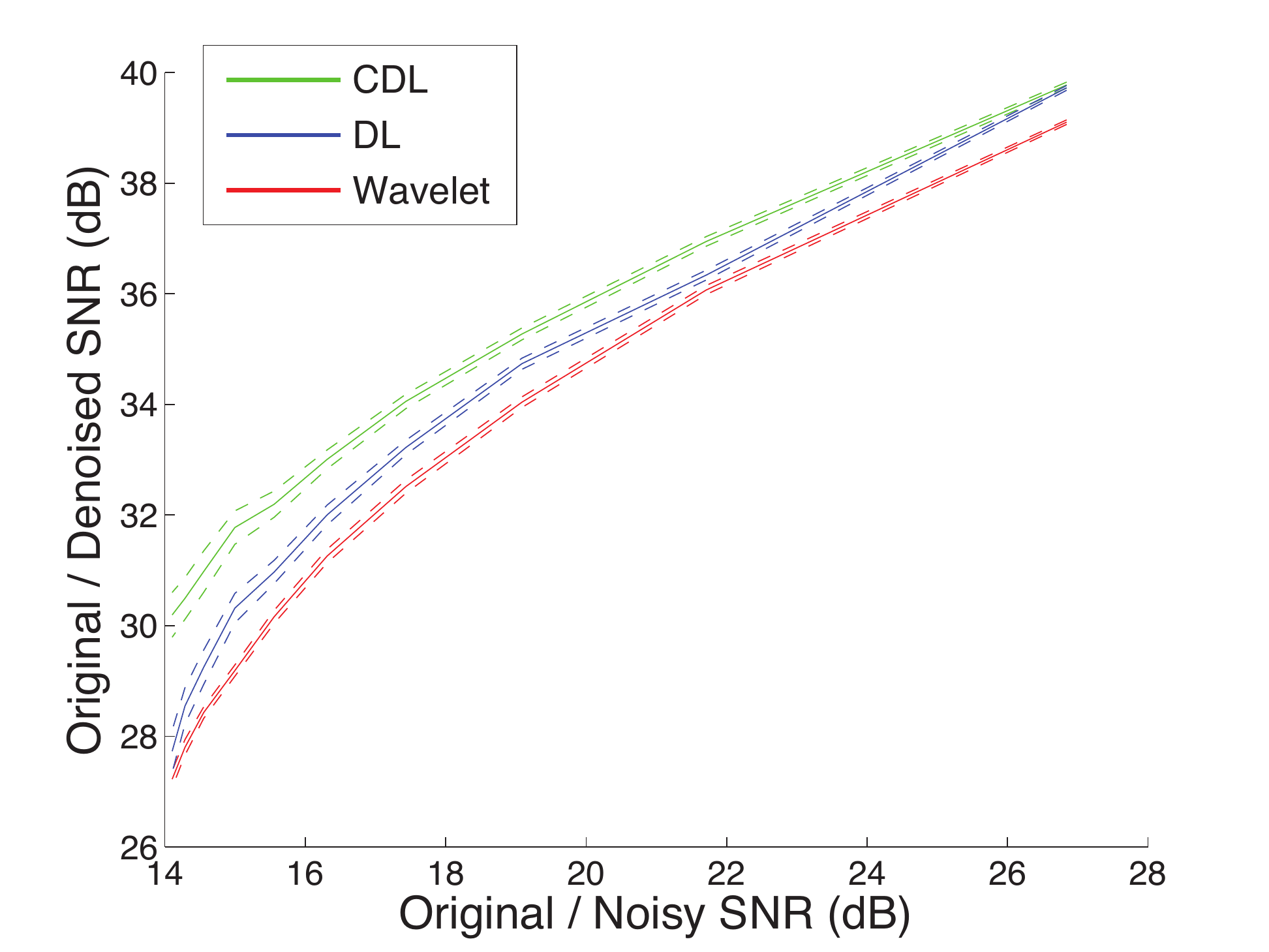}}
}}
\caption{Benchmark for galaxy cluster image from Figure \ref{fig:exp_images} comparing CDL to DL and wavelet denoising methods. \subref{pics:dico} shows a centered learned dictionary learned on a second, noiseless image and used for denoising. \subref{pics:curve} shows the PSNR curve for the three methods. Centered dictionary learning method is represented by the green curve, the classic dictionary learning in blue and the wavelet-based method in red. The horizontal axis represents the PSNR (dB) between the image before and after adding noise. For denoising, we use OMP with a stopping criterion fixed depending on the level of noise that was added. 100 experiments were repeated for each value of noise. }
\label{fig:centered_galaxies}
\end{figure*}

\begin{figure*}
\centerline{
\hbox{
\subfigure{
\label{pics:dico}
\includegraphics[width=0.8\columnwidth]{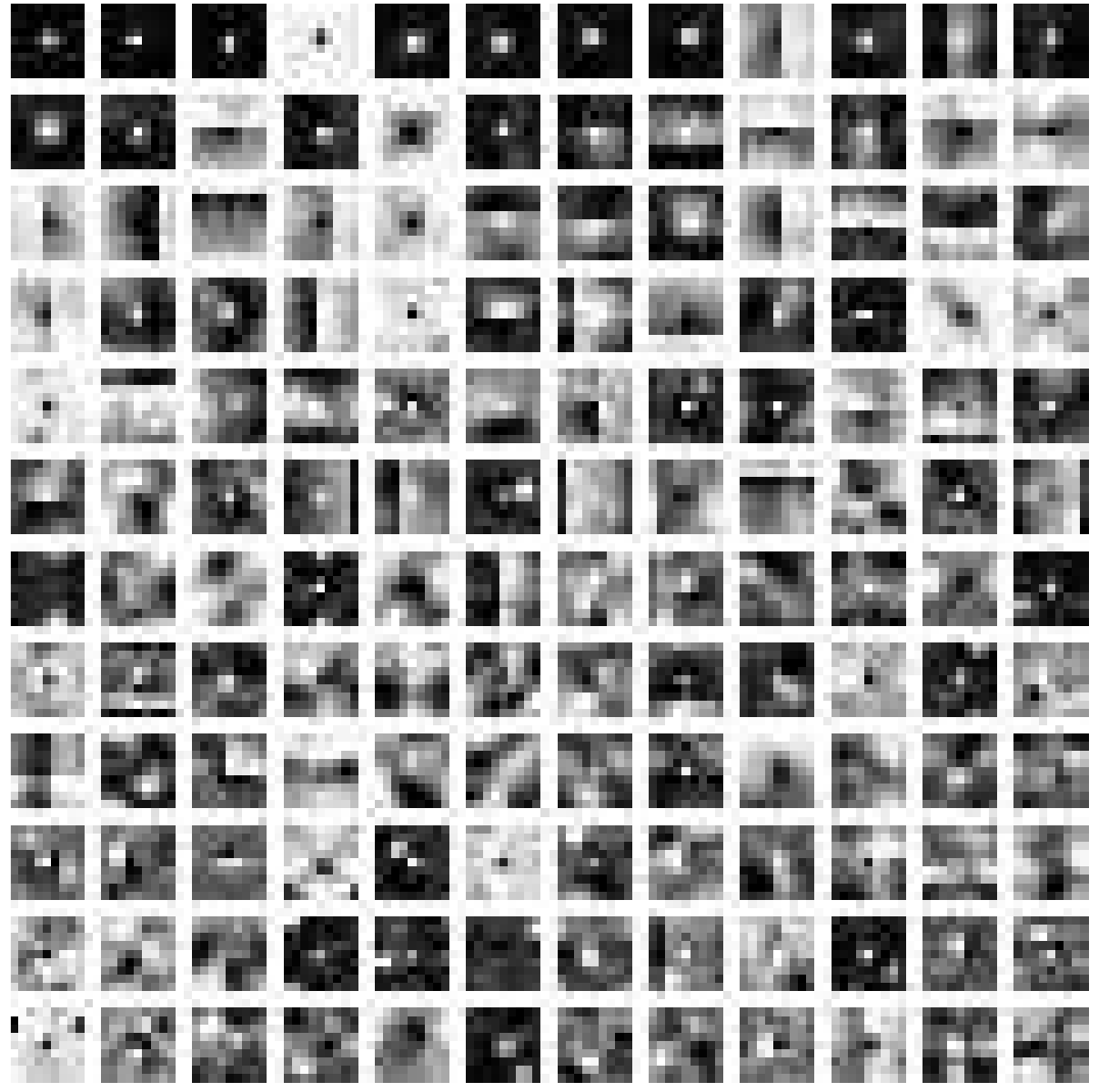}}
\subfigure{
\label{pics:curve}
\includegraphics[width=1\columnwidth]{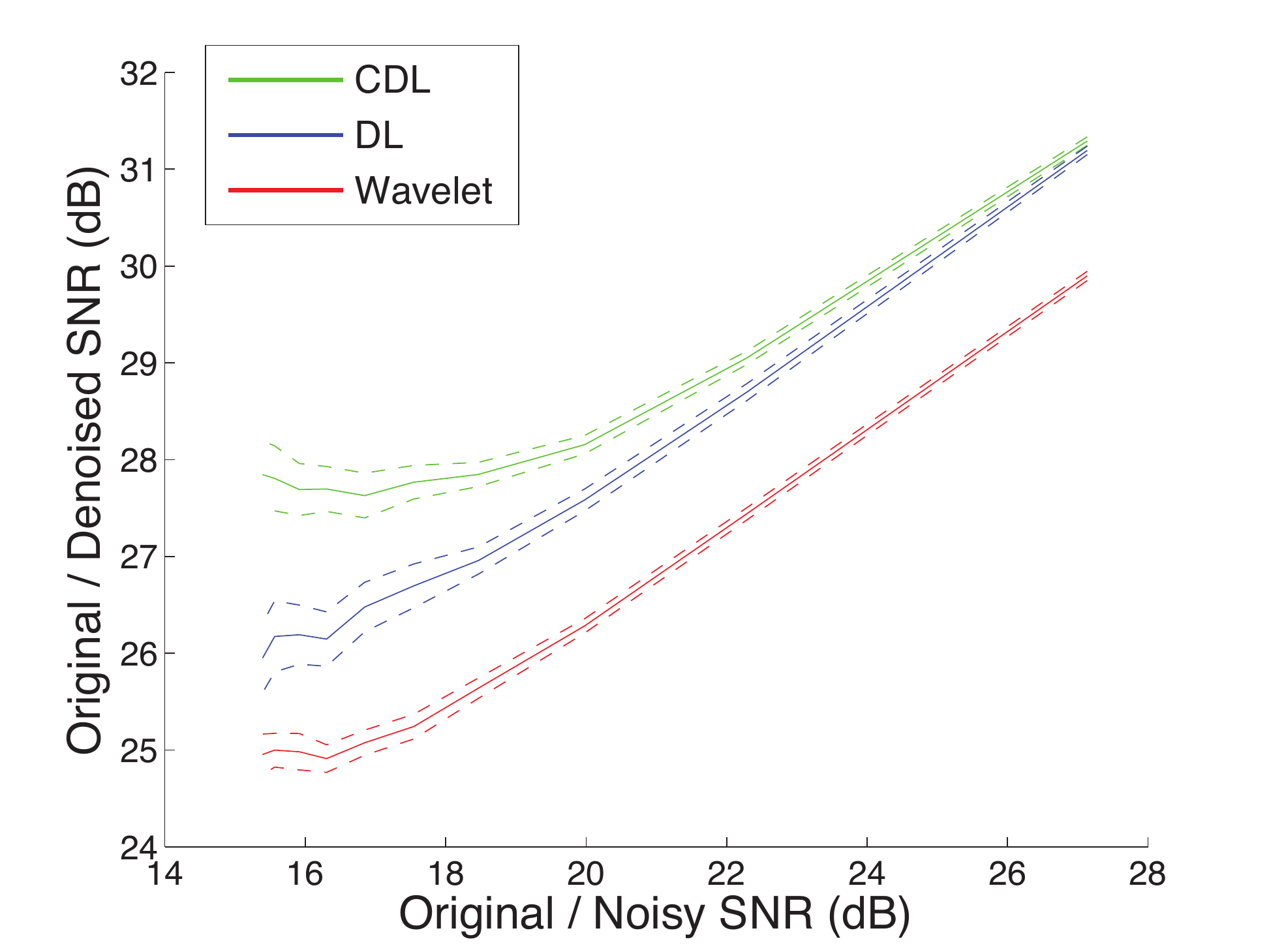}}
}}
\caption{Benchmark for star-making region image from Figure \ref{fig:exp_images2} comparing CDL to DL and wavelet denoising methods. \subref{pics:dico} shows a centered learned dictionary learned on a second, noiseless image and used for denoising. \subref{pics:curve} shows the PSNR curve for the three methods. Centered dictionary learning method is represented by the green curve, the classic dictionary learning in blue and the wavelet-based method in red. The horizontal axis represents the PSNR (dB) between the image before and after adding noise. For denoising, we use OMP with a stopping criterion fixed depending on the level of noise that was added. 100 experiments were repeated for each value of noise. }
\label{fig:centered_star_gen}
\end{figure*}

\begin{figure*}
\centerline{
\hbox{
\includegraphics[width=0.8\columnwidth]{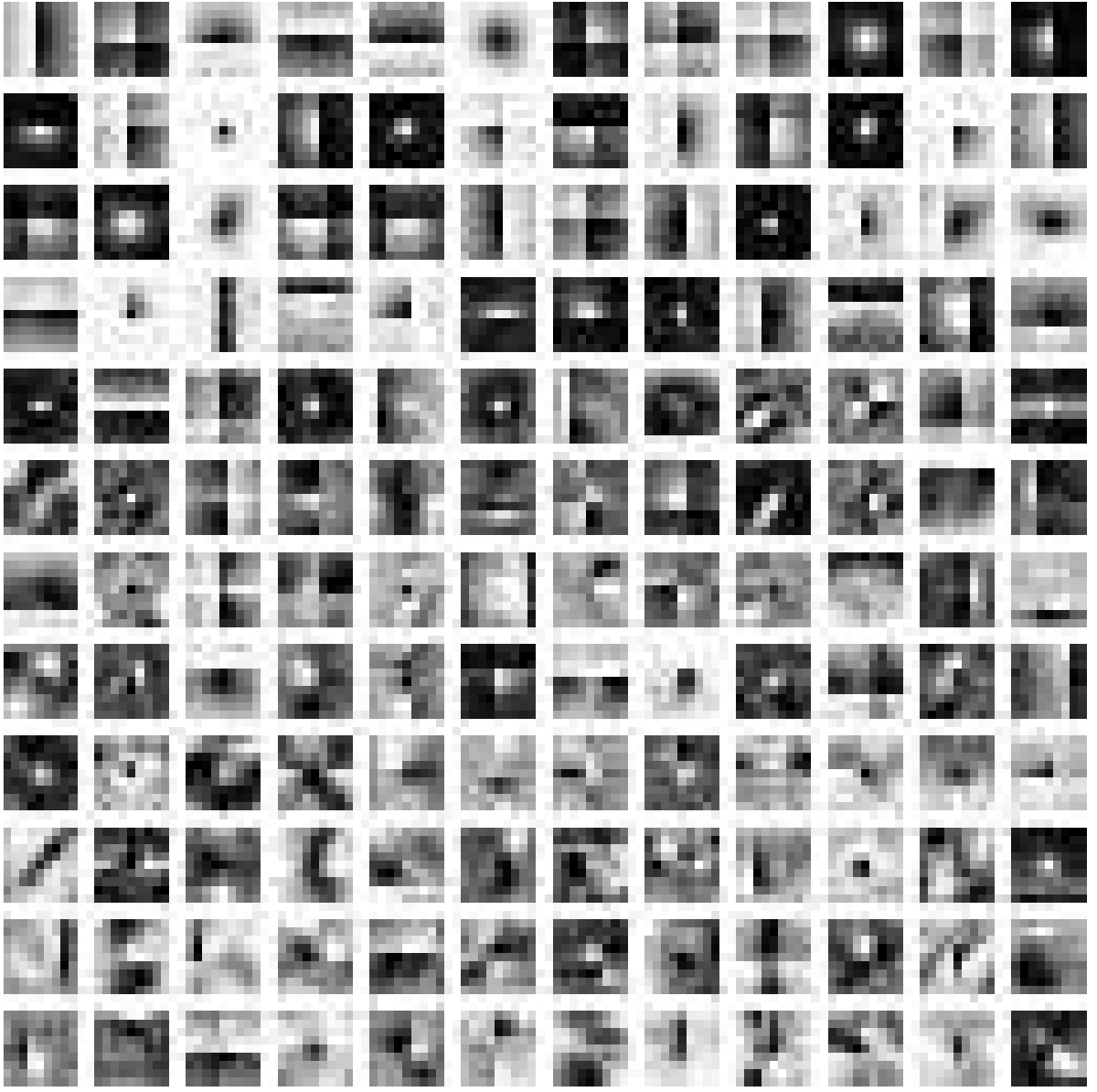}
\includegraphics[width=1\columnwidth]{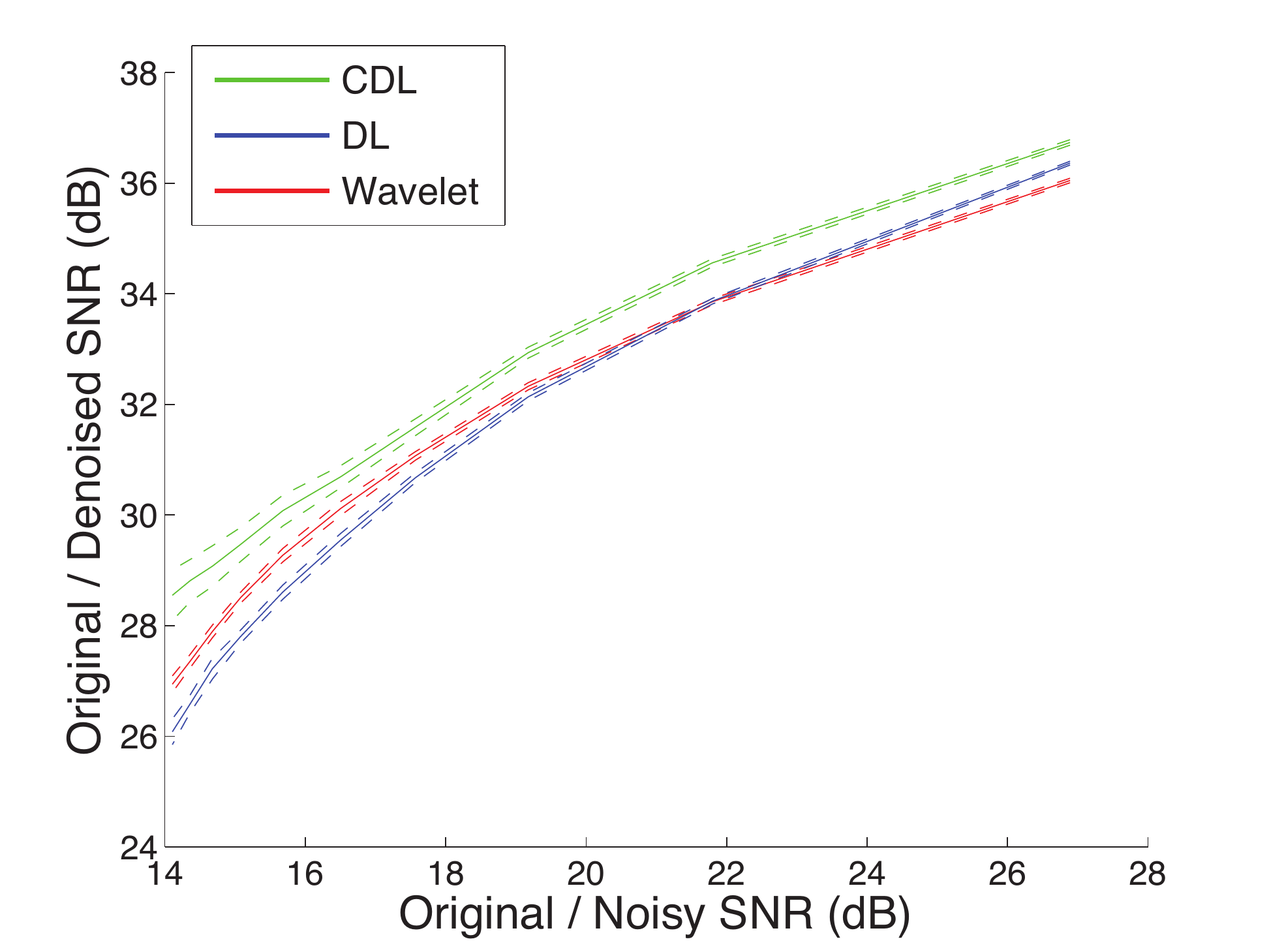}
}}
\caption{Benchmark for Abell 1689 image from Figure \ref{fig:exp_images2} comparing CDL to DL and wavelet denoising methods. \subref{pics:dico} shows a centered learned dictionary learned on a second, noiseless image and used for denoising. \subref{pics:curve} shows the PSNR curve for the three methods. Centered dictionary learning method is represented by the green curve, the classic dictionary learning in blue and the wavelet-based method in red. The horizontal axis represents the PSNR (dB) between the image before and after adding noise. For denoising, we use OMP with a stopping criterion fixed depending on the level of noise that was added. 100 experiments were repeated for each value of noise. }
\label{fig:centered_lensing}
\end{figure*}

We show now how DL behaves when learning on real astronomical noiseless images, that is images that present an extremely low level of noise or that have been denoised and thus are considered noiseless. We give several benchmarks to show how the centered dictionary learning is able to outperforms the classic approach. We denoise two previously presented images, and two additional images shown in Figure \ref{fig:exp_images2}. We perform the learning step on similar noiseless images, see Figure \ref{fig:second_set}. The benchmark results are presented in Figures \ref{fig:centered_nebula}, \ref{fig:centered_galaxies}, \ref{fig:centered_star_gen} and \ref{fig:centered_lensing}. Figure \ref{fig:centered_lensing} illustrates a particular case where the classical dictionary learning becomes less efficient than the wavelet-based denoising algorithm while using the centered learning and denoising yields better results at any noise level. For each benchmark, we added a white Gaussian noise with a varying standard deviation to one image and learn a centered dictionary and a non-centered dictionary on a second similar noiseless image. We use the same set of parameters for both learning. CDL performs better than the classic DL method and wavelet-based denoising. A consequence of the better sparsifying capability of the centered dictionary is a faster computation during the sparse coding step. The noiseless dictionaries prove to be efficient for any level of noise.

\subsection{Photometry and source detection}
Although the final photometry is generally done on the raw data  \cite{xmmlss,fermi2012_catalog}, 
it is important that the denoising does not introduce a strong bias on the flux of the different sources because it would 
dump their amplitude and reduce the number of detected sources.

\begin{figure*}
\centerline{
\hbox{
\subfigure{
\label{pics:threshold}
\includegraphics[width=1.2\columnwidth]{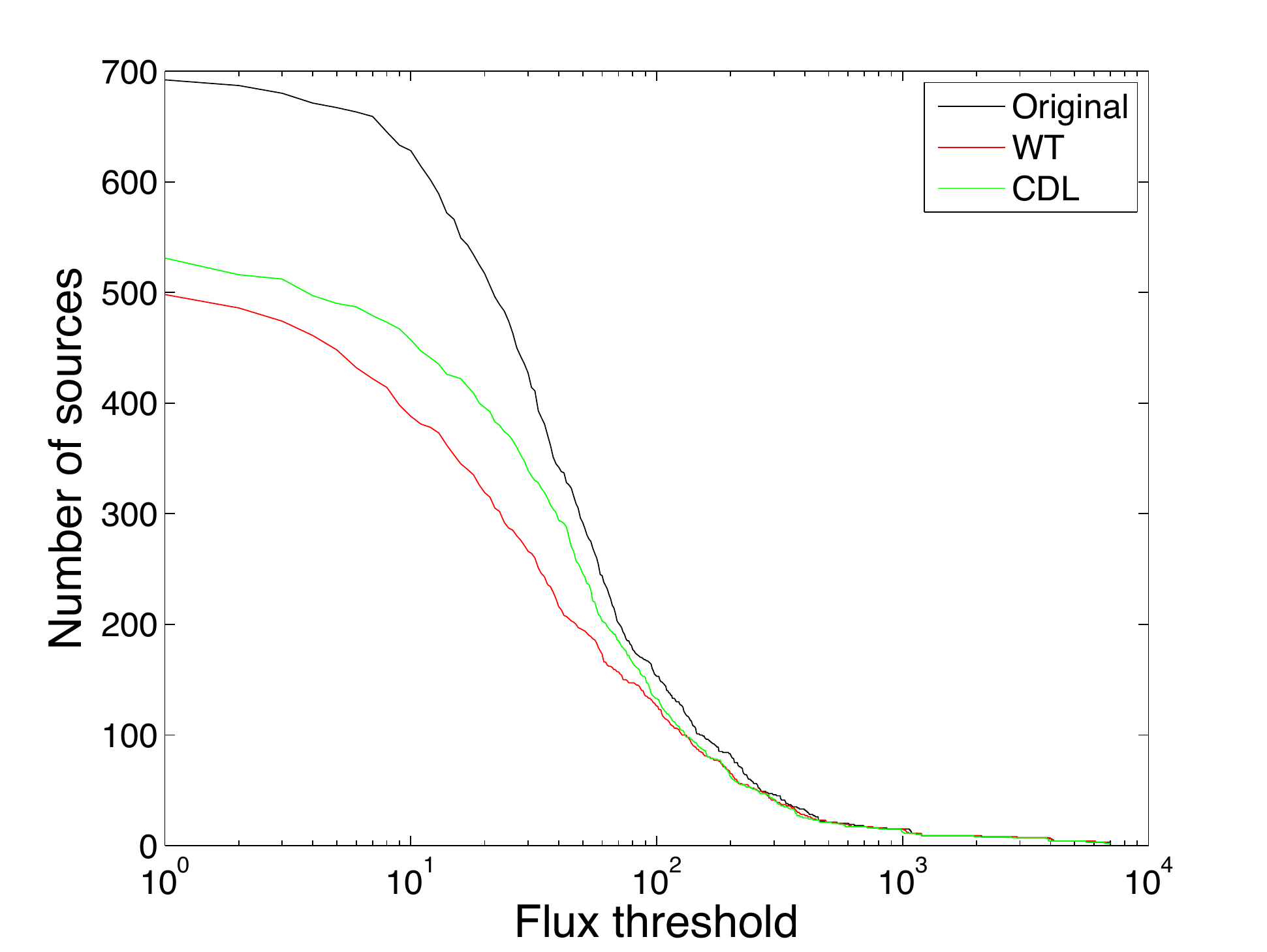}
}
\subfigure{
\label{pics:conservation}
\includegraphics[width=1.2\columnwidth]{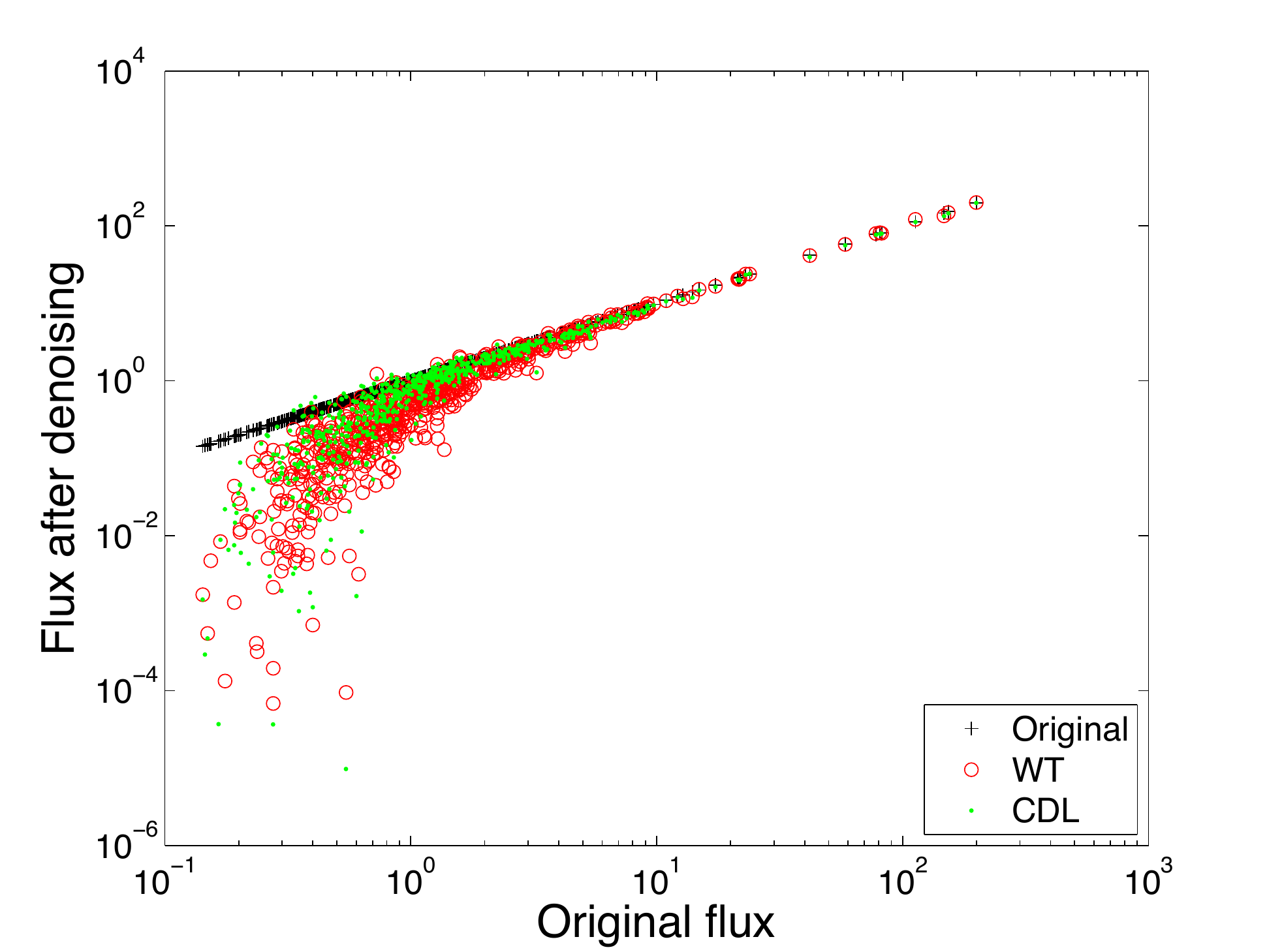}
}
}}
\caption{Source photometry comparison between CDL and wavelet denoising. Figure \subref{pics:threshold} shows how many sources have a flux above a varying threshold after denoising. Figure \subref{pics:conservation} shows how the flux is dampened by denoising, representing the source flux after denoising as a function of the source flux before denoising}
\label{fig:photometry_curves}
\end{figure*}

\begin{figure*}
\centerline{
\vbox{
\hbox{
\subfigure[][]{%
\label{pics:original}
\includegraphics[width=0.6\columnwidth]{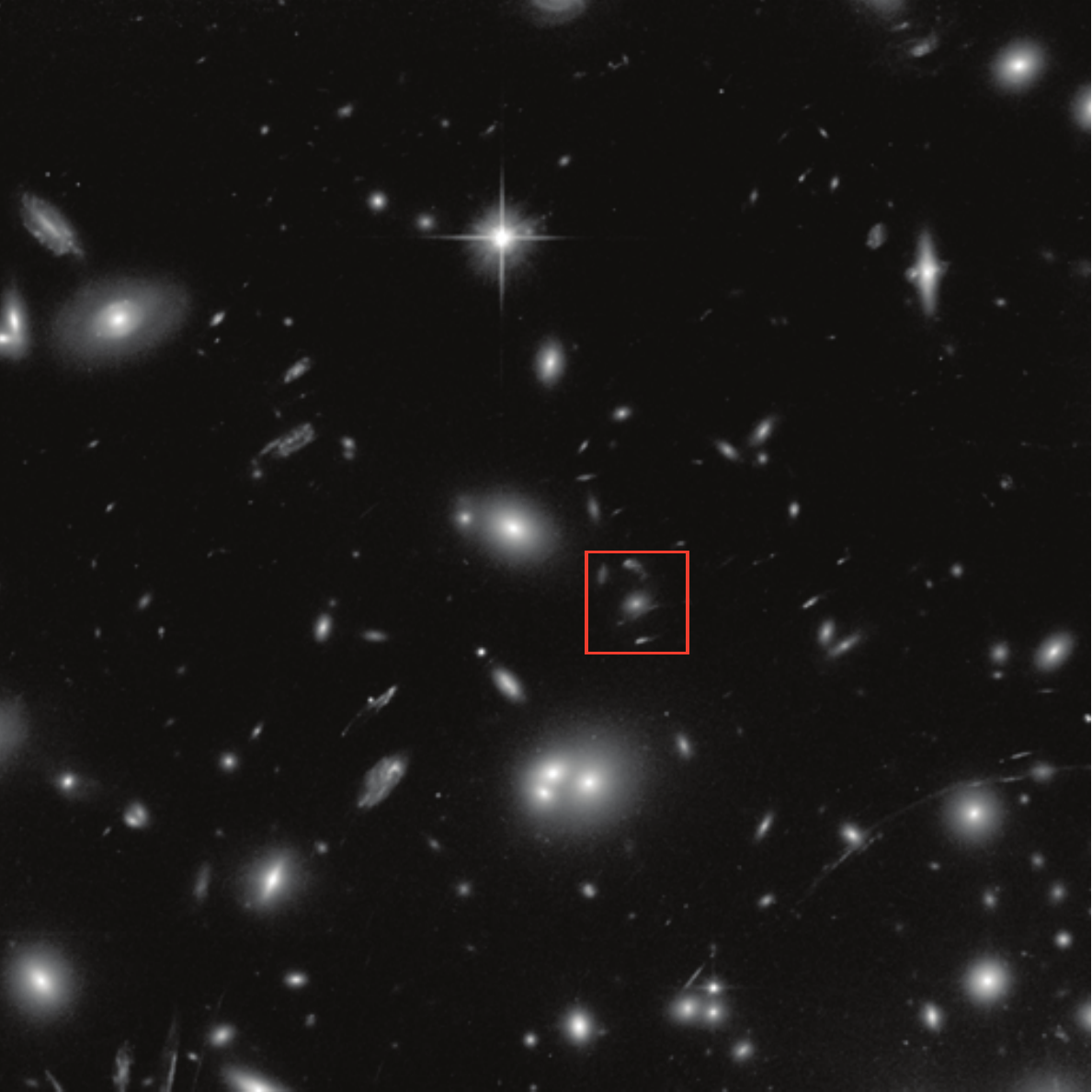}}
\subfigure[][]{%
\label{pics:src}
\includegraphics[width=0.6\columnwidth]{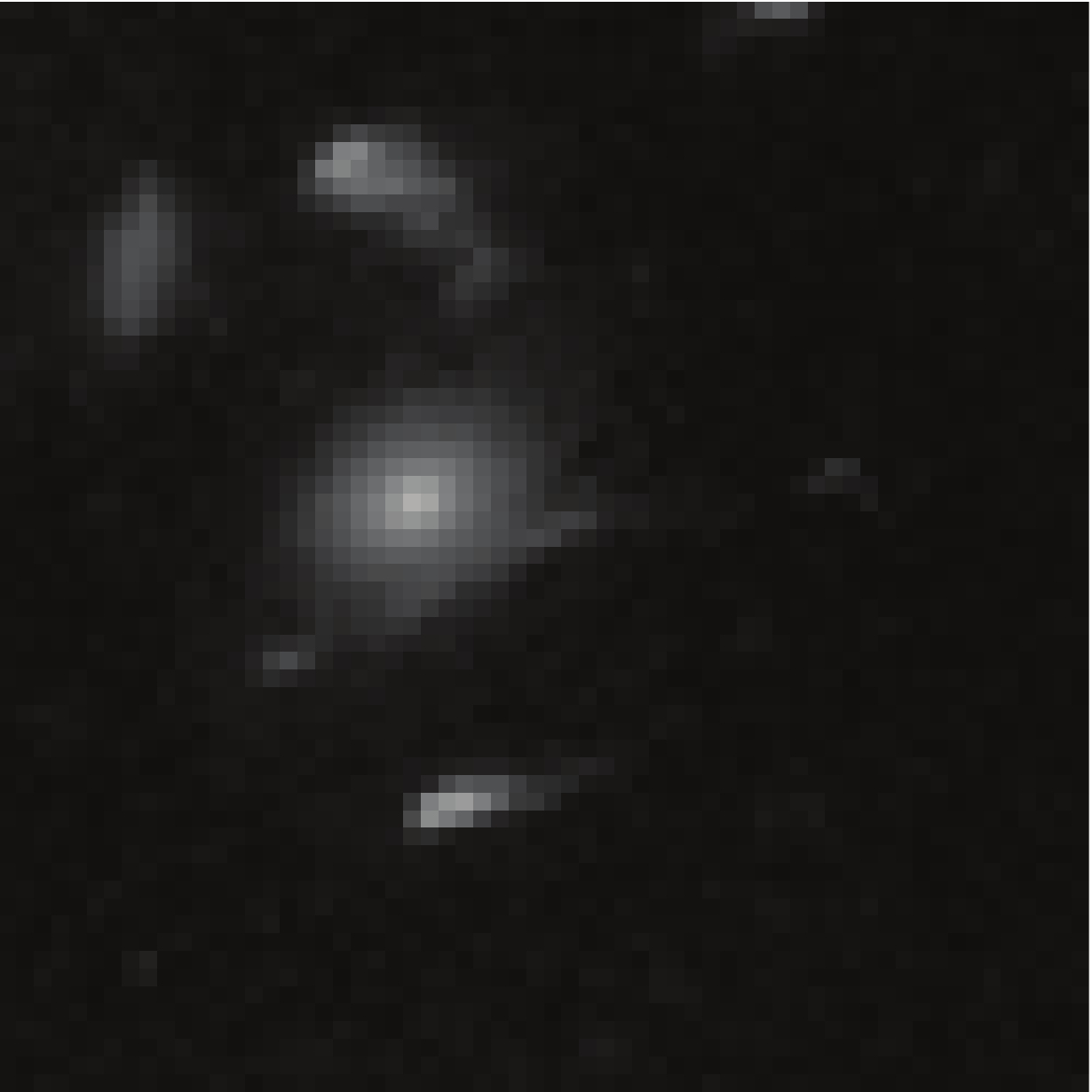}}
\subfigure[][]{%
\label{pics:n}
\includegraphics[width=0.6\columnwidth]{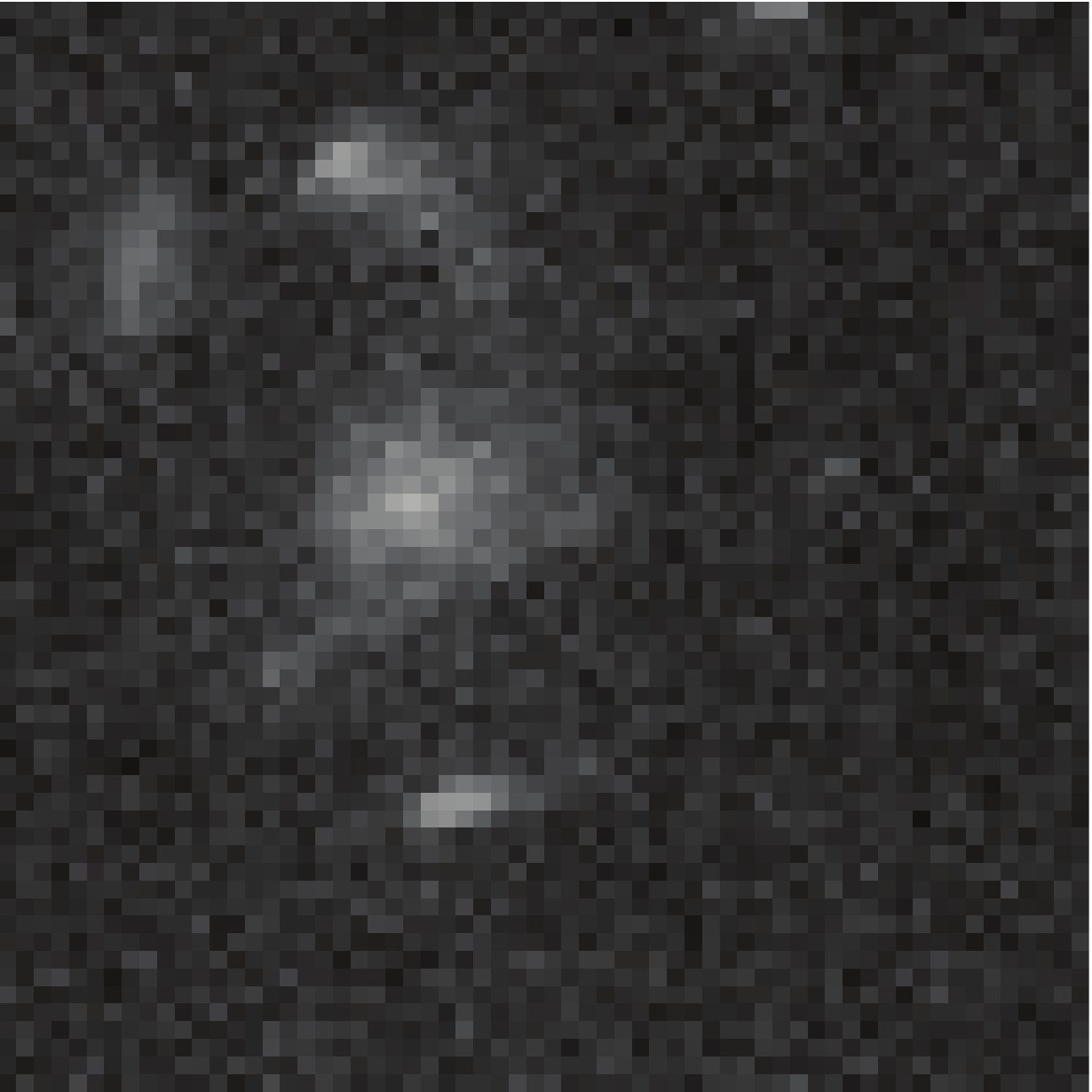}}
}
\hbox{
\subfigure[][]{%
\label{pics:w}
\includegraphics[width=0.6\columnwidth]{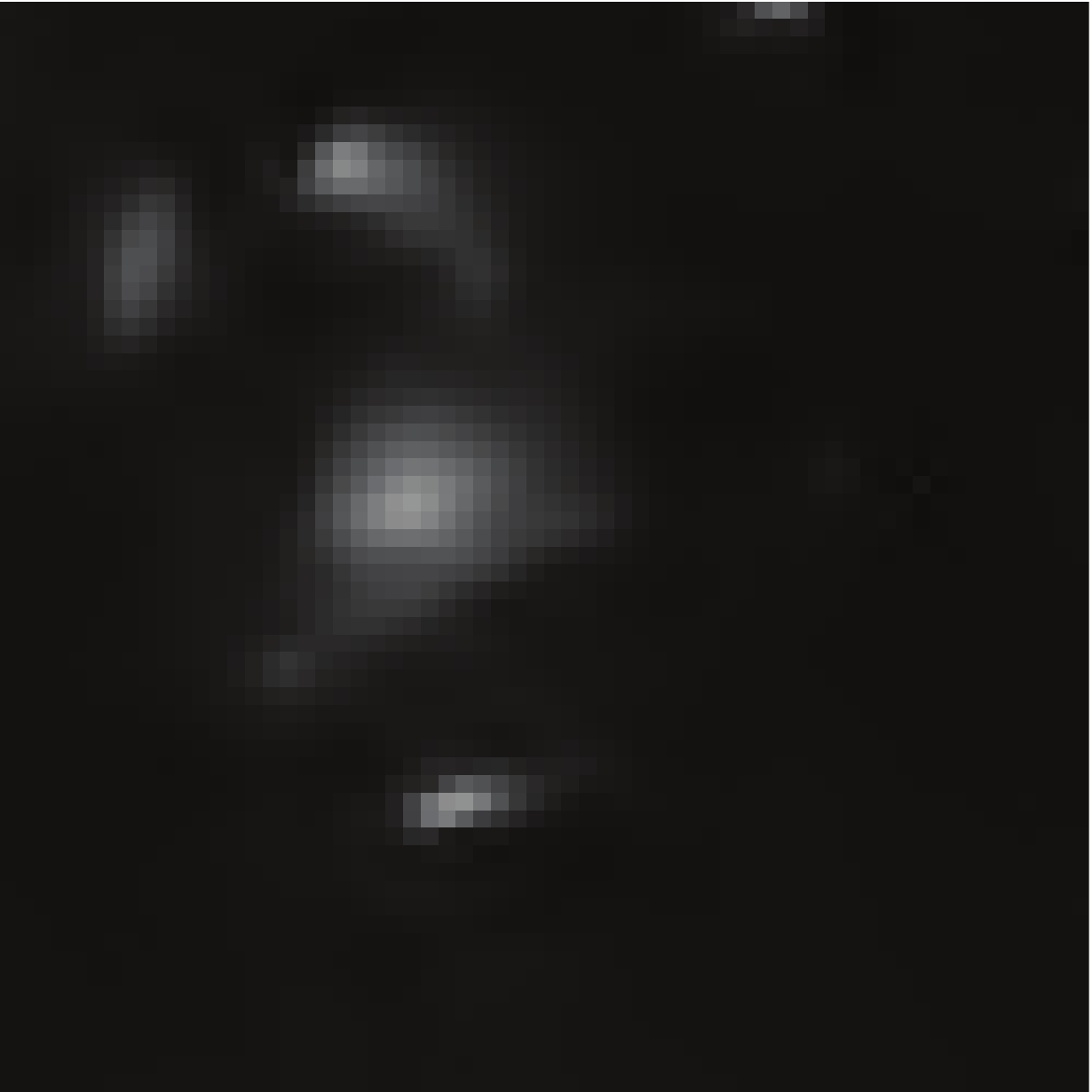}}
\subfigure[][]{%
\label{pics:dl}
\includegraphics[width=0.6\columnwidth]{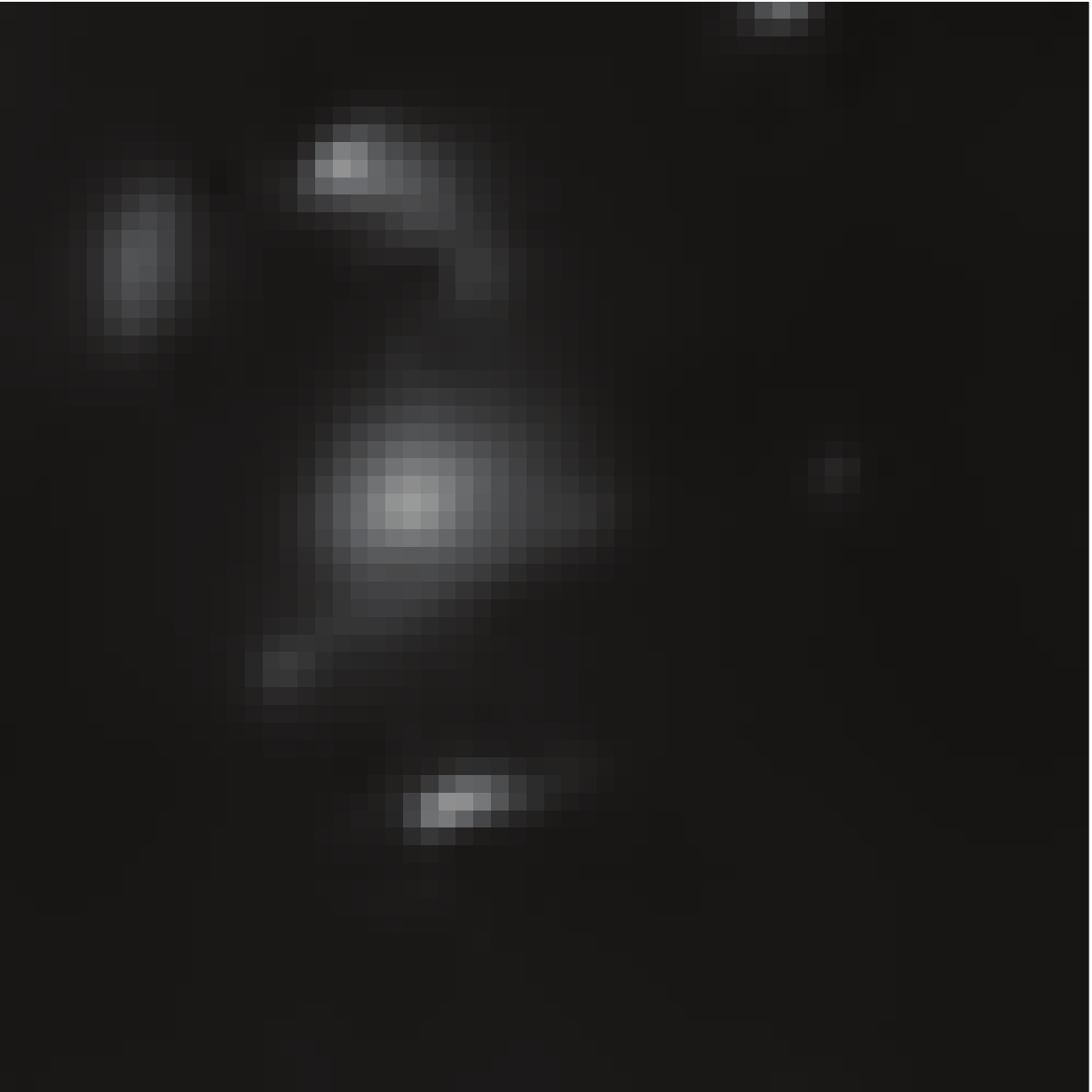}}
\subfigure[][]{%
\label{pics:dlc}
\includegraphics[width=0.6\columnwidth]{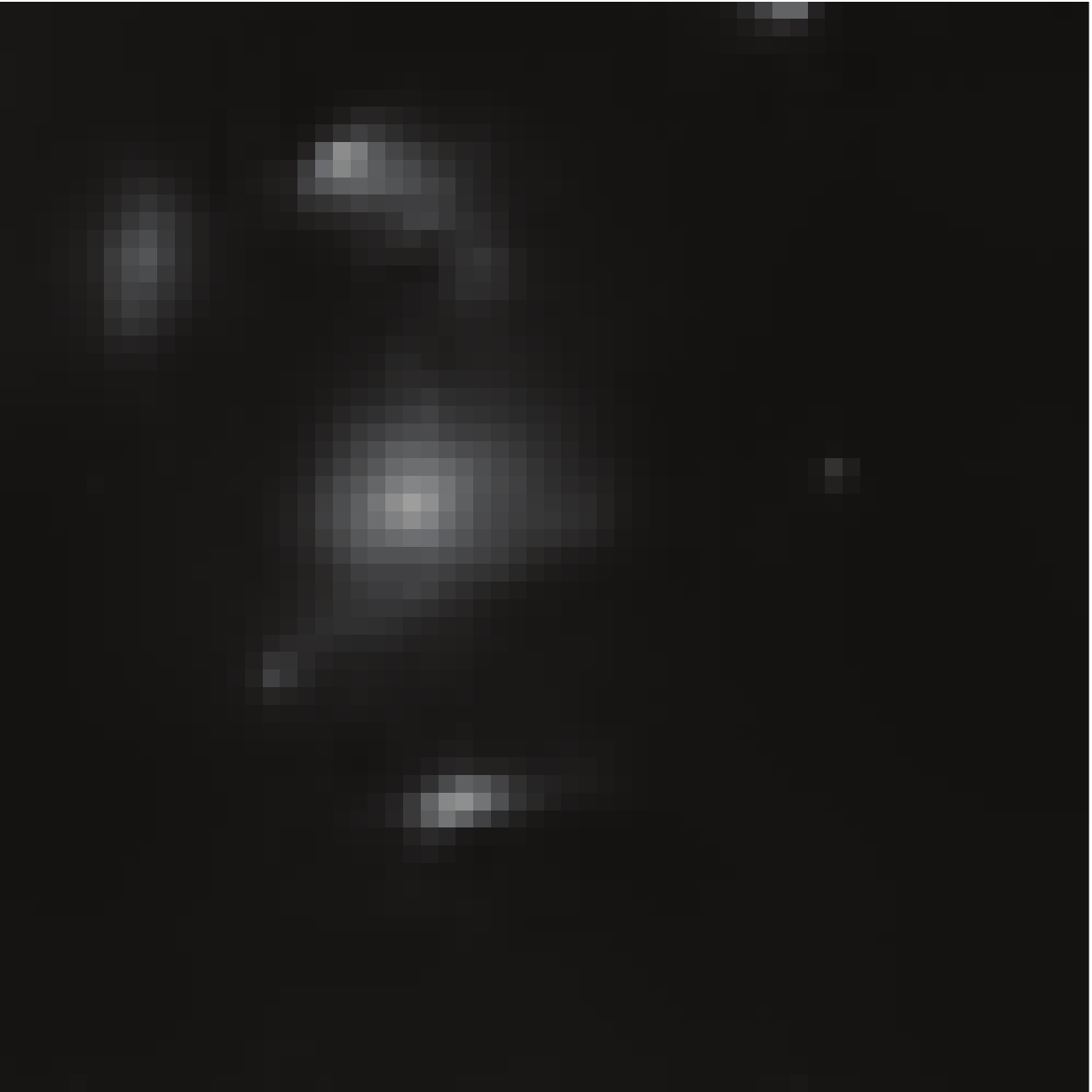}}
}}}
\caption{Zoomed features extracted from a galaxy cluster image. \subref{pics:original} shows the full source image before adding noise, \subref{pics:src} shows the noiseless source, \subref{pics:n} shows the noisy version, and \subref{pics:w}, \subref{pics:dl} and \subref{pics:dlc} respectively show the denoised feature using wavelet denoising, classic dictionary learning and centered dictionary learning.}
\label{fig:zoom_lensing}
\end{figure*}

\begin{figure*}
\centerline{
\vbox{
\hbox{
\subfigure[][]{%
\label{pics:original}
\includegraphics[width=0.642\columnwidth]{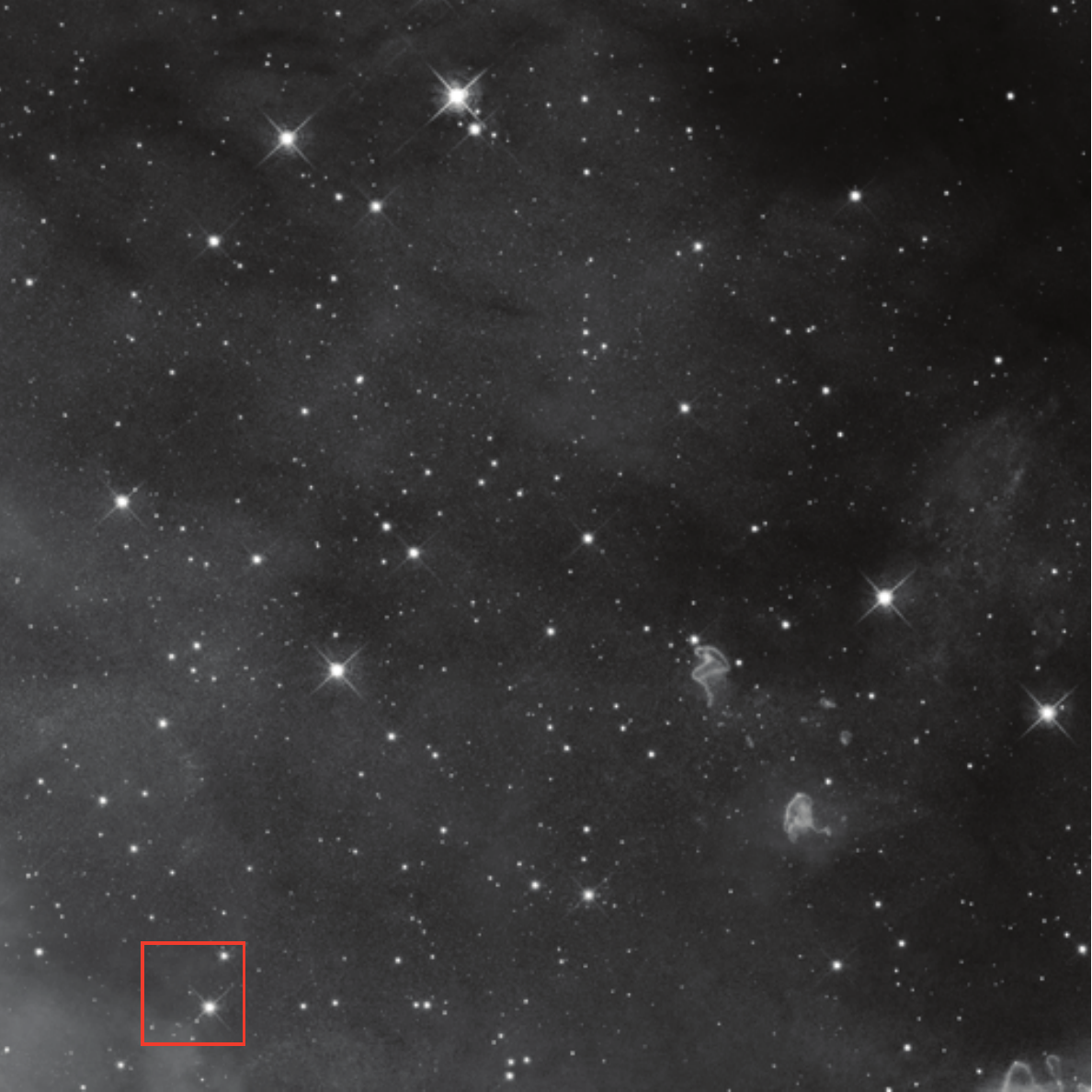}}
\subfigure[][]{%
\label{pics:src}
\includegraphics[width=0.6\columnwidth]{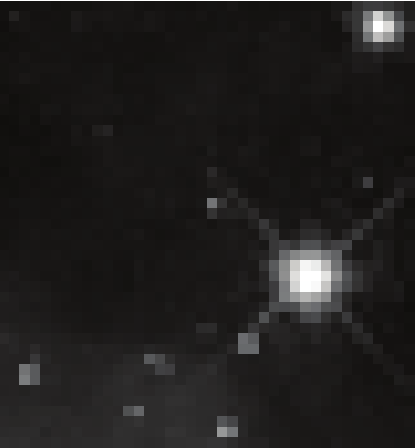}}
\subfigure[][]{%
\label{pics:n}
\includegraphics[width=0.6\columnwidth]{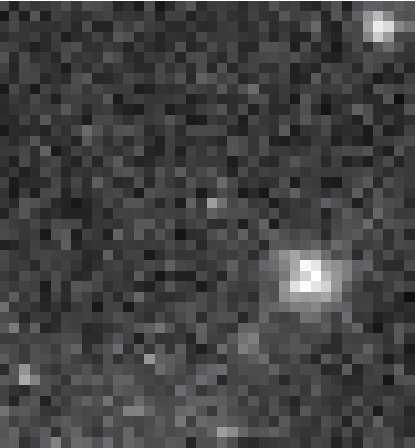}}
}
\hbox{
\subfigure[][]{%
\label{pics:w}
\includegraphics[width=0.6\columnwidth]{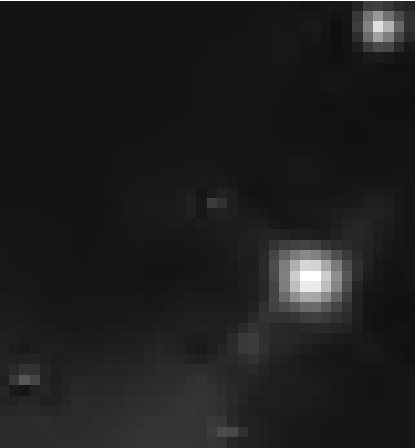}}
\subfigure[][]{%
\label{pics:dl}
\includegraphics[width=0.6\columnwidth]{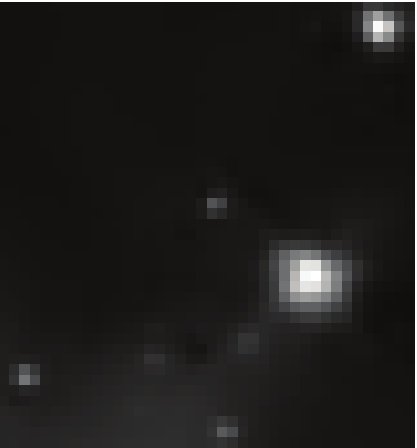}}
\subfigure[][]{%
\label{pics:dlc}
\includegraphics[width=0.6\columnwidth]{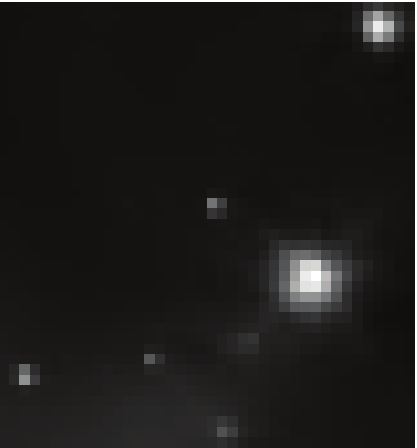}}
}}}
\caption{Zoomed features extracted from the previously shown nebular image. \subref{pics:original} shows the full source image before adding noise, \subref{pics:src} shows the noiseless source, \subref{pics:n} shows the noisy version, and \subref{pics:w}, \subref{pics:dl} and \subref{pics:dlc} respectively show the denoised feature using wavelet denoising, classic dictionary learning and centered dictionary learning.}
\label{fig:zoom_neb_1}
\end{figure*}

\begin{figure*}
\centerline{
\vbox{
\hbox{
\subfigure[][]{%
\label{pics:original}
\includegraphics[width=0.5\columnwidth]{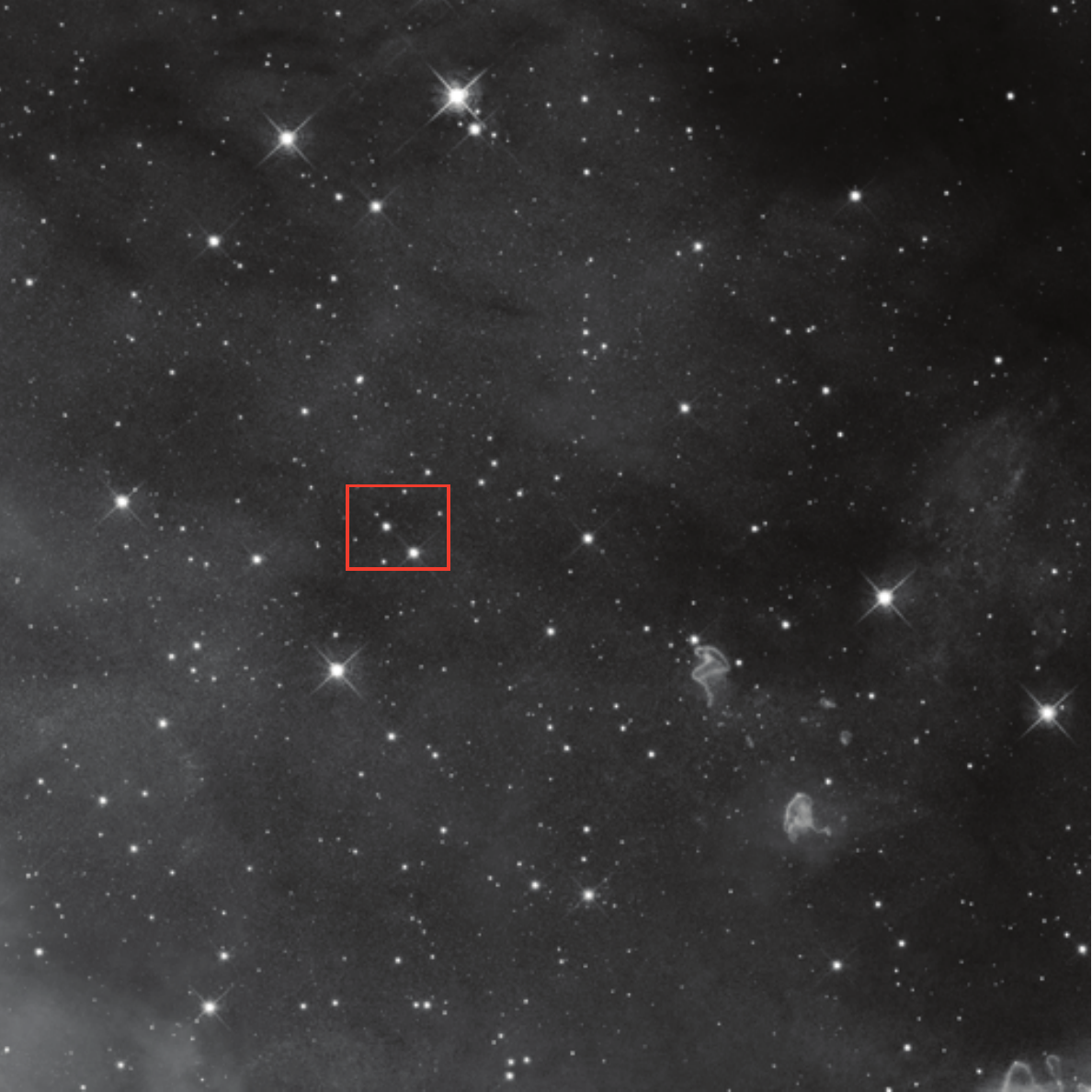}}
\subfigure[][]{%
\label{pics:src}
\includegraphics[width=0.6\columnwidth]{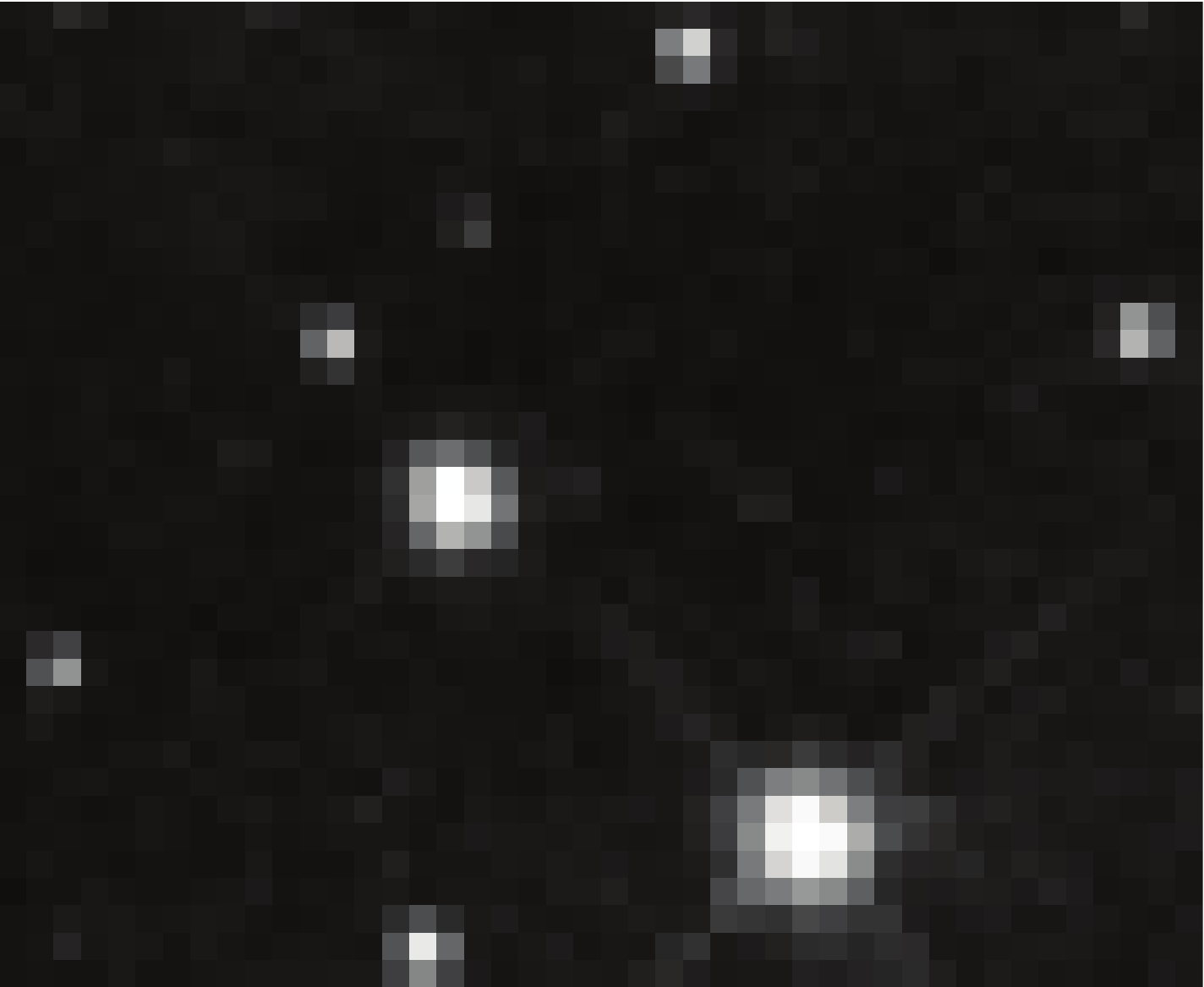}}
\subfigure[][]{%
\label{pics:n}
\includegraphics[width=0.6\columnwidth]{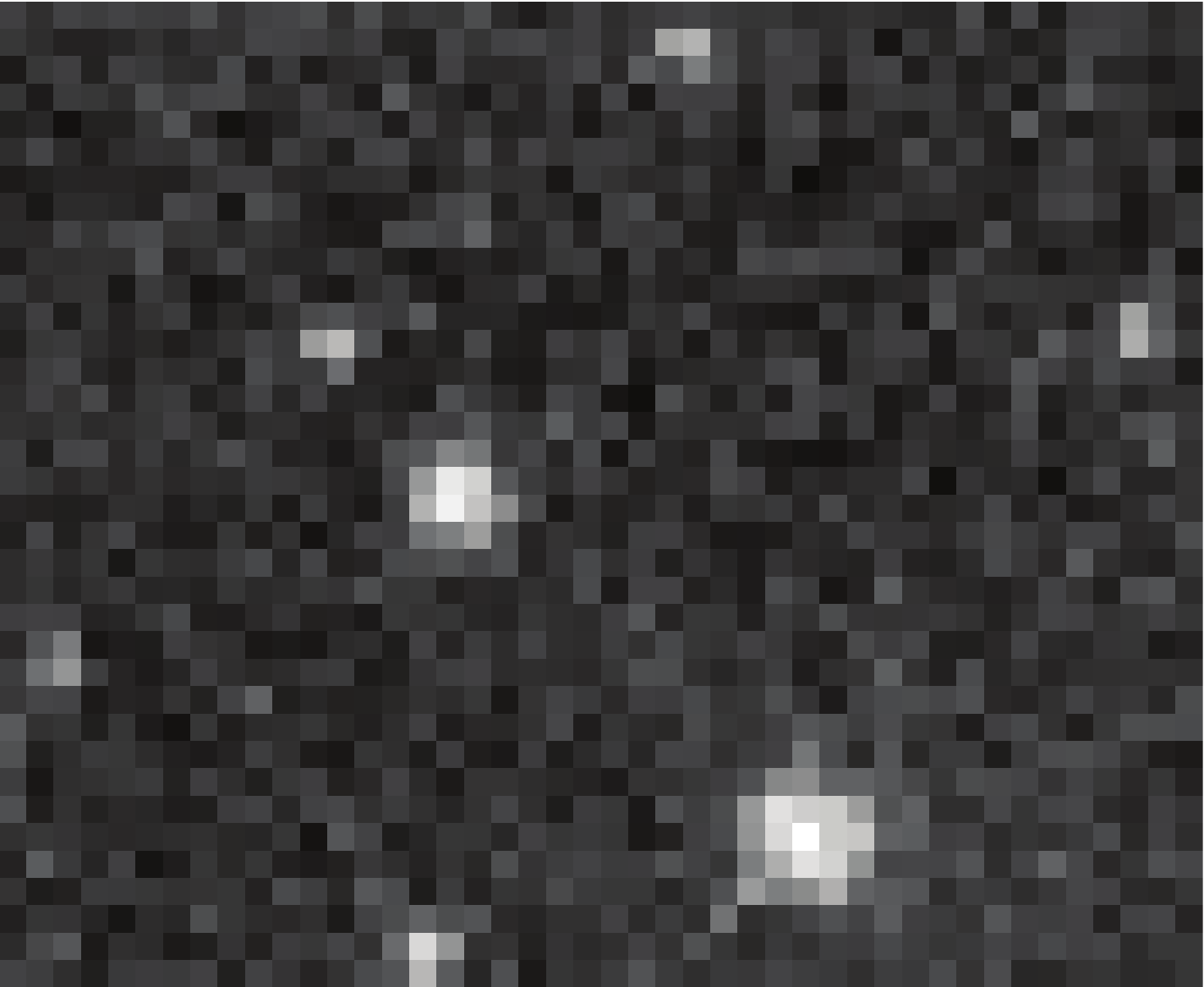}}
}
\hbox{
\subfigure[][]{%
\label{pics:w}
\includegraphics[width=0.6\columnwidth]{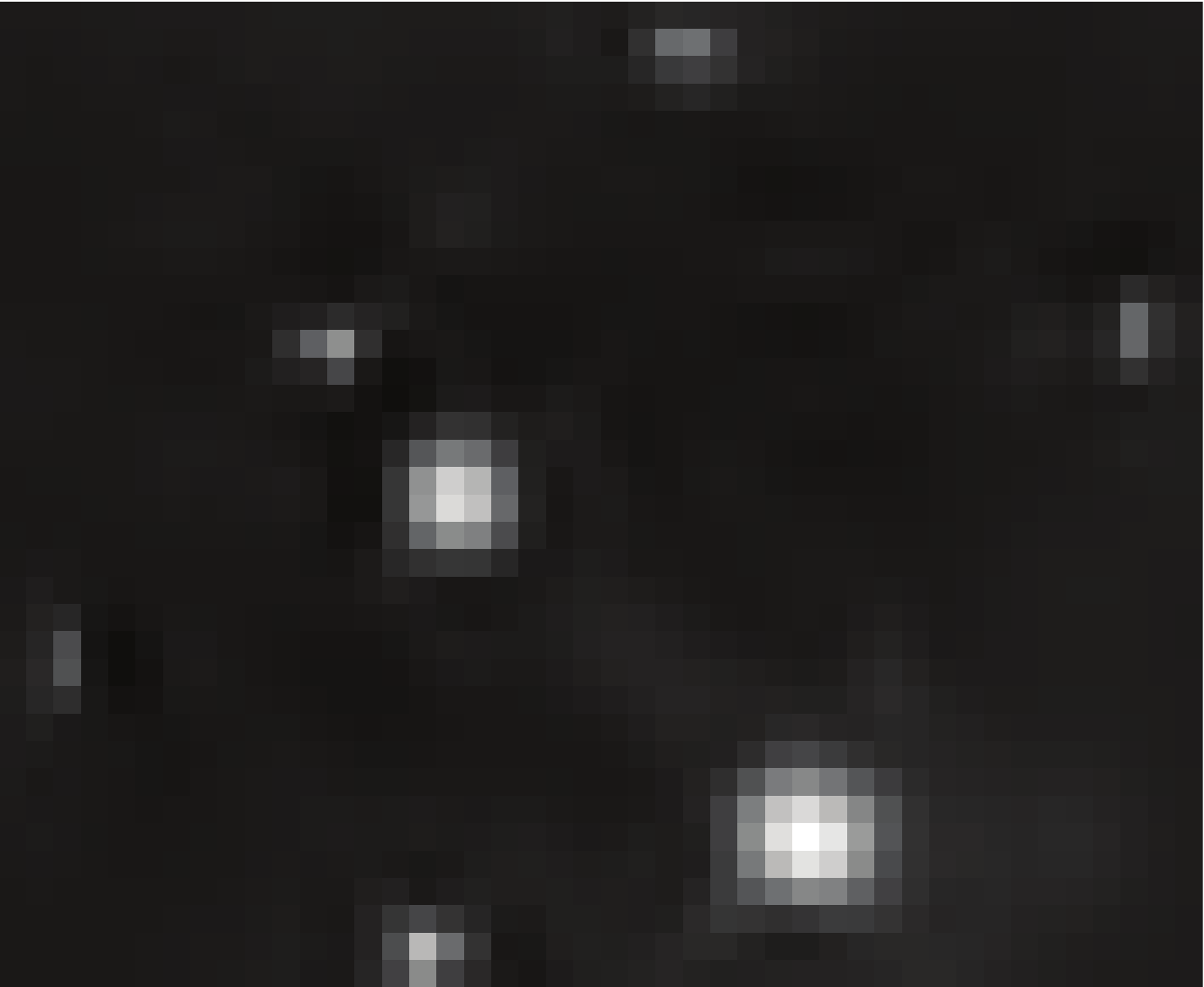}}
\subfigure[][]{%
\label{pics:dl}
\includegraphics[width=0.6\columnwidth]{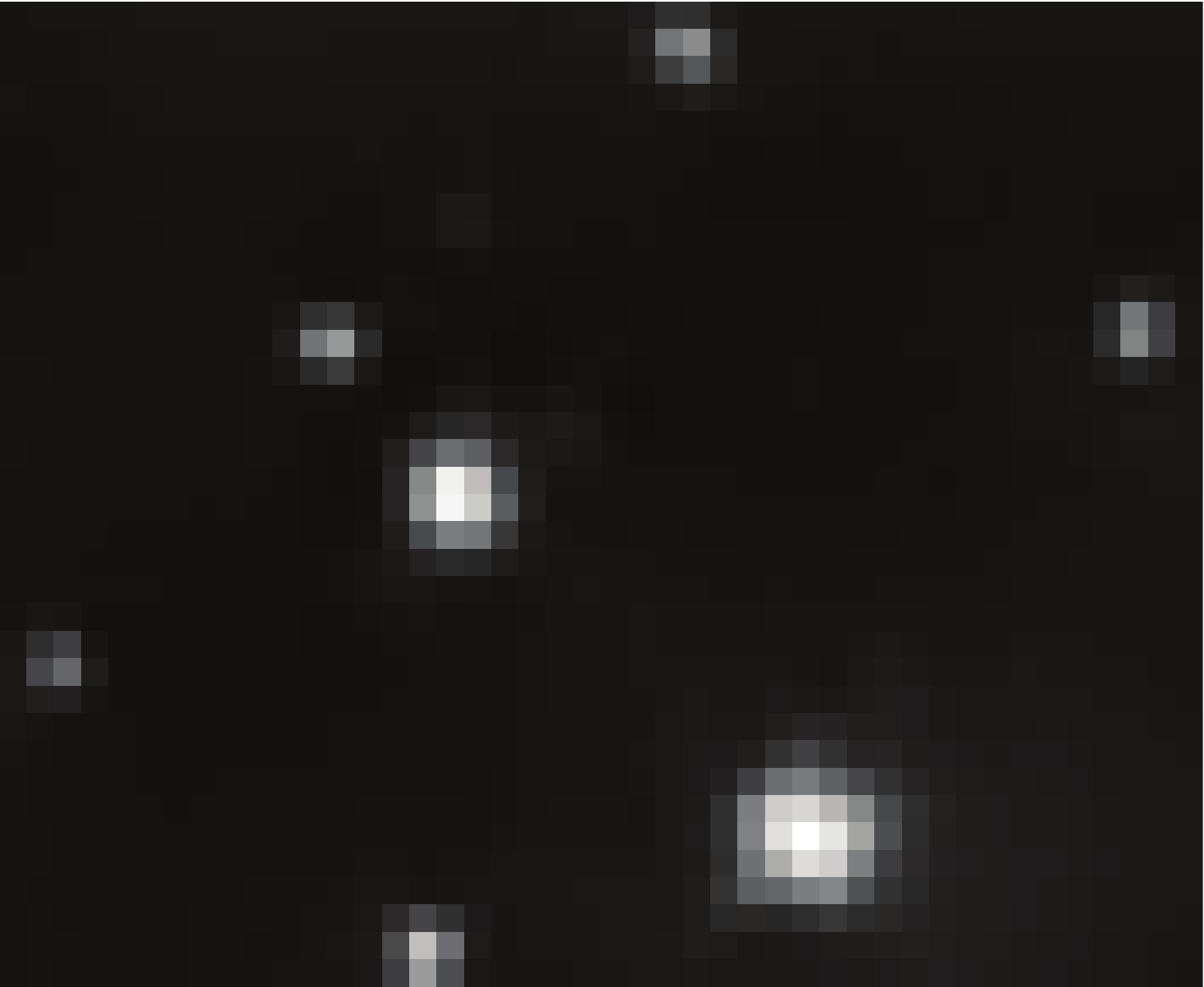}}
\subfigure[][]{%
\label{pics:dlc}
\includegraphics[width=0.6\columnwidth]{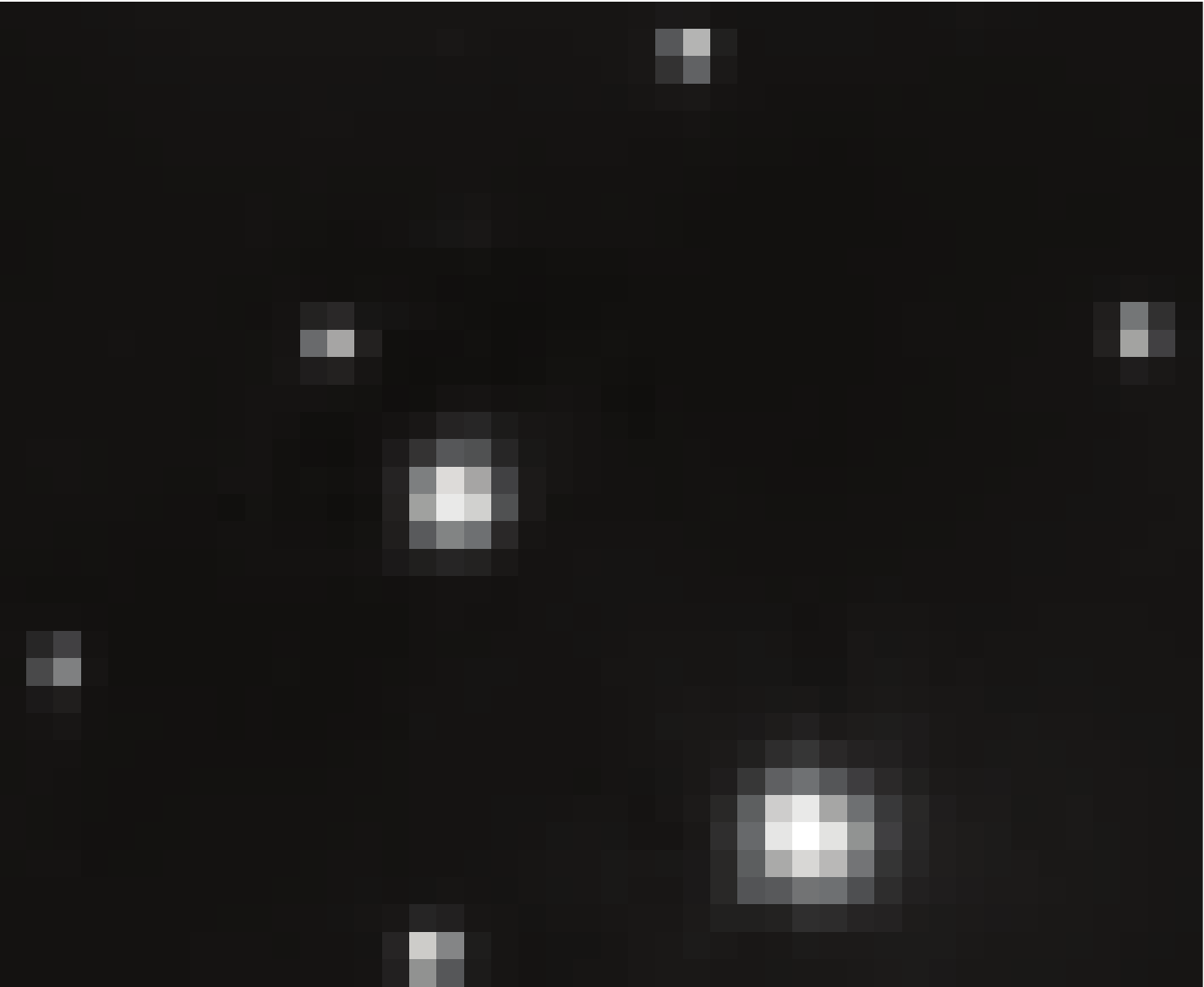}}
}}}
\caption{Zoomed features extracted from the previously shown nebular image. \subref{pics:original} shows the full source image before adding noise, \subref{pics:src} shows the noiseless source, \subref{pics:n} shows the noisy version, and \subref{pics:w}, \subref{pics:dl} and \subref{pics:dlc} respectively show the denoised feature using wavelet denoising, classic dictionary learning and centered dictionary learning.}
\label{fig:zoom_neb_2}
\end{figure*}

We provide in this section a photometric comparison of the wavelet and dictionary learning denoising algorithms. We use the top left quarter of the nebula image from \ref{fig:nebula_denoising}. We run Sextractor \citep{sextractor} using a $3\sigma$ detection threshold on the noiseless image, and we store the detected sources with their respective flux. We then add a white Gaussian noise with a standard deviation of $0.07$ to the image which has a standard deviation of 0.0853 (SNR = $10.43$ dB), and use the different algorithms to denoise it. We then use Sextractor, using the source location stored from the clean image analysis and processing the denoised images. We show in Figure \ref{fig:photometry_curves} two curves. The first one is the number of sources in the image with a flux above a varying threshold, for the original, wavelet denoised and CDL denoised images. The second curve shows how the flux is dampened by the different denoising methods. We also show in Figures \ref{fig:zoom_lensing}, \ref{fig:zoom_neb_1} and \ref{fig:zoom_neb_2} several features after denoising the galaxy cluster images using the different methods. It appears that the centered dictionary learning denoising restores objects with better contrast, less blur, and is more sensitive to small sources. We finally give several benchmarks to show how the centered dictionary learning is able to overcome the classic approach.

The learned dictionary based techniques show a much better behavior in term of flux comparison. This is consistent with the aspect of the features showed in Figures \ref{fig:zoom_lensing}, \ref{fig:zoom_neb_1} and \ref{fig:zoom_neb_2}. The CDL method induces less blurring of the sources  and is more sensitive to point-like features.

\section{Software}
We provide the matlab functions and script related to our numerical experiment at the URL http://www.cosmostat.org/software.html. 
\section{Conclusion}
We introduce a new variant of dictionary learning, the centered dictionary learning method, for denoising astronomical observations. Centering the training patches yields an approximate translation invariance inside the patches and lead to significant improvements in terms of global quality as well as photometry or feature restoration. We conduct a comparative study of different dictionary learning and denoising schemes, as well as comparing the performance of the adaptive setting to the state of the art in this matter. The dictionary learning appears as a promising paradigm that can be exploited for many tasks. We showed its efficiency in astronomical image denoising and how it overcomes the performances of stat-of-the-art denoising algorithms that use non-adaptive dictionaries. The use of dictionary learning requires to chose several parameters like the patch size, the number of atoms in the dictionary or the sparsity imposed during the learning process. Those parameters can have a significant impact on the quality of the denoising, or the computational cost of the processing. The patch-based framework also brings additional difficulties as one have to adapt it to the problem to deal with. Some tasks require a more global processing of the image and might require a more subtle use of the patches than the sliding window used for denoising. 

\section*{acknowledgements}
The authors thank Gabriel Peyre for useful discussions.
This work was supported by the French National Agency for Research (ANR -08-EMER-009-01) and 
the European Research Council grant SparseAstro (ERC-228261).

\bibliographystyle{plain} 
\bibliography{DL2012_double} 
\end{document}